\documentclass[pra,aps,amsmath, amssymb, twocolumn, superscriptaddress]{revtex4-1}

\usepackage{graphicx}
\usepackage{hyperref}
\usepackage{physics}
\usepackage{color}
\usepackage{enumerate}
\usepackage{bm}
\usepackage[normalem]{ulem}

\usepackage[caption=false]{subfig}

\usepackage{natbib}
\usepackage{graphicx}
\usepackage{braket}
\usepackage{amsmath}

\usepackage{amsthm}
\usepackage{amsfonts, amssymb}
\usepackage{bbm}
\usepackage{hyperref}
\newtheorem{proposition}{Proposition}
\newtheorem{corollary}{Corollary}

\newtheorem{lemma}{Lemma}
\usepackage{mathtools}
\usepackage{mathrsfs}
\usepackage{verbatim}

\DeclarePairedDelimiterX{\infdivx}[2]{(}{)}{%
	#1\;\delimsize\|\;#2%
}

\begin{document}

	\title{Extracting work from correlated many-body quantum systems}
	\author{Raffaele Salvia} 
	\email[Corresponding author: ]{raffaele.salvia@sns.it}
	\affiliation{Scuola Normale Superiore and University of Pisa, I-56127 Pisa, Italy}
	
	\author{Vittorio Giovannetti}
	\affiliation{NEST, Scuola Normale Superiore and Istituto Nanoscienze-CNR, I-56127 Pisa, Italy}

\begin{abstract}
The presence of correlations in the input state of a non-interacting many-body quantum system can lead to an increase in the amount of work we can extract from it under global unitary
 processes (ergotropy). 
The present work explore  such effect on translationally invariant systems relaying on the Matrix Product Operator formalism to define 
 a measure of how much they are  correlated. 
 We observe that in the thermodynamic limit of large number of sites, complete 
 work extraction can be attained for relatively small correlation strength (a reduction of a 2 factor in dB unit). Most importantly such an effect appears not to be associated with the presence of quantum correlations  (e.g. entanglement) in the input state (classical correlation suffices), and to be attainable by only using 
 {\it incoherent ergotropy}. As a byproduct of our analysis we also present a 
 rigorous formulation 
 of the heuristic typicality argument first formulated in [Alicki and Fannes, 2013], which gives the maximum work extractable for a set of many identical quantum systems in the asymptotic limit. \end{abstract}

	\maketitle
	
	\section{Introduction}
For purely classical models,  the maximum work $\Delta W^{(\max)}_{\rm{clas}}$ that can be extracted from a closed system by means of 
reversible, iso-entropic cycles 
is equal to the difference between the input energy $E_{in}$  and the energy $E_{th}$  
of the thermal equilibrium state that has the same entropy of the initial configuration, i.e. $\Delta W^{(\max)}_{\rm{clas}}=E_{in} - E_{th}$~\cite{Thomson1852,Fermi1955}. The corresponding quantity for a quantum system with Hamiltonian $H$, initialised in a state $\rho$, is called~\emph{ergotropy} $\mathcal{E}(\rho; H)$~\cite{Allahverdyan2004} and is 
 formally defined as the maximal amount of energy one can recover from the system by means of reversible coherent (i.e. unitary) processes. In general $\mathcal{E}(\rho; H)$ is strictly smaller than 
 the threshold one would get from a naive translation of the classical optimal term $\Delta W^{(\max)}_{\rm{clas}}$, i.e. the difference 
 $E(\rho; H) - \mathfrak{E}_{S(\rho)}$  between the mean energy $E(\rho; H):=\mbox{Tr}[H \rho]$ of the input state $\rho$, and the mean energy  $\mathfrak{E}_{S(\rho)}:=\mbox{Tr}[H \omega_\beta ]$ of
 the  thermal Gibbs state $\omega_\beta := e^{-\beta H}/Z_{\beta}$  that has the same von Neumann entropy $S(\rho)$. 
Nonetheless, thanks to the fact that $\mathcal{E}(\rho; H)$ is an extensive super-additive functional, the gap between the  work extractable  at the quantum level and $E(\rho; H) -  \mathfrak{E}_{S(\rho)}$   can  be progressively reduced by allowing
joint operations on an increasing collection of identical copies of the system. 
In particular,  indicating 
with 
 $H^{(N)}$ the global Hamiltonian of the $N$ copies written in terms of a sum of independent, uniform,
local terms (see Eq.~(\ref{HMDEF}) below),  it turns out that the  ergotropy per-site ${\mathcal{E}(\rho^{\otimes N}; H^{(N)})}/{N}$ of the model 
 is a non-decreasing function of $N$ whose asymptotic value (a quantity called from time-to-time {\it total
ergotropy}~\cite{Niedenzu2019})
  verifies the identity~\cite{AlickiFannes2013} 
 \begin{eqnarray}
\mathcal{E}_{\text{tot}}(\rho; H) &:=& 
\lim_{N \to \infty} \frac{\mathcal{E}(\rho^{\otimes N}; H^{(N)})}{N}
\nonumber \\
&=& E(\rho; H) -  \mathfrak{E}_{S(\rho)}\;.
\label{asymptotic_ergotropyy}
\end{eqnarray}
The limit~(\ref{asymptotic_ergotropyy}) corresponds to the saturation of the classical threshold for 
reversible, iso-entropic work extraction processes involving a quantum system. In all non trivial cases it
is explicitly attained  for an infinite number of copies, the only exceptions being associated to the cases where
$\mathcal{E}(\rho^{\otimes N}; H^{(N)})=0$ for all $N$.
A direct consequence of this observation is that when operating on 
non correlated input quantum system $\rho^{\otimes N}$ we can extract less work than in the classical case.
Such limitation however does not necessary hold if we allowed quantum correlations in the input state of the system: indeed  if we permit  
 operations that act on the entire many-body system,   
 the classical threshold~(\ref{asymptotic_ergotropyy}) can be overcame  allowing for the extraction of additional work~\cite{Huber2015, Francica2017}  (Notice that this is  the opposite of what happens if instead we are  forced to act only on
 individual parts of the quantum state: in this scenario in fact 
 correlations typically are detrimental for work extraction see e.g. Refs.~\cite{AlickiFannes2013,OPPE2002,VITA2019,GOOLD2017,BERA2017,MANZ2019,ANDO2019}).
Aim of  the present article is to investigate this issue by studying the ergotropy  $\mathcal{E} (\rho^{(N)}; H^{(N)})$  of correlated multipartite quantum states $\rho^{(N)}$ of 
 translationally invariant many-body quantum systems composed by $N$ sites. 
To express how much the state $\rho^{(N)}$ is correlated, we will use as a measure the minimum bond link rank $M$ necessary to represent $\rho^{(N)}$ as a Matrix Product Operator~\cite{Cirac2004MPO}: indicating with 
$r$  the rank (number of strictly positive eigenvalues) of the single-site density matrix $\rho$ of the model, 
the quantity $M$
 can range from $1$ (corresponding to factorized state $\rho^{(N)} =\rho^{\otimes N}$), to $r^2$ 
 (corresponding to the case of where $\rho^{(N)}$ is a pure GHZ entangled state).  Via constructive examples, we shall hence produce lower bounds for the maximum 
 ergotropy per site 
$\mathcal{E}(\rho^{(N)}; H^{(N)})/N$ attainable in the system for fixed values of the correlation parameter $M$,
comparing them to the classical  threshold limit~(\ref{asymptotic_ergotropyy}) reachable when the correlations are removed from the model.  

Our main finding is to show that, while forcing  $\rho^{(N)}$ to be pure by taking $M=r^2$ 
allows one to trivially boost $\mathcal{E}(\rho^{(N)}; H^{(N)})/N$ to its natural upper bound  (i.e. the single-site mean energy $E(\rho;H)$),
in the limit of large $N$ one can asymptotically reach the same result for much smaller values of $M$. In particular we prove that setting $M=r$ is enough, corresponding to a reduction of a factor 2 in dB unit. Most importantly it turns out that while for $M=r^2$ the saturation to $E(\rho;H)$ relays explicitly on the presence of the quantum correlations (entanglement) that forces  $\rho^{(N)}$ to be pure, for $M=r$ the same can be attained by only exploiting classical correlations. Our analysis shows also that  this effect  
can be attained through operations which do no affect the coherence of the input state~\cite{Baumgratz2014}: using the notation introduced in
~\cite{Francica2020, Cakmak2020,Touil2021}, this means that 
the asymptotic saturation of $\mathcal{E}(\rho^{(N)}; H^{(N)})/N$ to $E(\rho;H)$ we report here for $M=r$ can be obtained by just using~\emph{incoherent ergotropy}. 

We conclude mentioning that as a byproduct of our study we provide 
 a rigorous derivation of the heuristic typicality argument first formulated in \cite{AlickiFannes2013}, which gives the maximum work extractable for a set of identical quantum systems in the asymptotic limit of a large number of copies $N$.
 \\

The manuscript is organized as follows.
In Sec.~\ref{sec:prelim} we set the problem introducing the model, presenting the
notation, and giving the basic definitions that will be used in the remaining of the paper.
Also in an effort to improve readability, in  Sec.~\ref{sec:sum} we present a brief technical summary of the main results, with comments and general considerations. 
With Sec.~\ref{FFDA} we enter into the technical part of the manuscript: here in 
Sec.~\ref{prima} and \ref{seconda} 
we give a
detailed account of the Matrix Product Operator representation of translationally invariant states, while in Sec.~\ref{terza} and \ref{Quartanew} we focus on special families of the states
that, in the forthcoming sections will be adopted  to provide estimations for the attainable values of  $\mathcal{E}(\rho^{(N)}; H^{(N)})/N$ for given bond link rank $M$. 
The proofs of the results  
are presented in Sec.~\ref{sec:claims}, and conclusions are drawn in Sec.~\ref{CONCLUSIONS}. The paper contains also a series of technical appendices. 

\section{The problem}
\label{sec:prelim}
\begin{figure}
	\includegraphics[width=\columnwidth]{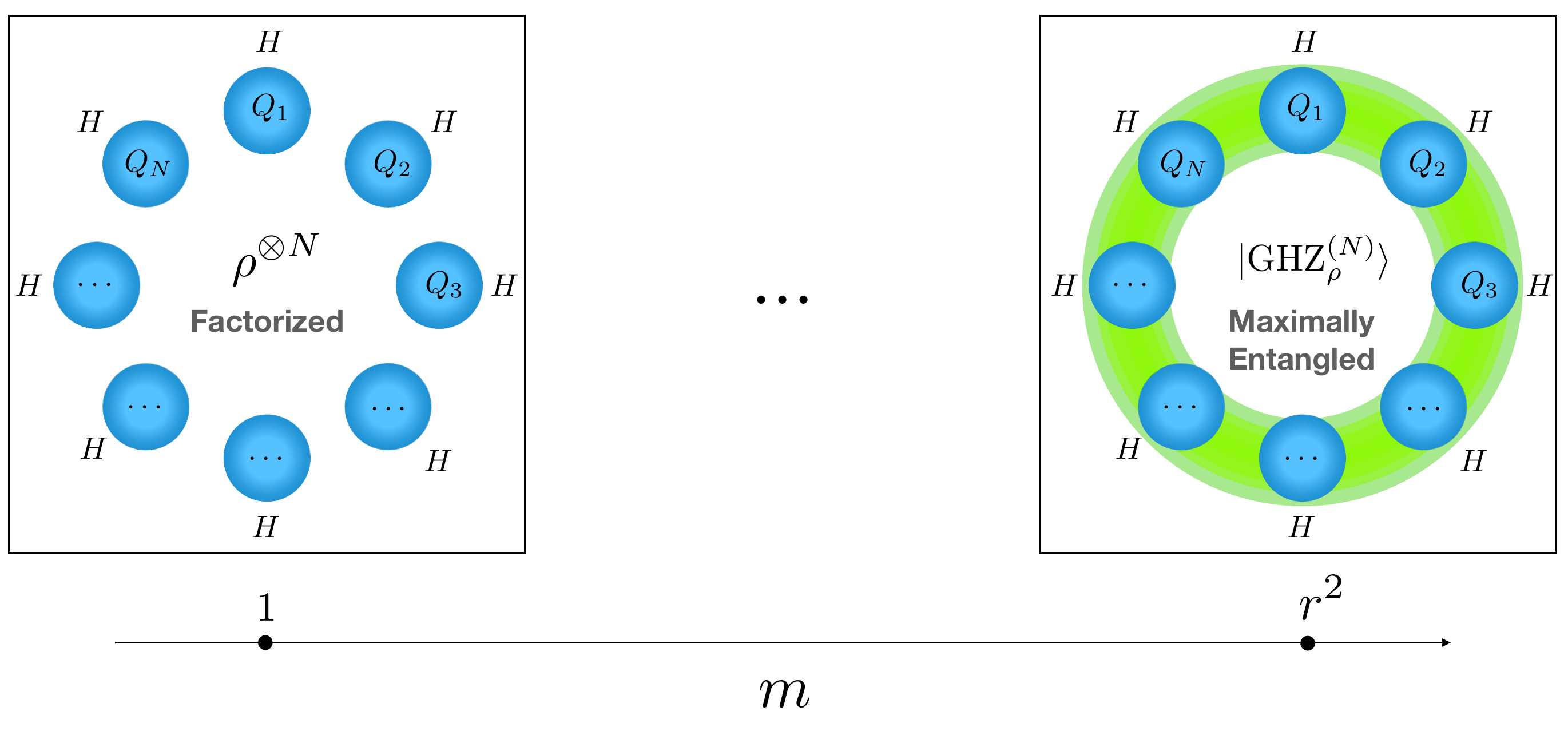}
	\caption[Pictorial rep of model]{Pictorial representation of the model: 
	we consider a collection of $N$ identical quantum systems $Q_1$, $Q_2$, 
	$\cdots$, $Q_N$ initialized into (possibly correlated) density matrices $\rho^{(N)}$ that are invariant under translation of the subsystem indexes. 
	The system Hamiltonian consists into a sum of homogeneous local terms 
	with no interactions. 
	Fixing the single-site density matrix $\rho$ of the model 
	(see Eq.~(\ref{cond_traccia_parziale}),
	the degree of intra-site correlations of $\rho^{(N)}$ are measured
	in terms of  the  parameter $m$ which defines the
	state BLR value~(\ref{MPOBLR}): for $m=1$, $\rho^{(N)}$ corresponds to the factorized state $\rho^{\otimes N}$, while for $m$ equal to the square of the rank $r$
	of $\rho$, we obtain the maximally entangled state $|\mbox{GHZ}^{(N)}_\rho\rangle$ of Eq.~(\ref{GHZ}). 
	}
	\label{figuramod}
\end{figure}

In our analysis we shall focus on a many-body quantum system consisting on an ordered collection $Q_1$, $Q_2$, $\cdots$, $Q_N$ of $N$ $d$-dimensional sites -- see Fig.~\ref{figuramod}. Indicating with ${\cal H}$ and ${\cal H}^{\otimes N}$ the
single-site and many-body Hilbert spaces, we represent with 
$\mathfrak{L}$ and $\mathfrak{L}^{(N)}$ the associated spaces of the 
linear operators, and with 
 $\mathfrak{S}$ and $\mathfrak{S}^{(N)}$
the sets of the corresponding density matrices. 
Assuming no interactions among the various sub-systems, 
we write the 
joint Hamiltonian of the model as a sum of homogenous local terms 
\begin{eqnarray} \label{HMDEF} 
H^{(N)} := \sum_{\ell =1}^N H_\ell\;, 
\end{eqnarray}  
with $H_\ell$ representing the same single-site operator $H\in \mathfrak{L}$ acting on the  $\ell$-th subsystem $Q_\ell$. Without loss of generality in what follows  we shall take $H$ (hence also $H^{(N)}$) to be positive semidefinite and to  have zero ground state eigenvalue; in particular we shall use the symbols $\epsilon_j$ to represent its eigenvalues that we order via the inequalities,
\begin{eqnarray} \label{ENERGYSPECTRUM} 
\epsilon_d \geq \epsilon_{d-1} \geq \cdots\geq  \epsilon_2 \geq \epsilon_1 =0\;, 
\end{eqnarray}  
and the symbols $|\epsilon_i\rangle$ to represent their associated eigenvectors, i.e 
\begin{eqnarray}
H |\epsilon_j\rangle = \epsilon_j |\epsilon_j\rangle. 
\end{eqnarray} 
Given hence $\rho^{(N)}\in \mathfrak{S}^{(N)}$ a generic (possible correlated) quantum state of the compound we 
write  its ergotropy as~\cite{Allahverdyan2004}
\begin{equation}
\label{correlated_ergotropy}
{{\cal E} (\rho^{(N)}; H^{(N)})} := E(\rho^{(N)}; H^{(N)}) - \min_{U } E(U\rho^{(N)}U^\dagger; H^{(N)})\;,
\end{equation}
where the minimization is performed over all transformation $U\in {\bf U}({\cal H}^{\otimes N})$  of the unitary set on ${\cal H}^{\otimes N}$, and where \begin{equation}
E(\rho^{(N)}; H^{(N)}):= \Tr[\rho^{(N)}H^{(N)}] \;,
\end{equation}
represents the energy expectation values on $\rho^{(N)}$ which by construction is non-negative.

In this paper we will be concerned on 
 $\rho^{(N)}$ belonging to the special subset $\mathfrak{S}_T^{(N)}$ of 
 $\mathfrak{S}^{(N)}$ formed by density matrices 
that are invariant under ciclic translations of the indexes sites,~i.e. 
 \begin{eqnarray} \label{dfdf} 
 \rho^{(N)} = T \rho^{(N)} T^\dag \;, 
 \end{eqnarray} 
 where $T$ is the transformation on ${\cal H}^{\otimes N}$ which for all $\ell=1,\cdots, N$ maps $Q_\ell$ into $Q_{\ell\oplus 1}$ (here $\oplus$ represents the sum modulus $N$).
 Since by construction  these states  
 are locally uniform, we can associate to each one of them the same
  single-site reduced density matrix \begin{equation}
\label{cond_traccia_parziale}
\Tr_{Q_1 \dots Q_{\ell-1}Q_{\ell+1} \dots Q_N}[\rho^{(N)}] = \rho\;,  \quad \forall \ell \in [1, N]\;, 
\end{equation}
  that allows us to express the global mean energy of the system as
\begin{equation} \label{LOCH} 
E \left( \rho^{(N)}; H^{(N)} \right) = N E(\rho; H) \; . 
\end{equation}
Factorized density matrices  $\rho^{(N)} = \rho^{\otimes N}$ are special instances of 
  $\mathfrak{S}_T^{(N)}$ which, due to the absence of  
  correlations among the various subsystems,  
  are fully characterized by their single-site counterpart~(\ref{cond_traccia_parziale}). 
A proper representation of $\mathfrak{S}_T^{(N)}$ can be obtained 
in terms of the \emph{translationally invariant matrix product operators} (TI-MPO) formalism~\cite{Cirac2004MPO}.
The key observation here is that given
$\{ \ket{i_1i_2 \dots i_N} = | i_1\rangle \otimes 
 | i_2\rangle\otimes  \cdots \otimes |i_N\rangle\}$   an orthonormal basis of ${\cal H}^{\otimes N}$ constructed from  the elements of reference, single-site basis $\{ \ket{i}; i=1,\cdots,d \}$, 
 for each  $\rho^{(N)}\in \mathfrak{S}_T^{(N)}$ 
 it is possible to associate a four rank tensor $\mathbb{A}\in \mathbb{C}^{d \times d \times M \times M}$ 
via the identity 
\begin{eqnarray}
\rho^{(N)} &=&
\sum_{i_1,j_1,\dots,i_N,j_N}   \ket{i_1 \dots i_N}\bra{j_1  \dots j_N} \nonumber \\
&&\;  \times 
\Tr[{A}^{i_1,j_1} \dots {A}^{i_N,j_N}] 
 \;,\label{MPO_rhoN}
\end{eqnarray}
with   $\{ A^{i,j}| {i,j\in1,\cdots d}\}$ being 
$d^2$ matrices of dimension $M \times M$ that we use to represent $\mathbb{A}$ -- see Sec.~\ref{prima} for an explicit derivation of this fact. 
In this construction 
the single-site counterpart~(\ref{cond_traccia_parziale}) of $\rho^{(N)}$ rewrites 
 \begin{equation}
\rho =
\sum_{i,j} \ket{i}\bra{j}  \;  \Tr[{A}^{i,j}{\bar{A}}^{N-1}]
 \;, \qquad \bar{A} :=  \sum_{i}{A}^{i,i},  \label{MPO_rhoN_red}
\end{equation}
while the  value of the parameter $M$  can be used to gauge the strength of the correlations among the various 
sites. 
For instance, as we shall see, setting $M=d^2$ one can construct multi-site GHZ entangled pure states (see Sec.~\ref{terza}), while to get  factorized states $\rho^{\otimes N}$ 
 it is sufficient to have $M=1$.  Notice however the correspondence between $\mathfrak{S}_T^{(N)}$ and the set of tensors $\mathbb{A}$ is  not
one to one. In particular determining whether the right-hand-side term of Eq.~(\ref{MPO_rhoN}) 
would produce a proper density matrix for a given $\mathbb{A}$ is a NP-hard problem, and it becomes undecidable
	in the thermodynamic limit~\cite{Kliesch2014}. 
	Most importantly for our purposes, multiple inequivalent choices of 
 $\mathbb{A}$, characterized by different values of $M$, can be assigned  to each $\rho^{(N)}\in \mathfrak{S}_T^{(N)}$.
To remove this ambiguity
 we define the \emph{TI-MPO bond link rank} (BLR) $\mathfrak{br}[\rho^{(N)}]$
  as the minimum $M$ required to represent the state $\rho^{(N)}\in \mathfrak{S}_T^{(N)}$ in the form~(\ref{MPO_rhoN})~\cite{Navascues2018}:
\begin{multline} \label{MPOBLR} 
\mathfrak{br} [\rho^{(N)} ]  := \min \left\{ M\geq 1 \; \middle\vert \; \exists \; \mathbb{A} \in \mathbb{C}^{d \times d \times M \times M} 
\right. \\ 
\left.
\mbox{s.t. Eq.~(\ref{MPO_rhoN}) holds true}
  \right\}\;.
\end{multline}
This quantity  is a proper functional of $\rho^{(N)}$: in particular, 
as explicitly shown in Sec.~\ref{seconda},  it does not depend on the specific choice of 
the local basis $\{\ket{i}; i=1,\cdots,d \}$ that we use to define the TI-MPO representation.
Unfortunately determining the exact value of $\mathfrak{br} [\rho^{(N)} ]$ is a rather challenging task. 
For our analysis however it will be sufficient to identify educated bounds for a special sub-set of states.

Equipped with the above definitions we can  now classify  the elements of $\mathfrak{S}_T^{(N)}$ 
in terms of their local properties and of their BLR.
Specifically, given $\rho \in \mathfrak{S}$ we first
define $\mathfrak{S}_T^{(N)}(\rho)$  as the subset of $\mathfrak{S}_T^{(N)}$ formed by 
states $\rho^{(N)}$ that admits $\rho$ as  single-site density matrix~(\ref{cond_traccia_parziale}), i.e. 
\begin{eqnarray} \label{NEWSETS1} 
&&\mathfrak{S}_T^{(N)}(\rho):= \Big\{ \rho^{(N)} \in 
\mathfrak{S}_T^{(N)}  \; \big\vert \;  \nonumber \\
&&\qquad \Tr_{Q_1 \dots Q_{\ell-1}Q_{\ell+1} \dots Q_N}  [\rho^{(N)}] =\rho\;,\;   \forall \ell \in [1,N] \Big\}\;, \nonumber 
\end{eqnarray} 
then we decompose $\mathfrak{S}_T^{(N)}(\rho)$ in terms of
the associated BLR, introducing the partitions 
\begin{eqnarray} \label{NEWSETS} 
&&\mathfrak{S}_T^{(N,m)}(\rho):= \Big\{ \rho^{(N)} \in 
\mathfrak{S}_T^{(N)}(\rho)   \; \big\vert \;
   \mathfrak{br}[\rho^{(N)}] \leq m 
 \Big\}\;. \nonumber 
\end{eqnarray} 
Notice that  for $m=1$, $\mathfrak{S}_T^{(N,1)}(\rho)$ contains only the factorized
state $\rho^{\otimes N}$, i.e.
\begin{eqnarray} \label{m1} 
&&\mathfrak{S}^{(N,1)}_{T}(\rho)= \Big\{ \rho^{\otimes N} \Big\}\;.
\end{eqnarray} 
On the contrary it can be shown that given $\mbox{r}[\rho]$ the rank of the local density matrix $\rho$, the set  $\mathfrak{S}_T^{(N,\mbox{r}^2[\rho])}(\rho)$  includes 
  the maximally entangled pure state 
 \begin{eqnarray} \label{GHZ} 
 |\mbox{GHZ}^{(N)}_\rho\rangle := \sum_{i=1}^{\mbox{r}[\rho]} \sqrt{\lambda_i} \; |  \lambda_i  \dots \lambda_i\rangle\;,
 \end{eqnarray} 
 with $\lambda_i> 0$ being the non-zero eigenvalues of $\rho$ and 
 $|\lambda_i\rangle$ being the associated eigenvectors -- an explicit proof of this fact is presented in Sec.~\ref{terza}. 
More generally  for fixed $\rho$, the $\mathfrak{S}_T^{(N,m)}(\rho)$'s  form a family of sets 
of increasing size which, in the non trivial case $\mbox{r}[\rho]\geq 2$, 
partition the collection of density matrices $\rho^{(N)}$ fulfilling the local
constraint~(\ref{cond_traccia_parziale}) into groups of  increasing multi-site correlation strength, i.e. 
\begin{equation} 
 \mathfrak{S}_T^{(N,m-1)}(\rho) \subseteq \label{hier} 
\mathfrak{S}_T^{(N,m)}(\rho) \;,  \quad \forall m\geq 2\;,
\end{equation} 
(for $\mbox{r}[\rho]=1$, i.e. when the local state is a pure, Eq.~(\ref{hier}) makes no sense
as in this case only $\mathfrak{S}_T^{(N,m=1)}(\rho)$ exists). 
 Our goal is to characterize  
the maximum ergotropy value that can be achieved 
 for a given degree $m$ of the correlation. For this purpose we define the functional  
\begin{eqnarray}\label{problema_rhoN_Mfissato}
&&\mathcal{E}^{(N,m)}_{\max}(\rho; H):=  \max_{\rho^{(N)} \in \mathfrak{S}_T^{(N,m)}(\rho)} \frac{ \mathcal{E}(\rho^{(N)}; H^{(N)}) }{N}\;,
\end{eqnarray}
that represents the maximum value of the ergotropy per-site that one can extract  from the system when it is in a joint state 
$\rho^{(N)}$ of $\mathfrak{S}_T^{(N,m)}(\rho)$. 
From (\ref{hier}) it follows that (\ref{problema_rhoN_Mfissato}) inherits the monotonic behaviour 
\begin{equation}
\mathcal{E}^{(N,m-1)}_{\max}(\rho; H)\leq \label{hier2} 
\mathcal{E}^{(N,m)}_{\max}(\rho; H) \;,  \quad \forall m\geq 2\;.
\end{equation}
Notice also that from (\ref{m1}) we trivially get  
\begin{equation}\label{lowerb} 
\mathcal{E}^{(N,1)}_{\max}(\rho; H)= \frac{ \mathcal{E}(\rho^{\otimes N}; H^{(N)}) }{N}  \;, 
\end{equation}
that represents a lower bond for all the other $\mathcal{E}^{(N,m)}_{\max}(\rho; H)$'s and which,
in the asymptotic limit of large $N$ approaches from below the 
classical threshold for the work extraction from the  single-site state $\rho$, given by 
the 
total ergotropy function~(\ref{asymptotic_ergotropyy}), i.e. 
\begin{equation}
\mathcal{E}^{(\infty,1)}_{\max}(\rho; H) := \lim_{N\rightarrow \infty} 
\mathcal{E}^{(N,1)}_{\max}(\rho; H)
  = \mathcal{E}_{\text{tot}}(\rho; H) \;.
  \label{asymptN1} 
\end{equation}
An upper bound for $\mathcal{E}^{(N,m)}_{\max}(\rho; H)$ can instead be easily obtained from~(\ref{correlated_ergotropy}) and (\ref{LOCH}) which allows us to rewrite~Eq.~(\ref{problema_rhoN_Mfissato}) in the equivalent form 
\begin{eqnarray}\label{problema_rhoN_Mfissato2}
&&\mathcal{E}^{(N,m)}_{\max}(\rho; H):=  E(\rho;H)  \\
\nonumber  
&&\qquad \quad -\min_{\rho^{(N)}  \in \mathfrak{S}_T^{(N,m)}(\rho)} 
\max_U \frac{ E(U\rho^{(N)} U^\dag ; H^{(N)}) }{N}\;.
\end{eqnarray}
Remembering then that  $H^{(N)}$ is positive semidefinite,  
we can hence  write 
\begin{eqnarray} \label{UPPER} 
\mathcal{E}^{(N,m)}_{\max}(\rho; H) \leq  E(\rho;H)\;,   \quad \forall m \;. \end{eqnarray}
Observe next that, for fixed $N$ and $\rho$, the inequality (\ref{UPPER}) it is certainly saturated for all 
 $m$  larger than or equal to the square of the rank of $\rho$, i.e. 
\begin{equation} \label{upperb} 
\mathcal{E}^{(N,m)}_{\max}(\rho; H) = E(\rho; H) \;, \quad \forall m\geq \mbox{r}^2[\rho]\;.
\end{equation}
This is a consequence of the mononicity relation~(\ref{hier2}) and of the fact that $\mathfrak{S}_T^{(N,\mbox{r}^2[\rho])}(\rho)$ contains at least the pure state
$|\mbox{GHZ}^{(N)}_\rho\rangle$ of Eq.~(\ref{GHZ}) which allows one to set equal to zero the negative term on the right-hand-side
of Eq.~(\ref{problema_rhoN_Mfissato2}) by choosing $U$ to be unitary transformation that moves such vector into the ground state  of $H^{(N)}$ (remember that in our model the ground state eigenvalue is equal to zero).

Determining how for fixed $N$ and $\rho$, 
one passes from the lower threshold  (\ref{lowerb})  to the upper  threshold (\ref{upperb}) by increasing the correlation parameter  $m$, is the focus  of the present work. 
For this purpose, in the following sections, we shall produce a series of lower bounds for $\mathcal{E}^{(N,m)}_{\max}(\rho; H)$, that at least in same regimes allows one to  determine its exact value.

\subsection{Summary of the main findings} \label{sec:sum}

In this section 
we anticipate the main results of the paper postposing the 
derivations in the second part of the work.  

 \begin{figure}
	\includegraphics[width=\columnwidth]{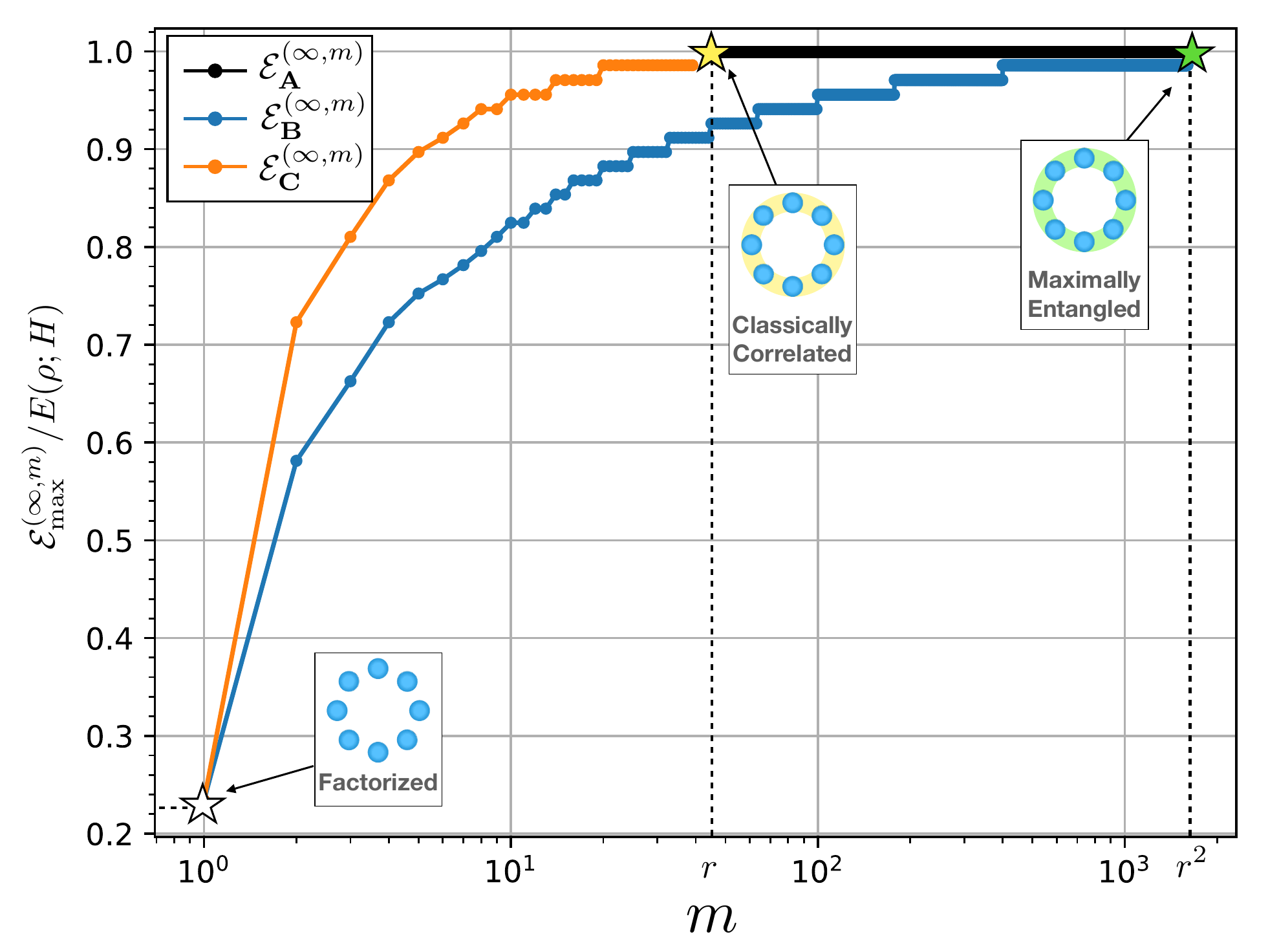}
	\caption[Asymptotic behaviour]{Plot of the lower bounds of the asymptotic ($N\rightarrow\infty$) per-site maximum many-body ergotropy  
	$\mathcal{E}^{(\infty,m)}_{\max}(\rho; H)$ rescaled by $E(\rho;H)$,
as a function of the BLR value $m$.
Notice that in agreement with~(\ref{flat}), $\mathcal{E}^{(\infty,m)}_{\max}(\rho; H)/E(\rho;H)$ saturates to the optimal value $1$ already when
$m$ coincide with the rank $r$ of the matrix $\rho$,  a with a reduction of a factor 2 in dB unit with respect to what one could expect from~(\ref{upperb}) that instead
predicts  this to happen for $m=\mbox{r}^2[\rho]$.  For smaller $m$  the our best estimation of the lower bound  (orange line in the plot) is the function $\mathcal{E}^{(\infty,m)}_{\bf C}(\rho; H)/E(\rho;H)$ of  Eq.~(\ref{toflat}); the blu curve is instead the lower bound $\mathcal{E}^{(\infty,m)}_{\bf B}(\rho; H)/E(\rho;H)$ of Eq.~(\ref{notflat}). 
	In this example data were obtained taking $H$ to be the Hamiltonian of a quantum system of dimension $d=40$ with randomly selected spaced level, and  $\rho$ a randomly selected, full rank ($r=d=40$) density matrix. 
	The white, yellow, and green stars in the plot represent, respectively,  the ergotropy values achieved on the factorized state $\rho^{\otimes N}$,
	the classically correlated state $\rho_{cc}^{(N)}$ (see Eq.~(\ref{rhoN_metodo2}) below), and the GHZ state $|\mbox{GHZ}^{(N)}_\rho\rangle$ of Eq.~(\ref{GHZ}). }
	\label{figure0}
\end{figure}
\subsubsection{Lower bounds for $\mathcal{E}^{(N,m)}_{\max}(\rho; H)$ and 
asymptotic saturation at $E(\rho;H)$ for $m=\mbox{r}[\rho]$} \label{seclower}
Our first observation is a lower bound for $\mathcal{E}^{(N,m)}_{\max}(\rho; H)$ which, for $N$ sufficiently large, cover the full spectrum of the BLR values, providing an interpolation between 
Eqs.~(\ref{lowerb}) and (\ref{upperb}): 
\begin{proposition}[Preliminary bound]
\label{prop:method1}
Given $\eta \in ]0,1[$ and a single-site density matrix $\rho$ of rank $\mbox{r}[\rho]=r$, for $m \leq {r}^2$ and $N$ sufficiently large the following inequality holds true
\begin{eqnarray} 
&& \mathcal{E}^{(N,m)}_{\max}(\rho; H)  \label{bound_ergotropia_prop1NEW} \\
&&\quad \geq E(\rho; H)-\mathfrak{E}_{s_{\bf B}} - (2 \sqrt{2.01 C \eta}
+ \epsilon_d 
e^{-\frac{2N\eta^2}{\alpha^2}}) \;,\nonumber
\end{eqnarray} 
where $\epsilon_d$ is maximum eigenvalue (\ref{ENERGYSPECTRUM}) of the single-site Hamiltonian $H$,  
 $\mathfrak{E}_{s}$ is the mean energy  of a single-site  thermal Gibbs state  with entropy $s\geq 0$  (see Eq.~(\ref{def_frakE}) in the Appendix), 
\begin{eqnarray}\label{sBdef} 
s_{\bf B} :=  \ln \left \lceil{r / \sqrt{m}} \;\right \rceil + N^{-1} \ln \sqrt{m} \;, 
\end{eqnarray}
while finally $C$ and $\alpha$ are finite, positive quantities which depend upon $\rho$ and $H$ only.
\end{proposition}

 The derivation of this result  is reported in Sec.~\ref{sec:method1} by focusing on a special sub-class of TI-MPO states (the  {\it block-wise purified} density matrices $\rho^{(N)}_{{\bf B}}$ introduced in Sec.~\ref{sec:method1_stato}). 
Taking  the $N\rightarrow \infty$ limit for $\eta$ assigned and then sending the latter to zero,
from Eq.~(\ref{bound_ergotropia_prop1_2NEW}) we can extrapolate the following simplified inequality
\begin{equation}\label{notflat} 
\mathcal{E}^{(\infty,m)}_{\max}(\rho; H) := \lim_{N\rightarrow \infty} 
\mathcal{E}^{(N,m)}_{\max}(\rho; H) \geq  \mathcal{E}^{(\infty,m)}_{\bf{B}}(\rho; H) 
  \;, \end{equation}
 which applies for all $m \leq  \mbox{r}^2[\rho]$ with 
 \begin{eqnarray}  \label{defEB} 
 \mathcal{E}^{(\infty,m)}_{\bf{B}}(\rho; H)  &:=& E(\rho;H) - \mathfrak{E}_{ \ln \left\lceil{\mbox{r}[\rho]  / \sqrt{m}}\;\right \rceil}\;.
\end{eqnarray} 
Notice that  since $\mathfrak{E}_{s}=0$ when $s=0$, for $m=\mbox{r}^2[\rho]$, $\mathcal{E}^{(\infty,m)}_{\bf{B}}(\rho; H)$ reduces to $E(\rho;H)$ and 
  Eq.~(\ref{notflat})  leads to~(\ref{upperb}).

Our next result is a refinement of {\bf Proposition~\ref{prop:method1}}. 
To begin with we observe that 
an improved version of Eq.~(\ref{notflat}) can be found 
 in the low BLR regime $m \leq \mbox{r}[\rho]$:
 \begin{proposition}[Low BLR regime] 
 \label{prop:method2}
	Given $\eta \in ]0,1[$ and a single-site density matrix $\rho$ of rank $r$, for 	 $m \leq {r}$ and $N$ sufficiently large the following inequality holds true
\begin{eqnarray} 
&& \mathcal{E}^{(N,m)}_{\max}(\rho; H)  \label{bound_ergotropia_prop1_2NEW} \\
&&\quad \geq E(\rho; H)-\mathfrak{E}_{s_{\bf C}} - (2 \sqrt{2.01 C \eta}
+ \epsilon_d 
e^{-\frac{2N\eta^2}{\alpha^2}}) \;,\nonumber
\end{eqnarray} 
where $\epsilon_d$, $C$, and $\alpha$ are as in {\bf Proposition~\ref{prop:method1}} while now 
\begin{eqnarray}\label{DEFSC} 
s_{\bf C}:= \ln \left \lceil{r / {m}} \;\right \rceil + N^{-1} \ln {m}\;.
\end{eqnarray} 
\end{proposition}
 The derivation of this result  is reported in Sec.~\ref{sec:method2}: interestingly enough this is done by focusing on a special class of TI-MPO states 
 (i.e. the set 
 $\rho^{(N)}_{{\bf C}}$ of Sec.~\ref{sec:method2_stato}) which represent classically correlated (not entangled) density matrices of the system which can be chosen
 to be diagonal in the energy eigenbasis. 
Similarly to what done for {\bf Proposition~\ref{prop:method1}} the analysis simplifies when 
taking   the  proper $N\rightarrow \infty$ limit.
In this case from Eq.~(\ref{bound_ergotropia_prop1_2NEW}) we get the  inequality
\begin{equation}\label{toflat} 
\mathcal{E}^{(\infty,m)}_{\max}(\rho; H) \geq 
 \mathcal{E}^{(\infty,m)}_{\bf{C}}(\rho; H) \;, \qquad \forall m \leq  \mbox{r}[\rho] \;, 
\end{equation} 
which for 
 \begin{eqnarray}  \label{defEC} 
 \mathcal{E}^{(\infty,m)}_{\bf{C}}(\rho; H)  &:=&E(\rho;H) - \mathfrak{E}_{ \ln \left\lceil{\mbox{r}[\rho]  / {m}}\right \rceil}\;,
\end{eqnarray} 
represents a clear improvement with respect to the constraint imposed by Eq.~(\ref{defEB}) -- see Fig.~\ref{figure0}. 
Notice in particular that since $\mathcal{E}^{(\infty,\mbox{r}[\rho])}_{\bf{C}}(\rho; H) = E(\rho;H)$,  
 Eq.~(\ref{toflat})  predicts a saturation of the upper bound (\ref{UPPER}) already for 
  $m=\mbox{r}[\rho]$, i.e. for values of $m$ which are much smaller than those suggested by (\ref{upperb}). Even more interesting is the fact that such asymptotic saturation is attained with states (the density matrices $\rho^{(N)}_{{\bf C}}$) 
  which, as already mentioned, contain no quantum correlations and which can be
  chosen to the diagonal with respect to the energy eigenbasis. The first property
  of the $\rho^{(N)}_{{\bf C}}$'s implies that, at least in the asymptotic regime of infinitely many sites, classical correlations are sufficient to enable full work extraction from the system. The second property instead tell us that this can be done via incoherent ergotropy 
  ~\cite{Francica2020, Cakmak2020,Touil2021}.
  
  To clarify the asymptotic attainability of the upper bound (\ref{UPPER}) for 
  large $m$ we present a final statement that applies for 
 BLR values that larger than or equal to $\mbox{r}[\rho]$:

\begin{proposition}[High BLR regime] 
	\label{prop:grandi_blr} Given  $N$ integer and 
a single-site density matrix $\rho$ of rank $r$, for $m\in\{ r, \cdots ,r^2\}$ the following inequality holds
\begin{equation} \label{upperbmgrandi} 
\mathcal{E}^{(N,m)}_{\max}(\rho; H) 
\geq E(\rho; H) - \frac{\Delta}{N} \; \;,
\end{equation}
with $\Delta$
being a positive constant term which depends upon the spectra of $\rho$ and
 $H$. 
\end{proposition}
The proof of this result which at variance with {\bf Propositions~\ref{prop:method1}} and {\bf \ref{prop:method2}}
does not require to have $N$ large, 
is given in Sec.~\ref{dfds} by focusing on the special class of (non pure) quantum states $\rho_{\bf A}^{(N)}$ that correspond to proper mixtures of multi-sites GHZ states (see  
Sec.~\ref{quarta}). In that section the quantity $\Delta$ appearing in Eq.~(\ref{upperbmgrandi}) is also identified with the gap between the single-site mean energy, and the
single-site ergotropy, i.e. 
\begin{eqnarray}
 \Delta:= {E}(\rho,H) - {\cal E}(\rho,H) \;.
 \end{eqnarray} 
 A direct consequence of Propositions~\ref{prop:grandi_blr}   is  the universal  identity 
\begin{eqnarray}\label{flat} 
\mathcal{E}^{(\infty,m)}_{\max}(\rho; H)  = E(\rho;H) \;, \quad m\geq \mbox{r}[\rho]\;,  \end{eqnarray} 
which clearly 
superseds~(\ref{notflat}) in the high BRL regime, confirming the saturation of the bound~(\ref{UPPER})  observed in Eq.~(\ref{toflat}) for $m=\mbox{r}[\rho]$.
\\

As evident e.g. from Fig.~\ref{figure0}, {\bf Propositions~\ref{prop:method2}} and {\bf \ref{prop:grandi_blr}} provide our best estimations for $\mathcal{E}^{(N,m)}_{\max}(\rho; H)$. For large enough $N$ and $m\geq r$  they are  optimal since lead to the 
exact evaluation of $\mathcal{E}^{(\infty,m)}_{\max}(\rho; H)$. 
 For small $m$ on the contrary
the constraint posed by Eq.~(\ref{bound_ergotropia_prop1_2NEW}) is certainly suboptimal:
to see this observe for instance that for $m=1$ we get 
\begin{eqnarray} \label{ddfdsf1} 
 \mathcal{E}^{(\infty,1)}_{\bf{C}}(\rho; H)  =E(\rho;H) - \mathfrak{E}_{\ln \left\lceil{\mbox{r}[\rho]}\right \rceil}\;,
\end{eqnarray} 
which is clearly not larger than the exact value~(\ref{asymptN1}) due the fact that 
$\mathfrak{E}_{\ln \left\lceil{\mbox{r}[\rho]}\right \rceil}$ is always larger than or equal to the
thermal energy  $\mathfrak{E}_{S(\rho)}$  associated with the Gibbs state that 
is iso-entropic with $\rho$ (same considerations apply of course for $\mathcal{E}^{(\infty,1)}_{\bf{B}}(\rho; H)$). 
An improvement for small $m$   can  however be obtained by invoking an heuristic (not rigorous) argument that we shall discuss in {\bf Remark 3} of  Sec.~\ref{sec:method2}.
 This suggests that for $m\ll r$ 
it should be possible to identify a special sub-class of states $\rho^{(N)}_{{\bf C}}$ we employed in the derivation of {\bf Proposition \ref{prop:method2}} which allow us to replace $s_{\bf C}$ in Eq.~(\ref{bound_ergotropia_prop1_2NEW}) with the improved value 
	\begin{eqnarray} \label{sCheu} 
	\left. s_{\bf C}\right|_{\rm(heu)}:= S(\rho) - \frac{N-1}{N}\ln m \; ,
	\end{eqnarray}
and  (\ref{ddfdsf1}) with 
\begin{equation}
\left.  \mathcal{E}^{(\infty,1)}_{\bf{C}}(\rho; H)\right|_{\rm(heu)}  =E(\rho;H) -\mathfrak{E}_{S(\rho)} = \mathcal{E}_{\text{tot}}(\rho; H)\;. \end{equation}  
 
\subsubsection{Lower bound for $\frac{\mathcal{E}(\rho^{\otimes N}; H^{(N)})}{N}$} 
As mentioned in the introduction a secondary result our analysis is 
to set on  rigorous ground of the typicality argument first formulated by Alicki and Fannes~\cite{AlickiFannes2013} for  the maximum work extractable for a set of many identical quantum systems. Specifically we show that 

\begin{corollary} 
\label{prop:productstates}
Given  $\eta \in ]0,1[$ and a single-site density matrix $\rho$, for $N$ sufficiently large the following inequality holds true
\begin{equation}  \label{CORO1} 
 \frac{{\cal E}(\rho^{\otimes N}; H^{(N)})}{N} 
\geq \mathcal{E}_{\text{tot}}(\rho; H)
 - (2\sqrt{2.01 C \eta}
+ \epsilon_d 
  e^{-\frac{2N\eta^2}{\alpha^2}})\;,
\end{equation}
with $\epsilon_d$, $C$, and $\alpha$  as in {\bf Proposition~\ref{prop:method1}}.
\end{corollary}
The proof of this result can be seen as a special instance of
{\bf Proposition~\ref{prop:method1}} and it
 is given in Sec.~\ref{sec:claimsnew}.

\section{TI-MPO representation of translationally invariant states} \label{FFDA}
This section is dedicated to clarify some technical aspects of the TI-MPO representation and to present explicit examples of elements of $\mathfrak{S}_T^{(N,m)}(\rho)$ to be used 
 to construct our lower bounds for $\mathcal{E}^{(N,m)}_{\max}(\rho; H)$. 
First of all in Sec.~\ref{prima} we formally show that any translational invariant state of the $N$ sites,
admits a TI-MPO representation~(\ref{MPO_rhoN}) for a proper choice of the tensor $\mathbb{A}$;
then in Sec.~\ref{seconda} we show that the definition of the BLR $\mathfrak{br} [\rho^{(N)}]$ defined in Eq.~(\ref{MPOBLR}) is unaffected by the choice of the local basis~$\{\ket{i}; i=1,\cdots,d \}$ entering in the
TI-MPO representation~(\ref{MPO_rhoN}); 
in Sec.~\ref{terza} we prove instead that the
 GHZ-like states of the form~(\ref{GHZ}) belong to the set $\mathfrak{S}_T^{(N,\mbox{r}^2[\rho])}(\rho)$; finally in Sec.~\ref{Quartanew} we  provide an explicit TI-MPO representation for three different families of correlated states in $\mathfrak{S}_T^{(N)}(\rho)$.

\subsection{Existence of TI-MPO representation} \label{prima} 
In order to show that  the elements $\rho^{(N)}$ of $\mathfrak{S}_T^{(N)}$ can be expressed in the
 TI-MPO representation of Eq.~(\ref{MPO_rhoN}), let start from  a not necessarily TI - MPO representation of such an element,
 i.e. 
 \begin{eqnarray}
\rho^{(N)} &=&
\sum_{i_1,j_1,\dots,i_N,j_N}  \ket{i_1 \dots i_N}\bra{j_1  \dots j_N} \nonumber \\
&&\;  \times  \Tr[{A}_1^{i_1,j_1} \dots {A}_N^{i_N,j_N}]
 \;,\label{MPO_rhoNnotTI}
\end{eqnarray}
where for $k=1,\cdots, N$,  $\{ A_k^{i,j}| {i,j\in1,\cdots d}\}$ are a set of $d^2$, $\bar{M}\times \bar{M}$ matrices associated
to the $k$-th site, which always exists for $\bar{M}$ sufficiently large~\cite{Parker2020}. 
Now exploiting the translational invariance property~(\ref{dfdf}) and the ciclicity of the trace,  it follows that
for all $\ell=0,\cdots, N-1$  we can also write
 \begin{eqnarray}
\rho^{(N)} &=&
\sum_{i_1,j_1,\dots,i_N,j_N} \ket{i_1\dots i_N}\bra{j_1  \dots j_N}  \nonumber \\
&&\;  \times \Tr[{A}_{1\oplus \ell }^{i_1,j_1} \dots {A}_{N\oplus \ell}^{i_N,j_N}]
 \;,\label{MPO_rhoNnotTI2}
\end{eqnarray}
with $\oplus$ representing the sum modulus $N$. Hence we get 
 \begin{eqnarray}
\rho^{(N)} &=&
\sum_{i_1,j_1,\dots,i_N,j_N}  \ket{i_1 \dots i_N}\bra{j_1\dots j_N} \nonumber \\
&&\;  \times\frac{1}{N}  \sum_{\ell=0}^{N-1}  \Tr[{A}_{1\oplus \ell }^{i_1,j_1} \dots {A}_{N\oplus \ell}^{i_N,j_N}] 
 \;,\label{MPO_rhoNnotTI3}
\end{eqnarray}
which is of the form Eq.~(\ref{MPO_rhoN}) by identifying $M= \bar{M} N$ and taking the matrices ${A}^{i,j}$ as 
\begin{eqnarray} \label{fdfsfa} 
A^{i,j} =\frac{1}{{N}^{1/N}} \sum_{\ell=1}^{N}  {A}_{\ell }^{i,j}\otimes |v_\ell\rangle\langle v_{\ell \oplus 1}|\;,
\end{eqnarray} 
with $\{ |v_\ell\rangle ; \ell =1,\cdots, N-1\}$ being orthonormal vectors of an auxiliary  $N$-dimensional vector space (hereafter, in the writing of the MPO matrices  we adopt the  Dirac notation). 

\subsection{Invariance of the BLR with respect to the choice of the local basis} \label{seconda}

 Here we show that the choice of the local basis $\{ |i\rangle\}$ entering the TI-MPO representation~(\ref{MPO_rhoN}) 
 does not affect the definition 
BLR~(\ref{MPOBLR}) of a state $\rho^{(N)} \in \mathfrak{S}_T^{(N)}$.
To see this consider $\{ \ket{\phi_i}; i=1,\cdots,d \}$ a new basis of ${\cal H}$ connected to 
$\{ \ket{i}; i=1,\cdots,d \}$ via a unitary transformation $U$, i.e. 
\begin{eqnarray} 
\ket{\phi_i} = U \ket{i} \;, \forall i=1,\cdots,d \;. 
\end{eqnarray} 
Replacing this into~(\ref{MPO_rhoN})  we get 
\begin{eqnarray}
\rho^{(N)} &=&
\sum_{i_1,j_1,\dots,i_N,j_N} U^{\dag \otimes N} \ket{\phi_{i_1} \dots \phi_{i_N}}\bra{\phi_{j_1}  \dots \phi_{j_N}} U^{\otimes N}\nonumber \\
&&\;  \times  \Tr[{A}^{i_1,j_1} \dots {A}^{i_N,j_N}] 
\;,
\nonumber \\
 &=&
\sum_{i'_1,j'_1,\dots,i'_N,j'_N} \ket{\phi_{i'_1}\dots \phi_{i'_N}}\bra{\phi_{j'_1}  \dots \phi_{j'_N}}
 \nonumber \\
&&\;  \times \Tr[{B}^{i'_1,j'_1} \dots {B}^{i'_N,j'_N}]
\;,
\label{MPO_rhoNnew12}
\end{eqnarray}
where for $i',j' = 1, \cdots, d$ 
\begin{eqnarray} \label{fdfd} 
{B}^{i',j'} = \sum_{i,j=1}^d \langle \phi_{i'} | U^\dag | \phi_i\rangle {A}^{i,j}   \langle \phi_{j} | U^\dag | \phi_{j'}\rangle 
\end{eqnarray} 
forming a new set of $d^2$, $M\times M$ matrices. Equation~(\ref{fdfd}) creates a one-to-one correspondence between
the TI-MPO representations of $\rho^{(N)}$ constructed with the basis $\{ \ket{\phi_i}; i=1,\cdots,d \}$ with bond link $M$, and the
TI-MPO representations  associated
with the basis $\{ \ket{i}; i=1,\cdots,d \}$ with the same bond link: accordingly
the BLR~(\ref{MPOBLR}) computed with respect to those two basis will produce the same value.
Notice in particular that this allows us to identify $\ket{i}$ appearing in~(\ref{MPO_rhoN})
with the eigenvectors $|\lambda_i\rangle$  of the local density matrix $\rho$ of Eq.~(\ref{cond_traccia_parziale}), a choice that, via Eq.~(\ref{MPO_rhoN_red}), will allow us to identify the 
associated eigenvalues as  
\begin{eqnarray} \Tr[{A}^{i,j}{\bar{A}}^{N-1}] = \lambda_i \; \delta_{i,j} \;. \label{formula}  \end{eqnarray}

\subsection{TI-MPO representation for GHZ-like states} \label{terza} 
Consider $\rho$ a single-site density matrix of rank $\mbox{r}[\rho]$
\begin{eqnarray}\label{rhodef} 
\rho = \sum_{i=1}^{\mbox{r}[\rho]} \lambda_i |\lambda_i\rangle\langle \lambda_i| \;, 
\end{eqnarray} 
with $\{ |\lambda_i\rangle, i=1,\cdots, d\}$ its eigenvectors, and $\{ \lambda_i; i =1,\cdots, \mbox{r}[\rho]\}$ 
the corresponding non-zero eigenvalues (for $i\geq \mbox{r}[\rho]+1$ of course $\rho |\lambda_i\rangle =0$). 
Consider also the associated 
 GHZ-like state~(\ref{GHZ}) 
 which is explicitly translationally invariant. Here we prove that $|\mbox{GHZ}^{(N)}_\rho\rangle$  
  belongs to the set $\mathfrak{S}_T^{(N,\mbox{r}^2[\rho])}(\rho)$, i.e.
 \begin{eqnarray} |\mbox{GHZ}^{(N)}_\rho\rangle\langle \mbox{GHZ}^{(N)}_\rho| \in \mathfrak{S}_T^{(N,\mbox{r}^2[\rho])}(\rho)\;,
 \label{result} 
 \end{eqnarray} 
or equivalently that its 
BLR~(\ref{MPOBLR})  is smaller or equal than $\mbox{r}^2[\rho]$, i.e.
 \begin{eqnarray} \mathfrak{br} \left[|\mbox{GHZ}^{(N)}_\rho\rangle\langle \mbox{GHZ}^{(N)}_\rho| \right] \leq \mbox{r}^2[\rho]\;. 
 \label{ineqbond} 
 \end{eqnarray} 
To show this fact it is sufficient to exhibit a TI-MPO representation~(\ref{MPO_rhoN}) of 
$|\mbox{GHZ}^{(N)}_\rho\rangle\langle \mbox{GHZ}^{(N)}_\rho|$ with a 
 tensor $\mathbb{A}\in \mathbb{C}^{d \times d \times M \times M}$ characterized by  $M=\mbox{r}^2[\rho]$. 
 Exploiting the observation of Sec.~\ref{seconda} we shall focus on representations 
which uses  the eigenvectors $\{ |\lambda_i\rangle; i=1,\cdots, d\}$  of  $\rho$ as local basis $\{ |i \rangle; i =1, \cdots, d\}$.
Then we notice that the solution is provided by the identity 
\begin{eqnarray} 
 |\mbox{GHZ}^{(N)}_\rho\rangle\langle \mbox{GHZ}^{(N)}_\rho| &=& \sum_{i,j=1}^{\mbox{r}[\rho]}  \sqrt{\lambda_i\lambda_j}\;
 \ket{\lambda_i\dots\lambda_i}   \bra{\lambda_j\dots\lambda_j}
 \nonumber \\
 &=& \sum_{i_1,j_1,\dots,i_N,j_N}  
  \ket{\lambda_{i_1} \dots \lambda_{i_N}}\bra{\lambda_{j_1} \dots \lambda_{j_N}} \nonumber \\
&&\;  \times 
\Tr[{A}^{i_1,j_1} \dots {A}^{i_N,j_N}] \;,
 \end{eqnarray} 
 where defining $\{ |v_i\rangle; i= 1,\cdots,  \mbox{r}[\rho]\}$ an orthonormal basis of an auxiliary vector space ${\cal H}_A$  of dimension $\mbox{r}[\rho]$,  we took ${A}^{i,j}$ to be the $\mbox{r}^2[\rho]\times \mbox{r}^2[\rho]$ matrices 
\begin{eqnarray}
{A}^{i,j} := \left\{ \begin{array}{ll} 
\left({\lambda_i\lambda_j}\right)^{\frac{1}{2N}}\;  |v_i\rangle\langle v_i| \otimes |v_j\rangle\langle v_j| \;, & \forall i,j \leq \mbox{r}[\rho]\;, \\ \\ 
0 \;, & \mbox{otherwise.} 
\end{array} \right.
\end{eqnarray} 
We remark that the argument we presented here doesn't ensure that
 $\mathfrak{br} [|\mbox{GHZ}^{(N)}_\rho\rangle\langle \mbox{GHZ}^{(N)}_\rho| ]$ coincides with $\mbox{r}^2[\rho]$, but
 (as implied by Eq.~(\ref{ineqbond})) only that
 the former is an upper bound for the latter. In other words, formally speaking, we cannot exclude that 
 $|\mbox{GHZ}^{(N)}_\rho\rangle\langle \mbox{GHZ}^{(N)}_\rho|$ belongs to  $\mathfrak{S}_T^{(N,m)}(\rho)$ for same $m< \mbox{r}^2[\rho]$ (we conjecture
 however that this is not the case).

   \subsection{TI-MPO representation of three families of correlated states of
   $\mathfrak{S}_T^{(N)}(\rho)$} \label{Quartanew} 
   
\begin{figure}
	\includegraphics[width=\columnwidth]{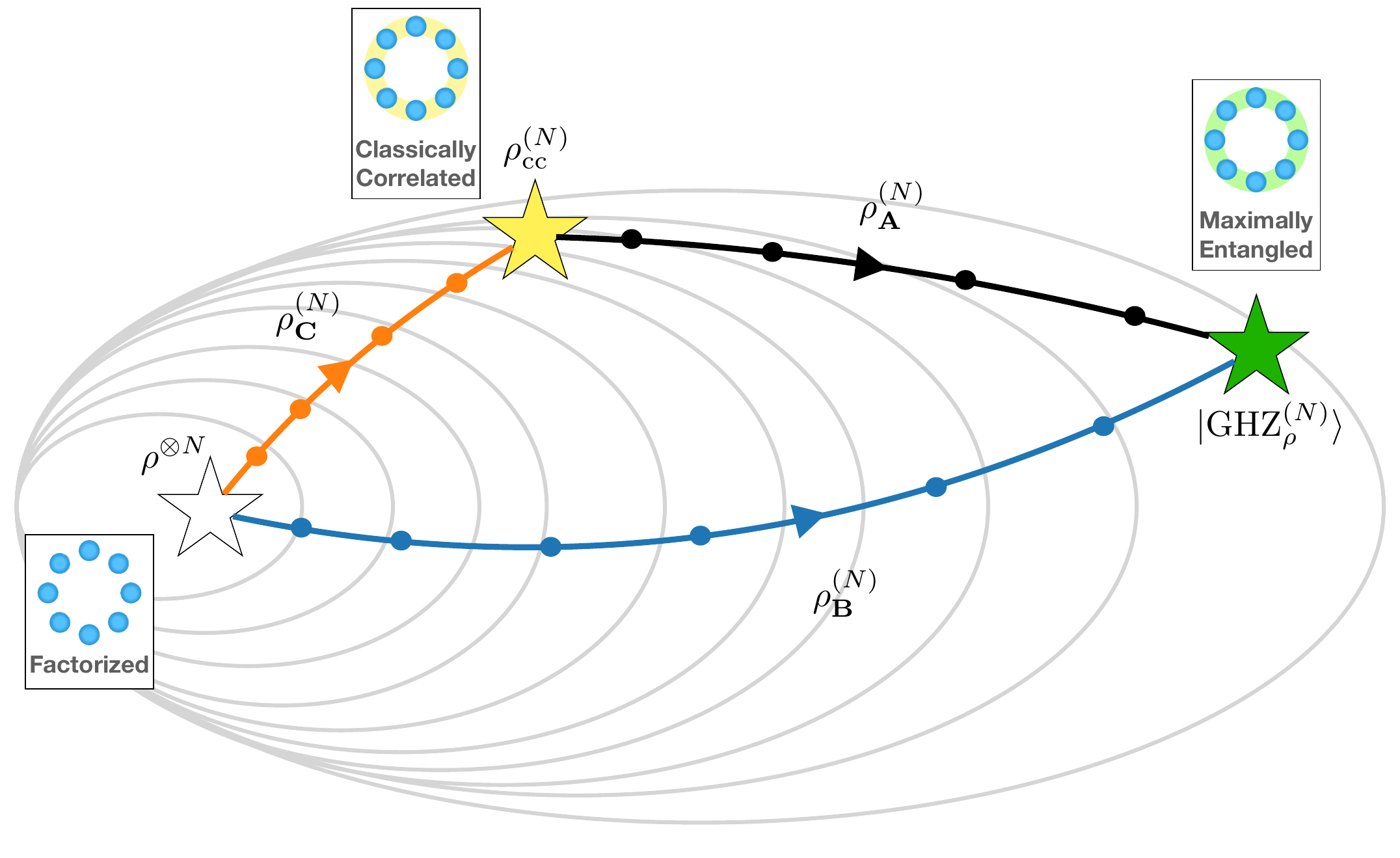}
	\caption[Pictorial rep of the states trajectories]{Graphical representation of the families of states 
	of $\mathfrak{S}_T^{(N)}(\rho)$ 
				defined in Sec.~\ref{Quartanew}.
	Specifically  by the black  trajectory represents the set of convex convolution of GHZ-like states $\rho^{(N)}_{\bf A}$ defined in Eq.~(\ref{DEFMIXGHZ}) of Sec.~\ref{quarta},
which connect  the pure state $|\mbox{GHZ}^{(N)}_\rho\rangle$ of Eq.~(\ref{GHZ}) (green star in the figure) with its classically correlated counterpart
$\rho^{(N)}_{\rm cc}$ defined in Eq.~(\ref{purest_separable_state0}) and represented by the
yellow star in the picture; the blue trajectory  is instead the set of block-wise purified states 
 $\rho^{(N)}_{\bf B}$ defined in Eq.~(\ref{NORM}) of Sec.~\ref{sec:method1_stato},
which
interpolates from $|\mbox{GHZ}^{(N)}_\rho\rangle$ to the fully factorized element $\rho^{\otimes N}$ (white star);
finally the orange trajectory  represents  the set of classically correlated  states 
$\rho^{(N)}_{\bf C}$ defined in Eq.~(\ref{rhoN_metodo2})  of Sec.~\ref{sec:method2_stato}, which
interpolates from $\rho^{\otimes N}$ to $\rho^{(N)}_{\rm cc}$.
	The grey lines represent the subsets $\mathfrak{S}_T^{(N,m)}(\rho)$ of Eq.~(\ref{NEWSETS}): the larger
	is the 
	set represent the higher is the associated values of  the BLR $\mathfrak{br} [\rho^{(N)} ]$ of the corresponding states.}
	\label{figura1}
\end{figure}

In this section we present an explicit TI-MPO representation for three different families of states in $\mathfrak{S}_T^{(N)}(\rho)$ which, as pictorially depicted in Fig.~\ref{figura1}, allows us span various BLR values. 
All these examples are constructed starting from the set of the non-null
 eigenvalues of the single-site density matrix $\rho$ of the model, i.e. 
\begin{eqnarray}
{\mathfrak{K}}[\rho]:=\{ \lambda_i; i= 1,\cdots, \mbox{r}[\rho]\}\;,
\end{eqnarray} 
with $\lambda_i$ and $\mbox{r}[\rho]$ defined as in  Eq.~(\ref{rhodef}), and dividing it into smaller, non-empty subsets. Specifically, given 
 $L$ integer greater than or equal to  $1$ and no larger than $\mbox{r}[\rho]$, we consider a collection
\begin{eqnarray}\label{PARTITION} 
{\mathfrak{P}}:=\{{\mathfrak{K}}_{1}, {\mathfrak{K}}_{2},\cdots, {\mathfrak{K}}_{L}\}\;,\end{eqnarray} 
 of $L$ non-overlapping, non-null subsets of ${\mathfrak{K}}[\rho]$, which provides a partition of such set, i.e.
 \begin{eqnarray} 
{\mathfrak{K}}_{\ell} \cap{\mathfrak{K}}_{\ell'} &=& \O \;, \qquad \forall \ell\neq \ell' \;, \label{exclusive}  \\
\bigcup_{\ell=1}^M  {\mathfrak{K}}_{\ell}&=& {\mathfrak{K}}[\rho] \;.  \label{overlapping} 
 \end{eqnarray} 
 In what follow we shall use the symbol  $\# {\mathfrak{K}}_{\ell}$ to represent the cardinality of
  ${\mathfrak{K}}_{\ell}$ and indicate with $\#_{\max}$ and $|{\mathfrak{P}}|^{(p)}$ 
  their maximum value and the sum of their $p$-powers, i.e. the quantities
\begin{eqnarray} \label{maxcard} 
\#_{\max} &:=& \max_{\ell=1,\cdots, L} \left( \#{\mathfrak{K}}_{\ell}\right) \;, \\
|{\mathfrak{P}}|^{(p)} &:=&    \sum_{\ell =1}^L (\# {\mathfrak{K}}_{\ell})^p \;,
 \label{ineqbondA} 
 \end{eqnarray} 
 which are related by the inequality  
   \begin{eqnarray} 
L \leq |{\mathfrak{P}}|^{(p)}  \leq (\#_{\max})^p L  \;.  \label{ineqbondAeq} 
 \end{eqnarray} 

 By construction the $\#{\mathfrak{K}}_{\ell}$'s  are all greater than or equal to $1$ and 
 sum up to  the rank of $\rho$, i.e.  
 \begin{eqnarray}
\sum_{\ell=1}^L \# {\mathfrak{K}}_{\ell} 
= \mbox{r}[\rho] \;, \label{normalizasum} 
\end{eqnarray}
which implies
\begin{eqnarray} \label{maxcard1} 
\frac{ \mbox{r}[\rho] }{L} \leq &\#_{\max}& \leq  \mbox{r}[\rho]  - L +1  \;,\\
\mbox{r}[\rho] \leq &|{\mathfrak{P}}|^{(p)}&  \leq \mbox{r}^p[\rho]  \;.  \label{ineqbondAeq1} 
\end{eqnarray} 
Notice that if $L$ is an exact  divisor of $\mbox{r}[\rho]$ 
one can force all the subsets $\mathfrak{K}_{\ell}$  to have the same cardinality, i.e. 
\begin{eqnarray} \label{uniformP} 
\#{\mathfrak{K}}_{\ell} ={\mbox{r}[\rho]}/{L} \;, \qquad \forall \ell \in 1, \cdots, L\;,
\end{eqnarray} 
corresponding to an exact saturation of the lower bound in (\ref{maxcard1}):
when this happens one gets $|{\mathfrak{P}}|^{(p)}=\mbox{r}^p[\rho]/{L^{p-1}}$, and we say that  the associated ${\mathfrak{P}}$ is 
 an {\it uniform} partition of ${\mathfrak{K}}[\rho]$.
  In case $L$ is not an exact divisor of   $\mbox{r}[\rho]$  we say instead that ${\mathfrak{P}}$ is an {\it almost uniform} partition of ${\mathfrak{K}}[\rho]$ if 
 \begin{eqnarray} \label{almuniformP} 
\#_{\max}  = \left \lceil{\mbox{r}[\rho] / L}\right \rceil 
 \;, 
\end{eqnarray} 
leading to $|{\mathfrak{P}}|^{(p)}\leq   (\left \lceil{\mbox{r}[\rho] / L}\right \rceil)^p L$. 

Next we 
 introduce a  labelling for the elements of  each individual 
 subsets of ${\mathfrak{P}}$; specifically, given $\ell=1,\cdots, L$, we indicate with 
 $\lambda_{k,\ell}$  the $k$-th element of
 ${\mathfrak{K}}_{\ell}$, so that 
 \begin{eqnarray}\label{notazione} 
 {\mathfrak{K}}_{\ell}=\{ \lambda_{k,\ell}; k=1,\cdots, \# {\mathfrak{K}}_{\ell} \}\;.\end{eqnarray}
Notice that, thanks to (\ref{exclusive}) and~(\ref{overlapping}), 
 for each $i= 1,\cdots, \mbox{r}[\rho]$ there is a unique choice of the indexes
$\ell=1,\cdots L$ and $k= 1,\cdots ,\# {\mathfrak{K}}_{\ell}$ that identifies the $i$-th eigenvalue
of $\rho$   as the $k$ element of the subset ${\mathfrak{K}}_{\ell}$: in the following we shall exploit this  one-to-one correspondence indicating with
the symbols $k^{[i]}$ and $\ell^{[i]}$ those special values via the mapping
\begin{eqnarray} \label{mapping} 
i \longmapsto k^{[i]}, \ell^{[i]} \quad \mbox{s.t.} \quad \lambda_i = \lambda_{k^{[i]},\ell^{[i]} }\;. 
\end{eqnarray} 
Finally for all $\ell=1,\cdots, L$, we introduce the quantity 
\begin{eqnarray} 
\label{sommeparziali}
\mathscr{S}_{\ell} := 
\sum_{k=1}^{\# {\mathfrak{K}}_{\ell}} \lambda_{k,\ell}\;,
\end{eqnarray} 
to gauge  the  statistical  
weight of the set $\mathfrak{K}_{\ell}$: these terms are clearly positive and
fulfils the  normalization condition
\begin{eqnarray}
\sum_{\ell=1}^L \mathscr{S}_{\ell} &=& 1 \;. \label{normaliza} 
\end{eqnarray}

\begin{table}[t!]
\begin{tabular}{c|c|c|c|}  
 & $\rho^{(N)}_{\bf A}$ & $\rho^{(N)}_{\bf B}$ & $\rho^{(N)}_{\bf C}$\\ 
 & & & \\
  \hline & & & \\
 {\rm rank} & \;  $L$ \; & $\left( \#_{\max}\right)^N$ & $|{\mathfrak{P}}|^{(N)}$ \\  
& & & \\
 \hline  & & & \\
  {\rm  BLR } &  $\leq |{\mathfrak{P}}|^{(2)}$ & $\leq L^2$ & \; $\leq L$ \; \\
 & & & \\ 
  \hline
\end{tabular}
				\caption{Rank and BLR~(\ref{MPOBLR}) for the states 
				$\rho^{(N)}_{\bf A}$, $\rho^{(N)}_{\bf B}$, and $\rho^{(N)}_{\bf C}$ of $\mathfrak{S}_T^{(N)}(\rho)$ 
				defined in Sec.~\ref{Quartanew}. 
				The integer parameter $L = 1,\cdots, \mbox{r}[\rho]$ counts the number of elements of the partition ${\mathfrak{P}}$ which we used
				to divide the set of positive eigenvalues ${\mathfrak{K}}[\rho]$ of the single-side state (see Eq.~(\ref{PARTITION})); $\#_{\max}$ is the maximum among the cardinalities of the  elements of 
				${\mathfrak{P}}$  (see Eq.~(\ref{maxcard})); while for $p$ integer, $|{\mathfrak{P}}|^{(p)}$ is the
				sum of the $p$-powers of the cardinalities (see Eq.~(\ref{ineqbondA})). For uniform partitions we
				get $\#_{\max}=  \mbox{r}[\rho]/L$ and $|{\mathfrak{P}}|^{(p)} = \mbox{r}^p[\rho]/L^{p-1}$ with 
				$\mbox{r}[\rho]$ being the rank of the single-site density matrix. 
							\label{tab1}} 
			\end{table}
\subsubsection{TI-MPO representation for convex convolutions of non overlapping GHZ-like states} \label{quarta} 
The first family of states of $\mathfrak{S}_T^{(N)}(\rho)$  we consider is 
a generalization
of the GHZ-like construction introduced in Sec.~\ref{terza} (this set will be used to
prove {\bf Proposition \ref{prop:grandi_blr}}). 
In particular for element 
$\mathfrak{K}_{\ell}$ of the
the partition~(\ref{PARTITION}) 
 we define a corresponding GHZ-like state via the identity
 \begin{eqnarray} \label{GHZell} 
 |\mbox{GHZ}^{(N)}_{\rho_{\ell}} \rangle : = \frac{1}{
\sqrt{\mathscr{S}_{\ell} }} \sum_{k=1}^{\# {\mathfrak{K}}_{\ell}} 
 \sqrt{\lambda_{k,\ell} } |\lambda_{k,\ell} \dots \lambda_{k,\ell}\rangle\;.
 \end{eqnarray} 
This a translational invariant state whose associated single-site local density matrix (\ref{cond_traccia_parziale})  
is given by 
\begin{eqnarray}
\rho_\ell:=  \frac{1}{\mathscr{S}_{\ell} }  \sum_{k=1}^{\# {\mathfrak{K}}_{\ell}} \lambda_{k,\ell}    |\lambda_{k,\ell}\rangle\langle_{k,\ell} \lambda_{k,\ell}| \;,
\end{eqnarray} 
which has a rank  $\mbox{r}[\rho_{\ell}]$ that corresponds to the cardinality of 
${\mathfrak{K}}_{\ell}$, i.e. 
\begin{eqnarray} \mbox{r}[\rho_{\ell}] = \#{\mathfrak{K}}_{\ell}\;. \end{eqnarray} 
From Eq.~(\ref{overlapping}) and (\ref{normaliza}) it follows that the vectors~(\ref{GHZell})  are orthonormal, i.e. 
\begin{eqnarray}
\langle \mbox{GHZ}^{(N)}_{\rho_{\ell}} |\mbox{GHZ}^{(N)}_{\rho_{\ell'}} \rangle = \delta_{\ell,\ell'} \;, 
\end{eqnarray} 
and provide a decomposition of the state~(\ref{GHZ}) via the identity 
\begin{eqnarray}
 |\mbox{GHZ}^{(N)}_{\rho} \rangle = \sum_{\ell=1}^L  \sqrt{\mathscr{S}_{\ell} } \; 
 |\mbox{GHZ}^{(N)}_{\rho_{\ell}} \rangle  \;. 
\end{eqnarray} 
Furthermore  by direct inspection one can easily verify that the  convex convolution of rank $L$ obtained by mixing the
vectors $|\mbox{GHZ}^{(N)}_{\rho_{\ell}} \rangle$ with statistical weight $\mathscr{S}_{\ell}$, i.e. the density matrix
\begin{eqnarray}  
\rho^{(N)}_{\bf A}  := \sum_{\ell=1}^L \mathscr{S}_{\ell}\;   |\mbox{GHZ}^{(N)}_{\rho_{\ell}} \rangle \langle
\mbox{GHZ}^{(N)}_{\rho_{\ell}} | \;,  \label{DEFMIXGHZ} 
\end{eqnarray} 
 is an element of  $\mathfrak{S}_T^{(N)}(\rho)$, i.e. it admits $\rho$ of Eq.~(\ref{rhodef})  as single-site reduced density matrix (ultimately this is a consequence of the identity $\sum_{\ell=1}^L \mathscr{S}_{\ell} \rho_\ell = \rho$), and has a rank equal to $L$ as reported in Table~\ref{tab1}.  
 Notice in particular that as $L$ varies from  the extremal cases  $L=1$ and $L=\mbox{r}[\rho]$ 
 the states (\ref{DEFMIXGHZ}) interpolate from the GHZ-like configuration to its classically correlated counterpart, i.e. 
  \begin{eqnarray}
 \rho^{(N)}_{{\bf A}}\Big|_{L=1}  &=&  |\mbox{GHZ}^{(N)}_\rho\rangle\langle \mbox{GHZ}^{(N)}_\rho| \;, 
\label{purest_separable_state1} \\ 
\rho^{(N)}_{{\bf A}}\Big|_{L=\mbox{r}[\rho]}  &=& \rho^{(N)}_{\rm cc}:= 
   \sum_{i=1}^{\mbox{r}[\rho]} \lambda_i \ket{\lambda_i \dots \lambda_i}\bra{\lambda_i \dots \lambda_i} ,
\label{purest_separable_state0}\end{eqnarray}
(see Fig.~\ref{figura1}).

 We now show that  the BLR of  $\rho^{(N)}_{\bf A}$ fulfils the inequality
 \begin{eqnarray} \mathfrak{br} [\rho^{(N)}_{\bf A} ] \leq |{\mathfrak{P}}|^{(2)} =  \sum_{\ell =1}^L (\# {\mathfrak{K}}_{\ell})^2
  \;,\label{ineqbound1} \end{eqnarray} 
or equivalently that  
  \begin{eqnarray}  \rho^{(N)}_{\bf A} \in \mathfrak{S}_T^{(N,|{\mathfrak{P}}|^{(2)})}(\rho)
  \;.\label{ineqbound12} \end{eqnarray}   
For the trivial choice $L=1$ where ${\mathfrak{P}}$   contains only ${\mathfrak{K}}[\rho]$ as unique element,
  $\rho^{(N)}_{\bf A}$ corresponds to the GHZ-like state~(\ref{GHZ})  which purifies $\rho$, $|{\mathfrak{P}}|^{(2)}=\mbox{r}^2[\rho]$, and Eq.~(\ref{ineqbound12}) reduces to the property (\ref{result}) which we proved in the previous section. 
To show that (\ref{ineqbound12}) holds also for  $L>1$ we adopt a similar scheme and present an explicit TI-MPO decomposition
 \begin{eqnarray}
\rho^{(N)}_{\bf A} &=& \sum_{i_1,j_1,\dots,i_N,j_N}  
  \ket{\lambda_{i_1} \dots \lambda_{i_N}}\bra{\lambda_{j_1} \dots \lambda_{j_N}} \nonumber \\
&&\;  \times 
\Tr[{A}^{i_1,j_1}_{{\bf A}}  \dots {A}^{i_N,j_N}_{{\bf A}} ] \;,  \label{target} 
 \end{eqnarray} 
  that employs  the eigenvectors $\{ |\lambda_i\rangle; i=1,\cdots, d\}$  of  $\rho$ as local basis
and which explicitly uses matrices  $|{\mathfrak{P}}|^{(2)}\times |{\mathfrak{P}}|^{(2)}$ matrices ${A}^{i,j}_{{\bf A}}$.
For this purpose observe first that given $\ell=1,\cdots, L$, the state $|\mbox{GHZ}^{(N)}_{\rho_\ell} \rangle$ admits
a TI-MPO representation
\begin{eqnarray} 
 |\mbox{GHZ}^{(N)}_{\rho_\ell} \rangle\langle \mbox{GHZ}^{(N)}_{\rho_\ell}|  &=& \sum_{i_1,j_1,\dots,i_N,j_N}  
  \ket{\lambda_{i_1} \dots \lambda_{i_N}}\bra{\lambda_{j_1} \dots \lambda_{j_N}} \nonumber \\
&&\;  \times 
\Tr[{A}_{\ell}^{i_1,j_1} \dots {A}_{\ell}^{i_N,j_N}] \;, \label{intnato} 
 \end{eqnarray} 
with
\begin{equation}
{A}_\ell^{i,j} := \left\{ \begin{array}{ll} 
\left({\frac{\lambda_{i} \lambda_{j}}{\mathscr{S}^2_{\ell}} 
}\right)^{\frac{1}{2N}}\!\! |v_{k^{[i]}}^{(\ell)} \rangle\langle v_{k^{[i]}}^{(\ell)} | \otimes |v_{k^{[j]}}^{(\ell)}\rangle\langle v_{k^{[j]}}^{(\ell)} |, 
& \forall \lambda_i,\lambda_j \in {\mathfrak{K}}_{\ell}, \\ \\ \\ \\
0 \;, & \mbox{otherwise,}
\end{array} \right.
 \label{matrA} 
\end{equation} 
  where  $k^{[i]}$ and $k^{[j]}$ are defined by the mapping~(\ref{mapping}),  and where 
$\{ |v_{k}^{(\ell)} \rangle; k= 1,\cdots,   \#{\mathfrak{K}}_{\ell}\}$ form an orthonormal basis of an auxiliary vector
space ${\cal H}_{\ell}$ of dimension $\mbox{r}[\rho_\ell]$ (notice that the $A_\ell^{i,j}$'s operate on two copies of 
 ${\cal H}_{\ell}$, i.e. on
${\cal H}_{\ell}^{\otimes 2}$, and are hence $(\#{\mathfrak{K}}_{\ell})^2 \times  (\#{\mathfrak{K}}_{\ell})^2$).
Equation~(\ref{target})  can now be obtained by identifying the ${A}^{i,j}_{{\bf A}}$'s as proper direct sums over the index $\ell$
 of the matrices~(\ref{matrA}).  Specifically,  let first observe that the auxiliary space ${\cal H}_{{\bf A}}:= \bigoplus_{\ell=1}^L {\cal H}_{\ell}^{\otimes 2}$ admits  $\bigcup_{\ell=1}^L\{ |v_{k}^{(\ell)} \rangle\otimes |v_{k'}^{(\ell)} \rangle; k,k'= 1,\cdots,  \#{\mathfrak{K}}_{\ell}\}$ as an orthonormal set and 
 has dimension  \begin{eqnarray}
 \mbox{dim}[{\cal H}_{{\bf A}}] = \sum_{\ell=1}^L \mbox{dim}^2[{\cal H}_{\ell}] = \sum_{\ell=1}^L 
 (\#{\mathfrak{K}}_{\ell})^2 = |{\mathfrak{P}}|^{(2)}\;. 
 \end{eqnarray}  On such space we then introduce the matrices 
\begin{eqnarray}
{A}^{i,j}_{{\bf A}} &:=& \bigoplus_{\ell=1}^L  {\mathscr{S}_{\ell}}^{\tfrac{1}{N}} \;  {A}_\ell^{i,j} \;, 
\end{eqnarray}
which can be equivalently expressed as 
\begin{eqnarray} \label{DEFAA} 
{A}^{i,j}_{{\bf A}} &=&(\sqrt{\lambda_{i} \lambda_{j}})^{1/N}  \;
\delta_{\ell^{[i]},\ell^{[j]}}\\ \nonumber 
&&\times 
|v_{k^{[i]}}^{(\ell^{[i]})} \rangle\langle v_{k^{[i]}}^{(\ell^{[i]})} | \otimes |v_{k^{[j]}}^{(\ell^{[j]})}\rangle\langle 
v_{k^{[j]}}^{(\ell^{[j]})} | \;, 
\end{eqnarray} 
with  $\ell^{[i]}$ and $\ell^{[j]}$   defined by the mapping~(\ref{mapping})  and with 
$\delta_{\ell^{[i]},\ell^{[j]}}$ being the Kronecker delta symbol  that forces $\ell^{[i]}=\ell^{[j]}$.
The identity ~(\ref{target}) finally follows by observing that 
\begin{equation} 
\Tr[{A}^{i_1,j_1}_{{\bf A}}  \dots {A}^{i_N,j_N}_{{\bf A}} ]= \sum_{\ell=1}^L  \mathscr{S}_{\ell}
\Tr[{A}^{i_1,j_1}_{\ell}  \dots {A}^{i_N,j_N}_{\ell} ] \;,
\end{equation} 
and from~(\ref{intnato}) and (\ref{DEFMIXGHZ}).

 \subsubsection{TI-MPO representation for {\it block-wise purified} density matrices}
\label{sec:method1_stato}
The second family of elements of $\mathfrak{S}_T^{(N)}(\rho)$  we consider 
 allow us to connect the completely uncorrelated  $\rho^{\otimes N}$ to the pure GHZ-like state
 configuration  
 $|\mbox{GHZ}^{(N)}_\rho\rangle$ -- see trajectory $\rho^{(N)}_{{\bf B}}$ in Fig.~\ref{figura1}.
We dub  these states 
 {\it block-wise purified} density matrices and  define them by means of an explicit TI-MPO representation. They will be used in the derivation of {\bf Proposition~\ref{prop:method1}}.

The starting point of the analysis is again the partition~(\ref{PARTITION}) 
of the non-null eigenvalues of $\rho$
which satisfies the properties~(\ref{exclusive}) and~(\ref{overlapping}). 
For the sake of simplicity we report here the explicit derivation for the special case 
 in which  $L$ is an exact divisor of $\mbox{r}[\rho]$ and 
 ${\mathfrak{P}}$ is an uniform partition of ${\mathfrak{K}}[\rho]$, so that Eq.~(\ref{uniformP}) holds true:
 the extension of this construction to the general case is given in Appendix~\ref{APPGENA}.
Introduce next 
 an auxiliary Hilbert space ${\cal H}_{\bf B}$ of dimension $L$, with an orthonormal basis $\left\{ \ket{v_{\ell}} ; \ell= 1,\cdots,  L \right\}$ and 
 replace the matrices  
$A^{ij}_ {\bf{A}}$ of Eq.~(\ref{DEFAA}) with the 
 following set of $L^2 \times L^2$ matrices:
\begin{eqnarray}
A^{ij}_ {\bf{B}}&:=& 
\sqrt{\lambda_i \lambda_j} \left(\mathscr{S}_{{\ell^{[i]}}}\mathscr{S}_{{\ell^{[j]}}}\right)^{\frac{1-N}{2N}}  \; \delta_{k^{[i]},k^{[j]}} \nonumber \\ \label{DEFAB} 
&& \times \ket{v_{\ell^{[i]}}}\bra{v_{\ell^{[i]}}} \otimes \ket{v_{\ell^{[j]}}}\bra{v_{\ell^{[j]}}}  \; ,
\end{eqnarray}
where $k^{[i]}$, $\ell^{[i]}$  and  $\mathscr{S}_{\ell}$  defined as in  Eqs.~(\ref{mapping})  and~(\ref{sommeparziali}), respectively. 
We now define the block-wise purified density operator $\rho^{(N)}_{{\bf B}}$ via the TI-MPO representation 
\begin{eqnarray}
\rho^{(N)}_{{\bf B}} &:=& \sum_{i_1,j_1,\dots,i_N,j_N}  
  \ket{\lambda_{i_1} \dots \lambda_{i_N}}\bra{\lambda_{j_1} \dots \lambda_{j_N}} \nonumber \\
&&\;  \times 
\Tr[{A}^{i_1,j_1}_{{\bf B}}  \dots {A}^{i_N,j_N}_{{\bf B}} ] \;.  \label{targetB} 
 \end{eqnarray} 
By construction we get that $\rho^{(N)}_{{\bf B}}$ 
 has non-zero entries
\begin{eqnarray}
&&\bra{\lambda_{j_1} \lambda_{j_2} \dots  \lambda_{j_N}}\rho^{(N)}_{{\bf B}}  \ket{ \lambda_{i_1}  \lambda_{i_2} \dots  \lambda_{i_N}} \\
&& \qquad \qquad \qquad = \label{rhoN_method1}
\sqrt{
	\frac{\lambda_{i_1}\lambda_{i_2} \cdots \lambda_{i_N}}{\mathscr{S}_{\ell^{[i_1]}}^{N-1}} }\sqrt{\frac{\lambda_{j_1}\lambda_{j_2} \cdots \lambda_{j_N}}{\mathscr{S}_{\ell^{[j_1]}}^{N-1}}
} \;, \nonumber 
\end{eqnarray}
if the following conditions are true:
\begin{itemize}
	\item $\ell^{[i_1]} = \ell^{[i_2]} = \cdots = \ell^{[i_N]}$,
	\item $\ell^{[j_1]} = \ell^{[j_2]} = \cdots = \ell^{[j_N]}$,
	\item $\left( k^{[i_1]} , k^{[i_2]} , \cdots k^{[i_N]} \right) = \left( k^{[j_1]} , k^{[j_2]} , \dots k^{[j_N]} \right)$;
\end{itemize}
while
\begin{equation}
\bra{\lambda_{j_1} \lambda_{j_2} \dots  \lambda_{j_N}} \rho^{(N)}_{{\bf B}}  \ket{ \lambda_{i_1}  \lambda_{i_2} \dots  \lambda_{i_N}}= 0\;,  \label{identita0} 
\end{equation}
otherwise.
Making   use of Eq.~(\ref{formula}) and the fact that in the present case we have 
$\bar{A}_{\bf B}  =
 \sum_{\ell=1}^L \mathscr{S}_{{\ell}}^{\frac{1}{N}}\ket{v_{\ell}}\bra{v_{\ell}} \otimes \ket{v_{\ell}}\bra{v_{\ell}}$, it is not difficult to verify that  $\rho^{(N)}_{{\bf B}}$ satisfies the partial trace condition~(\ref{cond_traccia_parziale}): hence
 we can claim that  $\rho^{(N)}_{{\bf B}}$ is an element of $\mathfrak{S}_T^{(N,L^2)}(\rho)$, i.e.
\begin{eqnarray} \label{propertynewMM} 
\rho^{(N)}_{{\bf B}} \in \mathfrak{S}_T^{(N,L^2)}(\rho) \;. \end{eqnarray} 

A more explicit form for the density matrix  $\rho^{(N)}_{{\bf B}}$ can be obtained
 exploiting correspondence~(\ref{mapping}) and the fact that, thanks to the choice
 of working with uniform partitions,  the index $k$ of ${\lambda}_{k_{a
},\ell}$ run from $1$ to  $\mbox{r}[\rho]/L$ irrespectively from the value of $\ell$.
This leads to 
 \begin{equation}
\rho^{(N)}_{{\bf B}} \label{rhoN_method11}=   \sum_{\vec{k}}
\sum_{\ell,\ell'=1}^L   \sqrt{
	\tfrac{{\lambda}^{(N)}_{\vec{k}, \ell}}{\mathscr{S}_{\ell}^{N-1}} }\sqrt{\tfrac{{\lambda}^{(N)}_{\vec{k}, \ell'}}{\mathscr{S}_{\ell'}^{N-1}}
} \ket{ {\lambda}^{(N)}_{\vec{k}, \ell}}\bra{{\lambda}^{(N)}_{\vec{k}, \ell'}}\;, 
\end{equation}
where  the first sum runs over the
$N$-uple $\vec{k} := \left( k_{1} , k_{2} , \dots k_{N} \right)$, and where we used the short-hand notation 
\begin{eqnarray}
{\lambda}^{(N)}_{\vec{k}, \ell} &:=& \prod_{a=1}^{N} {\lambda}_{k_{a
},\ell} \;,
 \\
\ket{\lambda^{(N)}_{\vec{k}, \ell}}  &:=& \ket{ \lambda_{k_{1
},\ell} \; \lambda_{k_{2
},\ell}\; \dots \; \lambda_{k_{N
},\ell} }    \;. \label{EIGENRHOC} 
\end{eqnarray} 
\begin{figure}
	\includegraphics[width=\columnwidth]{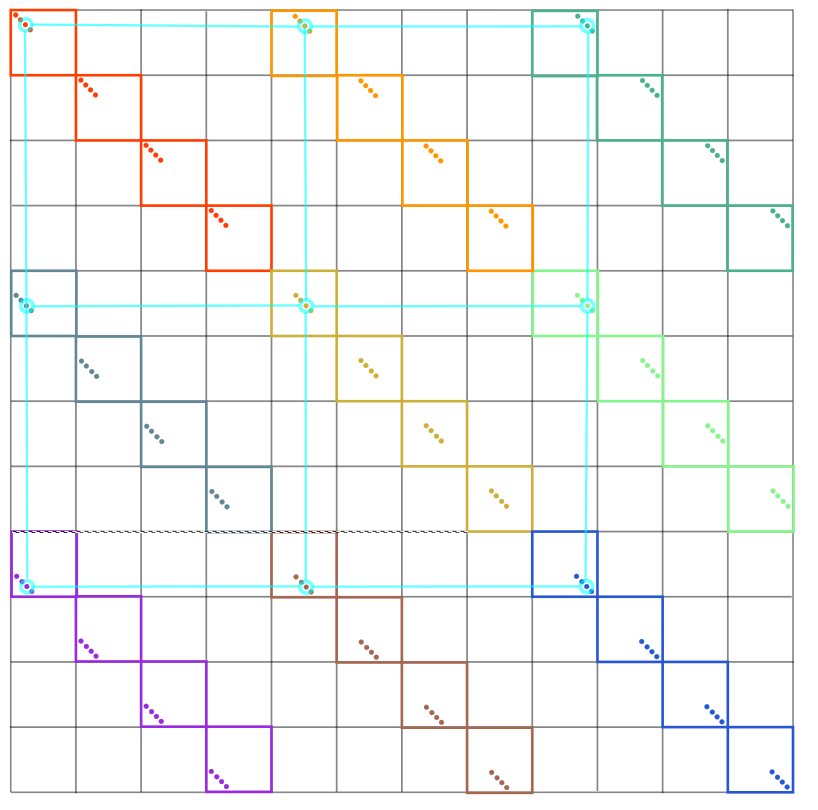}
	\caption[Decreasing the rank with correlations: first method]{Example of a
	block-wise purified state $\rho^{(N)}_{{\bf B}}$ of Eq.~(\ref{NORM}) with $N=2$, $\mbox{r}[\rho] =r =12$, $L=3$, $M=4$. The colored dots are the entries of the $r^N \times r^N$ density matrix which are different from zero. The dots connected by lines constitute a bloc of rank one.}
	\label{fig:method1}
\end{figure}
A close inspection of Eq.~(\ref{rhoN_method11}) reveals that 
the matrix $\rho^{(N)}_{{\bf B}}$ can be divided in a collection of
 uncoupled blocs, each one identified by a value of the $N$-uple $\vec{k}$, 
each having rank one, and with  non-zero eigenvalue given by 
\begin{equation}
\Lambda^{(N)}_{\vec{k}} := \sum_{\ell=1}^{L} \mathscr{S}_\ell \left(
\frac{\lambda^{(N)}_{\vec{k}, \ell} }{\mathscr{S}_\ell^{N} } \right) = 
 \sum_{\ell=1}^{L} 
\frac{\lambda^{(N)}_{\vec{k}, \ell} }{\mathscr{S}_\ell^{N-1} } 
 \;,
\label{autovalori_method1}
\end{equation}
(see Fig.~\ref{fig:method1}). 
 Specifically we get 
\begin{eqnarray} \label{NORM} 
\rho^{(N)}_{{\bf B}}  = \sum_{\vec{k}} \Lambda_{\vec{k}}^{(N)} \; 
|\Psi^{(N)}_{\vec{k}}\rangle \langle \Psi^{(N)}_{\vec{k}}| \;, 
\end{eqnarray} 
with 
\begin{equation} 
|\Psi^{(N)}_{\vec{k}}\rangle  :=\tfrac{1}{\sqrt{\Lambda_{\vec{k}}^{(N)}}}  \sum_{\ell=1}^{L} 
\sqrt{
	\tfrac{\lambda^{(N)}_{\vec{k}, \ell}}{\mathscr{S}_{\ell}^{N-1}} }\; 
 \ket{ \lambda^{(N)}_{\vec{k}, \ell}  } \;, 
\end{equation} 
being orthonormal elements of $\mathfrak{S}^{(N)}$.
It is worth stressing that
 (\ref{NORM}) is 
properly normalized thanks to the fact that 
\begin{eqnarray}
\sum_{\vec{k}} \Lambda_{\vec{k}}^{(N)}  &=&  \sum_{\vec{k}}
\sum_{\ell=1}^{L} \mathscr{S}_\ell^{1-N} \left(\prod_{a=1}^{N} \lambda_{k_{a
},\ell}\right) 
\\ \nonumber 
&=&\sum_{\ell=1}^{L} \mathscr{S}_\ell^{1-N} \left(\sum_{k=1}^{\#{\mathfrak{K}}_{\ell}} \lambda_{k,\ell}\right)^N = \sum_{\ell=1}^{L} \mathscr{S}_\ell = 1\;. 
\end{eqnarray} 
At variance with  the states
~(\ref{DEFMIXGHZ}) defined in the previous section, 
 the individual eigenvectors $|\Psi^{(N)}_{\vec{k}}\rangle$  of $\rho^{(N)}_{{\bf B}}$
 are in general not translationally invariant, 
while of course~(\ref{NORM}) obeys  to such symmetry. 
The rank of $\rho^{(N)}_{{\bf B}}$ is equal to the number of $N$-uple $\vec{k}$: 
 using the assumption~(\ref{uniformP}) this leads to
\begin{eqnarray} \label{rankrhob} 
\mbox{rank}[ \rho^{(N)}_{{\bf B}}] = ({\mbox{r}[\rho]}/{L})^N\;,
\end{eqnarray} 
corresponding to a reduction of a factor $L^{-N}$ with respect to  $\rho^{\otimes N}$, which instead has rank $\mbox{r}[\rho]^N$. More generally, as shown in Appendix~\ref{APPGENA}, when
$L$ in not an exact divisor of $\mbox{r}[\rho]$, Eq.~(\ref{rankrhob}) gets replaced by 
\begin{eqnarray} \label{rankrhob1} 
\mbox{rank}[ \rho^{(N)}_{{\bf B}}] = \left( \#_{\max}\right)^N\;,
\end{eqnarray} 
with $\#_{\max}$ the maximum cardinalities of the of the elements of ${\mathfrak{P}}$.
 In the extremal cases where $L=1$ and $L=\mbox{r}[\rho]$ we get 
 \begin{eqnarray}
 \rho^{(N)}_{{\bf B}}\Big|_{L=1}  &=& 
   \rho^{\otimes N}  \;, 
\label{purest_separable_state} \\ 
\rho^{(N)}_{{\bf B}}\Big|_{L=\mbox{r}[\rho]}  &=&  |\mbox{GHZ}^{(N)}_\rho\rangle\langle \mbox{GHZ}^{(N)}_\rho|  \;,
\label{purest_separable_state} 
\end{eqnarray}
with the last expression implying that 
the present TI-MPO representation assigns to 
$|\mbox{GHZ}^{(N)}_\rho\rangle$ the same BLR value 
as the the TI-MPO representation of Sec.~\ref{quarta} -- indeed in both cases
 predict the state to be an element of $\mathfrak{S}_T^{(N,\mbox{r}^2[\rho])}(\rho)$.

 \subsubsection{TI-MPO representation for classically correlated  density matrices}
\label{sec:method2_stato}
Our final example of TI-MPO quantum states is formed by a family of classically correlated density matrices 
represented by the elements $\rho^{(N)}_{{\bf C}}$ of Fig.~\ref{figura1},
which connect the completely uncorrelated  $\rho^{\otimes N}$ to the classical
correlated state $\rho^{(N)}_{\rm cc}$ of Eq.~(\ref{purest_separable_state0}).
As summarized in Table~\ref{tab1}  this family exhibits   behaviours in terms of 
 rank and BLR which is almost complementary with respect to those of
the family $\rho^{(N)}_{{\bf A}}$
of Sec.~\ref{quarta}. They will be used to derive {\bf Proposition 
\ref{prop:method2}}.

Starting again from the partition  ${\mathfrak{P}}$ of Eq.~(\ref{PARTITION})
we now consider the following TI-MPO operator 
\begin{eqnarray}
\rho^{(N)}_{{\bf C}} &:=& \sum_{i_1,j_1,\dots,i_N,j_N}  
  \ket{\lambda_{i_1} \dots \lambda_{i_N}}\bra{\lambda_{j_1} \dots \lambda_{j_N}} \nonumber \\
&&\;  \times 
\Tr[{A}^{i_1,j_1}_{{\bf C}}  \dots {A}^{i_N,j_N}_{{\bf C}} ] \;,  \label{targetC} 
 \end{eqnarray} 
with matrices  
\begin{equation}
A^{ij}_{{\bf C}} := \lambda_i \mathscr{S}_{\ell^{[i]}}^{\frac{1-N}{N}}
\; \delta_{i,j}\; 
\ket{{v}_{\ell^{[i]}}}\bra{{v}_{\ell^{[i]}}}   \; ,
\label{A_metodo2}
\end{equation}
operating on an auxiliary Hilbert space ${\cal H}_{\bf C}$ of dimension $L$ and characterized by an orthonormal basis $\left\{ \ket{{v}_{\ell}} ; \ell= 1,\cdots,  L\right\}$ (in the above
expressions  $\ell^{[i]}$  and  $\mathscr{S}_{\ell}$  are defined as in  Eqs.~(\ref{mapping})  and~(\ref{sommeparziali}), respectively).
By direct inspection one can verify 
that these states respect the partial trace condition~(\ref{cond_traccia_parziale})
and admit the following diagonal form 
\begin{eqnarray}
\rho^{(N)}_{{\bf C}} &=& \sum_{\ell=1}^{L} \mathscr{S}_{\ell}^{1-N} \left(  \sum_{k=1}^{\# \mathfrak{K}_\ell} \lambda_{ k,\ell} \ket{\lambda_{k,\ell}}\bra{ \lambda_{k,\ell}} \right)^{\otimes N} \nonumber \\
&=&  \sum_{\ell=1}^{L} \sum_{\vec{k}\in \{ 1, \cdots,\# \mathfrak{K}_\ell\}^N}  \Lambda_{\vec{k},\ell}^{(N)}
| \lambda_{\vec{k},\ell}^{(N)} \rangle \langle \lambda_{\vec{k},\ell}^{(N)}|\;,
\label{rhoN_metodo2}
\end{eqnarray}
with eigenvectors $| \lambda_{\vec{k},\ell}^{(N)} \rangle$  defined as in Eq.~(\ref{EIGENRHOC}),
and  associated eigenvalues given by   
\begin{eqnarray}
\Lambda_{\vec{k},\ell}^{(N)}  : = \mathscr{S}_\ell \left(
\frac{\lambda^{(N)}_{\vec{k}, \ell} }{\mathscr{S}_\ell^{N} } \right)= 
 \mathscr{S}_\ell^{1-N} \lambda_{\vec{k},\ell}^{(N)} \;. \label{autovalori_method2}
\end{eqnarray}
\begin{figure}
	\includegraphics[width=\columnwidth]{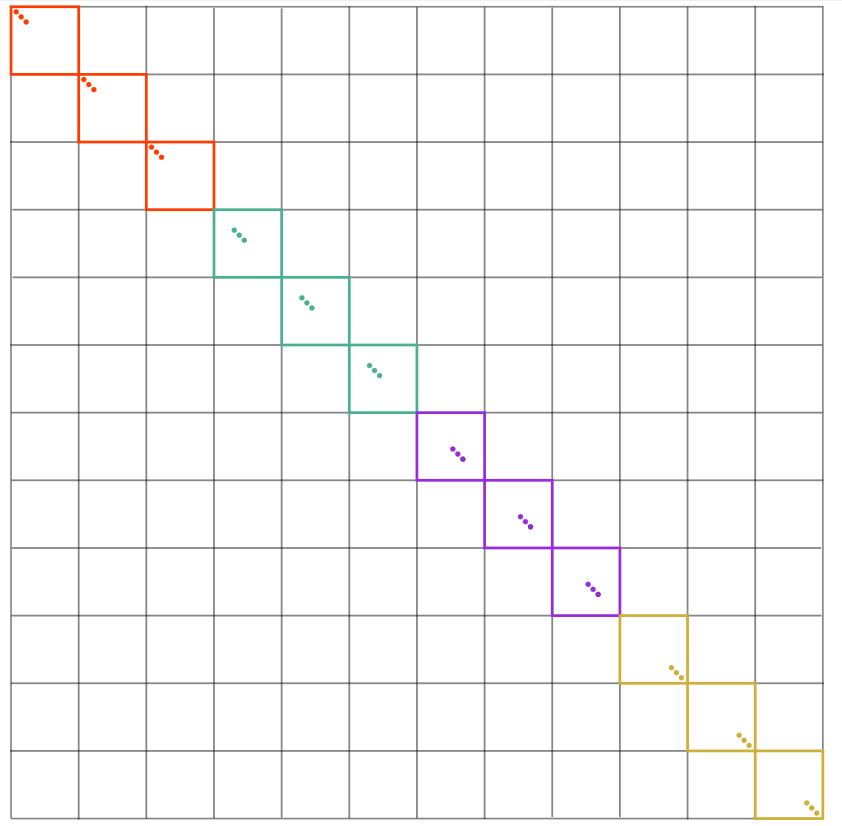}
	\caption[Decreasing the rank with correlations: second method]{Example of the state $\rho^{(N)}_{{\bf C}}$ defined in~(\ref{rhoN_metodo2}) with $N=2$, $\mbox{r}[\rho]=r=12$, $L=4$, ${M}=3$. The colored dots are the entries of the $r^N \times r^N$ density matrix which are different from zero.}
	\label{fig:MEPS_simplex}
\end{figure}
We stress that the states $\rho^{(N)}_{{\bf C}}$ are explicitly separable with respect to all possible partitions of the sites, and are diagonal in the same basis of the tensor
product state $\rho^{\otimes N}$. In particular, by choosing $\rho$ to be diagonal 
in the energy eigenbasis of $H$, we can force $\rho^{(N)}_{{\bf C}}$  to be
diagonal in the eigenbasis of $H^{(N)}$. 
By construction it also follows 
that the density matrix $\rho^{(N)}_{{\bf C}}$ 
 has BLR value that is upper bounded by $L$, so that \begin{eqnarray} \label{propertynew} 
\rho^{(N)}_{{\bf C}} \in \mathfrak{S}_T^{(N,L)}(\rho) \;, \end{eqnarray} 
(remember that for the case discussed in Sec.~\ref{quarta}, $L$ measured the rank of the state $\rho^{(N)}_{{\bf A}}$).
The  rank of  $\rho^{(N)}_{{\bf C}}$  is instead  equal to the total number of terms $| \lambda_{\vec{k},\ell}^{(N)} \rangle$
entering  (\ref{rhoN_metodo2}), i.e. 
\begin{eqnarray}\label{rankCC} 
\mbox{rank}[ \rho^{(N)}_{{\bf C}}] = \sum_{\ell=1}^L \left(\# \mathfrak{K}_\ell\right)^N =: |{\mathfrak{P}}|^{(N)} \;,
\end{eqnarray} 
where  in the last identity we invoked Eq.~(\ref{ineqbondAeq}) -- for $\rho^{(N)}_{{\bf A}}$, the term 
$|{\mathfrak{P}}|^{(2)}$
was instead an upper bound for the BLR. 
 In particular for uniform partitions Eq.~(\ref{rankCC}) corresponds to have
\begin{eqnarray} 
\mbox{rank}[ \rho^{(N)}_{{\bf C}}]\Big|_{{\mathfrak{P}}_{\mbox{\tiny{unif}}}} =  L ({\mbox{r}[\rho]}/{L})^N\;,
\end{eqnarray}  with  a reduction of a factor $L^{-N + 1}$ with respect to the rank of the completely uncorrelated state $\rho^{\otimes N}$.

 We conclude by observing that in the extremal cases where $L=1$ and $L=\mbox{r}[\rho]$ we get the tensor product and the the fully correlated classical state respectively, i.e. 
 \begin{eqnarray}
 \rho^{(N)}_{{\bf C}}\Big|_{L=1}  = 
 \rho^{\otimes N}  \;, 
\label{factorized} \qquad 
\rho^{(N)}_{{\bf C}}\Big|_{L=\mbox{r}[\rho]}  =   \rho^{(N)}_{\rm cc}\;. 
\end{eqnarray}
 Notice finally that  the  TI-MPO representation associated with the 
 family $\rho^{(N)}_{{\bf C}}$ assigns to 
$ \rho^{(N)}_{\rm cc}$ the same BLR value 
as the the TI-MPO representation of the family $\rho^{(N)}_{{\bf A}}$ of
Sec.~\ref{quarta} -- indeed in both cases
 predict the state to be an element of $\mathfrak{S}_T^{(N,\mbox{r}[\rho])}(\rho)$.

\section{Derivations}\label{sec:claims} 

This section is dedicated to derive  the results anticipated in Sec.~\ref{sec:sum}.

\subsection{Proof of  Propositions~\ref{prop:method1} and~\ref{prop:method2}}

Recalling the definition $\mathfrak{E}_s$ of the thermal energy associated with the single-site entropy $s$ given in Eq.~(\ref{def_frakE}) and 
of the corresponding effective heat  capacity $\mathfrak{C}_s$,
we start by proving the following statement:
\begin{proposition}
\label{prop:typicality} 
For any single-site entropy  value $s_0>0$ there exist $\eta_{\star}\in \; ]0,1[$ and a function
$N_{\star}(\eta)$ mapping $\eta \in \; ]0,\eta_{\star}[$ into 
$\mathbb{R}^+$, 
such that given $\eta \in\;  ]0,\eta_{\star}[$ and $N\geq N_{\star}(\eta)$, any state $\rho^{(N)}\in \mathfrak{S}_T^{(N)}$ admitting a eigenvalue subset 
${\mathfrak{T}}^{(N)}$ 
that has cardinality and associated total
population satisfying the constraints
\begin{eqnarray}
 \# {\mathfrak{T}}^{(N)} &\leq& e^{N \left( s_0 + \eta \right)} \; , 
\label{ipotesi_popsT}\\
\sum_{\lambda_i^{(N)} \in {\mathfrak{T}}^{(N)}} \lambda_i^{(N)} &\geq& 1 - e^{-\frac{2N\eta^2}{\alpha^2}}  \; ,
\label{ipotesi_cardT}
\end{eqnarray}
for some given $\alpha>0$, 
fulfils the following inequality 
\begin{eqnarray} 
&& \frac{{\cal E}(\rho^{(N)}; H^{(N)})}{N} \label{bound_ergotropia_prop1} \\
&&\quad \geq E(\rho; H)-\mathfrak{E}_{s_0} - (2\sqrt{2.01 \frak{C}_{s_0} \eta}
+ \epsilon_d 
  e^{-\frac{2N\eta^2}{\alpha^2}}) \;,\nonumber
\end{eqnarray} 
where $\epsilon_{d}$ is the largest eigenvalue of the single-site Hamiltonian $H$.
\end{proposition}
{\em Proof:--} 
Given  $s_0>0$ and $\alpha>0$, assume that 
 $\rho^{(N)}$ is an element of $\mathfrak{S}_T^{(N)}$  which 
admits an  subset 
${\mathfrak{T}}^{(N)}$ of eigenvalues fulfilling the conditions~(\ref{ipotesi_popsT}) and (\ref{ipotesi_cardT}) for some $\eta>0$ whose value will be determined in the following. Since the ergotropy, as defined in Eq.~(\ref{correlated_ergotropy}), is a maximum over the set of unitary transformation, to prove statement of the proposition 
 it would be sufficient to show an unitary transformation $U_\star \in {\bf U}({\cal H}^{\otimes N})$ such that the associated mean energy difference 
\begin{eqnarray}
W_{U_\star}(\rho^{(N)}; H^{(N)}) := E(\rho^{(N)}; H^{(N)}) - E(U_\star \rho^{(N)} U^\dagger_\star; H^{(N)})
\nonumber \\
\end{eqnarray}
is greater than the right-hand side of~(\ref{bound_ergotropia_prop1}). As suitable candidates for 
$U_\star$  we consider a unitary transformation which exchanges the eigenvectors associated with the elements of the set ${\mathfrak{T}}^{(N)}$ with a subset of the eigenvectors corresponding to the elements of the sets
 ${\mathfrak{A}_{\xi}^{(N)}}$ of  {\bf {Lemma} \ref{prop:insiemeA}} associated to the same value of $s_0$
and to a proper choice of $\xi$. 
  This construction might be regarded as a more general case of the one proposed in~\cite{PerarnauLlobet2019}, where it is applied for reducing the fluctuations in the extracted work - while we are interested in the mean values $W_{U_*} \left( \rho^{\otimes N}; H^{(N)}   \right)$.
 The  transformation we are targeting clearly can be identified if in ${\mathfrak{T}}^{(N)}$ there are more elements  than in ${\mathfrak{A}^{(N)}}$, i.e. 
\begin{equation}
\# {\mathfrak{A}_{\xi}^{(N)}}  \geq \# {\mathfrak{T}}^{(N)} \;. 
\label{condition_cardinalities}
\end{equation}
 The hypothesis~(\ref{ipotesi_cardT}) gives us the upper bound for the cardinality of ${\mathfrak{T}}^{(N)}$, while
 {\bf {Lemma} \ref{prop:insiemeA}} provides the lower bound~(\ref{cond_numelementiA}) for the cardinality of ${\mathfrak{A}_{\xi}^{(N)}}$ providing that the selected $\xi$ and $N$ 
 fulfils the constraints 
 \begin{eqnarray} \label{DFDA} 
 \xi \leq \xi_* \;, \qquad N\geq N_*(\xi)\;,
 \end{eqnarray} 
 with $\xi_*$ and $N_*(\xi)$ dependent upon $s_0$ and the structure of single-site Hamiltonian -- see 
  {\bf Remark 4} below {\bf {Lemma} \ref{prop:insiemeA}} in Appendix~\ref{applemma}.
 Combining these facts, we can see that the requirement~(\ref{condition_cardinalities}) can be satisfied by imposing
\begin{equation}
\label{cond_etaxi}
\eta=  \frac{\xi^2}{2.01 \frak{C}_{s_0}} \;, 
\end{equation} 
which we write incorporating the assumption~(\ref{fissozeta}). 
Notice now that with this choice all the eigenvalues of $\rho^{(N)}$ belonging to ${\mathfrak{T}}^{(N)}$ will be 
associated with energy eigenvalues $\epsilon_{\vec{i}}$ that can be upper bounded via
the  inequality~(\ref{cond_energiainsiemeA}); the remaining one instead can be bounded by 
the maximum eigenvalue of $H^{(N)}$, i.e. $N \epsilon_d$. Accordingly we can write 
\begin{eqnarray}
&&E(U_{\star} \rho^{(N)} U_{\star}^\dagger; H^{(N)}) 
\nonumber \\
&& \quad \leq N (\mathfrak{E}_{s_0} + 2\xi)
\left( \sum_{\lambda_i^{(N)} \in {\mathfrak{T}}^{(N)}} \lambda_i  \right) 
+  N \epsilon_d \left( \sum_{\lambda_i^{(N)} \notin {\mathfrak{T}}^{(N)}} \lambda_i^{(N)} \right)
 \nonumber \\
&& \quad \leq N (\mathfrak{E}_{s_0} + 2\xi) \nonumber 
 +  N \epsilon_d \left(1-  \sum_{\lambda_i^{(N)} \in {\mathfrak{T}}^{(N)}} \lambda_i \right)\\
 && \quad \leq 
 N (\mathfrak{E}_{s_0} + 2\xi) 
 + N \epsilon_d 
 e^{-\frac{2N\eta^2}{\alpha^2}}  \nonumber \\
 && \quad  = 
 N (\mathfrak{E}_{s_0} + 2\sqrt{2.01\frak{C}_{s_0} \eta} ) 
 + N \epsilon_d  \nonumber
  e^{-\frac{2N\eta^2}{\alpha^2}}  \;,
 \label{bound_passivo_new}
\end{eqnarray}
where in the last passage we used Eq.~(\ref{ipotesi_cardT}).
From this we can then obtain the thesis 
\begin{eqnarray}  
&& \frac{{\cal E}(\rho^{(N)}; H^{(N)})}{N} \geq W_{U_\star}(\rho^{(N)}; H^{(N)})\\
&&\quad \geq E(\rho; H)-\mathfrak{E}_{s_0} - (2.01 \sqrt{2\frak{C}_{s_0} \eta}
+ \epsilon_d 
  e^{-\frac{2N\eta^2}{\alpha^2}}) \;,\nonumber
\end{eqnarray}
which according to (\ref{cond_etaxi}) and (\ref{DFDA})  is valid for $\eta>0$ satisfying the constraint 
\begin{eqnarray} \label{con1} 
\eta \leq \eta_{\star} :=\min\left\{\frac{\xi_*^{2}}{2.01\frak{C}_{s_0}}, 1\right\}\;, 
\end{eqnarray} 
and for $N$ integers such that 
\begin{eqnarray}\label{con2} 
N \geq N_{\star}(\eta):= N_*(\sqrt{2.01 \frak{C}_{s_0} \eta})\;. 
\end{eqnarray} 
Notice in particular that  thanks to~(\ref{defNast}) we can fix the functional dependence 
of $N_{\star}(\eta)$ as 
 \begin{eqnarray} \label{defNstar} 
N_{\star}(\eta) = {K_{\star} }/{ \eta} \;,
\end{eqnarray} 
with the constant term 
\begin{eqnarray} 
K_{\star} := \frac{K_{*} }{2.01 \frak{C}_{s_0} }\;, 
\end{eqnarray} 
depending upon $s_0$ and $H$.  $\square$
\\

{\bf Remark 1:} We stress that  the coefficient $2.01$ appearing on the right-hand-side of Eq.~(\ref{bound_ergotropia_prop1}) 
is a byproduct of the choice~(\ref{fissozeta}):  generalization for arbitrary $\zeta\in ]0,1[$ can be obtained by simply replacing $2.01$  with $2/\zeta$. 
\\

{\bf Remark 2:} The thermal capacity $\frak{C}_{s_0}$ is always finite. In fact, as shown in Ref.\cite{Correa2015}, it can be bounded by a constant which depends only on the dimension $d$ of the Hilbert space. Therefore we can always rewrite~(\ref{bound_ergotropia_prop1}) in the weaker form
\begin{eqnarray} 
&& \frac{{\cal E}(\rho^{(N)}; H^{(N)})}{N} \label{bound_ergotropia_prop1_2} \\
&&\quad \geq E(\rho; H)-\mathfrak{E}_{s_0} - (2\sqrt{2.01 \frak{C}_{\max} \eta}
+ \epsilon_d 
e^{-\frac{2N\eta^2}{\alpha^2}}) \;,\nonumber
\end{eqnarray} 
with the definition 
\begin{eqnarray} \label{CMAXX} 
\frak{C}_{\max} := \max_{0 < s < \ln \mbox{r}[\rho]} \frak{C}_{s} \; .
\label{Cmax}
\end{eqnarray}

\subsubsection{Proof of {\bf Proposition \ref{prop:method1}}}
\label{sec:method1}
The result follows by showing that 
 family of density matrix $\rho^{(N)}_{{\bf B}}$ introduced in Sec.~\ref{sec:method1_stato} and Appendix~\ref{APPGENA}  fulfils the hypotheses of {\bf Proposition~\ref{prop:typicality}}.
  In particular we shall focus on those cases where the partition 
${\mathfrak{P}}$ of Eq.~(\ref{PARTITION}) is almost uniform, so that  Eq.~(\ref{almuniformP}) holds true.
Under this condition for every $\ell \in \{1,\cdots, L\}$, let us introduce the quantities 
\begin{equation}
s[\mathfrak{K}_\ell] := - \sum_{k = 1}^{\# \mathfrak{K}_\ell} \frac{\lambda_{k,\ell}}{\mathscr{S}_\ell} \ln{\frac{\lambda_{k,\ell}}{\mathscr{S}_\ell}} \, ,
\label{s_piccolo}
\end{equation}
which represents the Shannon entropy of the probability distribution 
$\{ \lambda_{ k,\ell} / \mathscr{S}_\ell \}_{k = 1,\dots,\# \mathfrak{K}_\ell}$.
Since 
each $s[\mathfrak{K}_\ell]$ is the sum of $\# \mathfrak{K}_\ell$ terms, 
from (\ref{maxcard}) and (\ref{almuniformP}) 
we have the trivial bound
\begin{equation}
s[\mathfrak{K}_\ell] \leq  \ln (\#_{\max}) = \ln \left \lceil{\mbox{r}[\rho] / L}\right \rceil \;.
\label{massimo_smax}
\end{equation} 
Identifying  then the probability distribution $\mathfrak{X}$ of  {\bf Lemma~\ref{lemma:chernoff_popolazioni}} with $\{ \lambda_{ k,\ell} / \mathscr{S}_\ell \}_{k = 1,\dots,\# \mathfrak{K}_\ell}$, it follows that 
the set 
\begin{equation}
{\mathfrak{X}}_\ell^{(N)} :=  \left\{ \vec{k} \; \middle\vert \; \mathscr{S}^{-N}_\ell \prod_{a=1}^{N} \lambda_{ k_a, \ell } \geq e^{-N(s[\mathfrak{K}_\ell] +\eta)} \right\}\;,
\label{def_mathfrakTi}
\end{equation}
has  cardinality bounded by
\begin{equation}
\# {\mathfrak{X}}_\ell^{(N)} \leq e^{N(s[\mathfrak{K}_\ell] + \eta)} \; ,
\label{card_mathfrakT}
\end{equation}
and satisfy the inequality 
\begin{equation}
\sum_{ \vec{k}
 \in {\mathfrak{X}}_\ell^{(N)}}  \prod_{a=1}^{N} \lambda_{k_a,\ell} 
\geq \mathscr{S}^N_\ell
\left( 1 - e^{-\frac{2N\eta^2}{ \alpha^2[\mathfrak{K}_\ell] } } \right) \; ,
\label{PoptotaleinTi_conalphai}
\end{equation}
with
\begin{equation}
\alpha[\mathfrak{K}_\ell]  := \max_{k =1,\cdots,  \# \mathfrak{K}_\ell}\ln\lambda_{k,\ell} - 
\min_{k =1,\cdots,  \# \mathfrak{K}_\ell}\ln\lambda_{k,\ell} \; .
\label{def_alphai}
\end{equation}
Notice that~(\ref{def_alphai}) is smaller than or equal to the  logarithmic spectral ratio $\alpha(\rho)$
of the single-site density matrix, i.e. the quantity 
\begin{eqnarray}  
 \alpha(\rho) := \ln\left( \lambda_{\max} / \lambda_{\min} \right) \label{def_alpha}\;,
\end{eqnarray}
with $\lambda_{\max}$ and $\lambda_{\min}$ being respectively the maximum and the minimum
positive eigenvalues of $\rho$.
We can hence replace (\ref{def_alphai}) with the inequality 
\begin{equation}
\sum_{ \vec{k}
 \in {\mathfrak{X}}_\ell^{(N)}}  \prod_{a=1}^{N} \lambda_{ k_a,\ell } 
\geq \mathscr{S}^N_\ell
\left( 1 - e^{-\frac{2N\eta^2}{\alpha^2(\rho)} } \right) \; .
\label{PoptotaleinTi}
\end{equation}
Observe next that  the union of the all the sets ${\mathfrak{X}}_\ell^{(N)}$, 
\begin{equation}
{\mathfrak{X}}^{(N)}:= \bigcup_{\ell=1}^L {\mathfrak{X}}_\ell^{(N)}\;,
\label{unione_T}
\end{equation}
has cardinality bounded by
\begin{equation}
\# {\mathfrak{X}}^{(N)} \leq \sum_{\ell=1}^L \# {\mathfrak{X}}^{(N)}_\ell \leq \sum_{\ell=1}^L e^{N(s[\mathfrak{K}_\ell] + \eta)} \leq L e^{N (  \ln \left \lceil{\mbox{r}[\rho] / L}\right \rceil + \eta)}\;,
\label{cardT_method1}
\end{equation}
where  in the third inequality we used (\ref{card_mathfrakT}) and 
in the last passage we invoked (\ref{massimo_smax}).

Let us recall that  the eigenvalues  of $\rho^{(N)}_{{\bf B}}$ are the quantities
$\Lambda^{(N)}_{\vec{k}}$  reported in Eq.~(\ref{autovalori_method1}) and  consider 
the subset ${\mathfrak{T}}_{\bf B}^{(N)}$ of such values characterized by  $\vec{k}$ vectors belonging to ${\mathfrak{X}}^{(N)}$, i.e. 
\begin{eqnarray} 
{\mathfrak{T}}^{(N)}_{\bf B} := \left\{ \Lambda^{(N)}_{\vec{k}} \big| \vec{k}\in {\mathfrak{X}}^{(N)}\right\}\;. 
\end{eqnarray} 
On one hand, using Eq.~(\ref{cardT_method1}) we can hence write
\begin{equation}
\# {\mathfrak{T}}^{(N)}_{\bf B} = 
\# {\mathfrak{X}}^{(N)} \leq e^{N (  \ln \left \lceil{\mbox{r}[\rho] / L}\right \rceil + N^{-1}\ln L+ \eta)}\;,
\label{cardT_method101}
\end{equation}
while, on the other hand, using~(\ref{unione_T}) and (\ref{PoptotaleinTi})  we can infer that the total population 
associated with such subset  is at least
\begin{eqnarray} \label{popsT_method1}
\sum_{ \Lambda^{(N)}_{\vec{k}} \in {\mathfrak{T}}^{(N)}_{\bf B}}  \Lambda^{(N)}_{\vec{k}}&=&
\sum_{\vec{k} \in {\mathfrak{X}}^{(N)}}
\sum_{\ell=1}^{L} \mathscr{S}_\ell^{1-N} \prod_{a=1}^{N} \lambda_{k_a,\ell}  \\
&\geq&
\sum_{\ell=1}^{L} \sum_{\vec{k} \in {\mathfrak{X}}_\ell^{(N)}} \mathscr{S}_\ell^{1-N} \prod_{a=1}^{N} \lambda_{k_a,\ell} \nonumber \\
&\geq&
\left( 1 - e^{-\frac{2N\eta^2}{\alpha^2(\rho)} } \right) \sum_{\ell=1}^{L} \mathscr{S}_\ell =   1 - e^{-\frac{2N\eta^2}{\alpha^2(\rho)} } \nonumber   \;.
\end{eqnarray}
Equations~(\ref{cardT_method1}) and~(\ref{popsT_method1}) certify that the set ${\mathfrak{T}}_{\bf B}^{(N)}$ fulfils the hypotheses~(\ref{ipotesi_cardT}) and~(\ref{ipotesi_popsT}) of {\bf Proposition~\ref{prop:typicality}}, with 
\begin{eqnarray} \label{defs0B} s_0 =  \ln \left \lceil{\mbox{r}[\rho] / L}\right \rceil  + N^{-1}\ln L\;. \end{eqnarray} Therefore we can conclude there exist  $\eta_{\star}\in \; ]0,1[$ and a function
$N_{\star}(\eta)$ mapping $\eta \in \; ]0,\eta_{\star}[$ into 
$\mathbb{R}^+$, 
such that given $\eta \in\;  ]0,\eta_{\star}[$ and $N\geq N_{\star}(\eta)$, we have 
\begin{eqnarray} 
&& \frac{{\cal E}(\rho^{(N)}_{{\bf B}}; H^{(N)})}{N} \label{ergotropia_correlata} \\
&& \quad \geq E(\rho; H)-\mathfrak{E}_{s_{\bf B}} 
 - (2\sqrt{2.01\frak{C}_{\max}  \eta}
+ \epsilon_d 
  e^{-\frac{2N\eta^2}{\alpha^2(\rho)}}) \;,\nonumber
\end{eqnarray}
where we used the weaker version~(\ref{bound_ergotropia_prop1_2}) of (\ref{bound_ergotropia_prop1}) discussed in {\bf Remark 2}, and recall 
 Eq.~(\ref{propertynewMM})  to set $m = L^2$ to identify
$s_0$ of Eq.~(\ref{defs0B}) with $s_{\bf B}$ of Eq.~(\ref{sBdef}). 
{\bf Proposition~\ref{prop:method1}} now finally follows by 
identifying the constants  $C$ and $\alpha$ of Eq.~(\ref{bound_ergotropia_prop1NEW})
with $\frak{C}_{\max}$ and 
$\alpha(\rho)$ respectively, and 
observing that for each $m$ and $\eta$, 
by choosing $N$ sufficiently large Eq.~(\ref{ergotropia_correlata}) is a trivial lower bound for the $\mathcal{E}^{(N,m)}_{\max}(\rho; H)$. 
$\square$
\\

\subsubsection{Proof of {\bf Proposition~\ref{prop:method2}}}
\label{sec:method2}

	The proof exploits again {\bf Proposition~\ref{prop:typicality}} and closely mimics the one we presented for   {\bf Proposition~\ref{prop:method1}}, the only difference being that this time we replace  the states $\rho^{(N)}_{{\bf B}}$ with the family of classically correlated states $\rho^{(N)}_{{\bf C}}$ defined in Sec.~\ref{sec:method1_stato} (orange trajectory of
	Fig.~\ref{figura1}) under the hypothesis that it is generated by an almost uniform partition ${\mathfrak{P}}$ (see Appendix~\ref{APPGENA}), so that ~(\ref{almuniformP}) is true.

To prove {\bf Proposition~\ref{prop:method2}}, we need to show that the state  $\rho^{(N)}_{{\bf C}}$ has a subset of eigenvalues ${\mathfrak{T}}_{\bf C}^{(N)}$ which satisfies~(\ref{ipotesi_cardT}) and~(\ref{ipotesi_popsT}). 
Observe that in this case the eigenvalues of $\rho^{(N)}_{{\bf C}}$ are given by the expressions $\Lambda_{\vec{k},\ell}^{(N)}$ of Eq.~(\ref{autovalori_method2}) which are labelled by $\ell$ and $\vec{k}$ (instead the eigenvalues of $\rho^{(N)}_{{\bf B}}$ where
identified only by the vectors $\vec{k}$). 
Accordingly we define ${\mathfrak{T}}_{\bf C}^{(N)}$ as 
\begin{equation}
{\mathfrak{T}}_{\bf C}^{(N)}:= \bigcup_{\ell=1}^L {\mathfrak{T}}_{{\bf C},\ell}^{(N)}\;,
\label{unione_Tc}
\end{equation}
where for $\ell \in \{ 1,\cdots, L\}$ we take 
\begin{eqnarray} \label{boundTNC} 
{\mathfrak{T}}^{(N)}_{{\bf C},\ell} := \left\{ \Lambda^{(N)}_{\vec{k},\ell} \big| \vec{k}\in {\mathfrak{X}}_{\ell}^{(N)}\right\}\;,
\end{eqnarray} 
with ${\mathfrak{X}}_{\ell}^{(N)}$ defined as in Eq.~(\ref{def_mathfrakTi}). 
Accordingly we have 
\begin{eqnarray}\label{cardT_method10101}
\# {\mathfrak{T}}^{(N)}_{\bf C} =  \sum_{\ell=1}^L \# {\mathfrak{T}}^{(N)}_{{\bf C},\ell}&=&
\sum_{\ell=1}^L \# {\mathfrak{X}}_{\ell}^{(N)} \\ \nonumber 
&\leq& e^{N (  \ln \left \lceil{\mbox{r}[\rho] / L}\right \rceil + N^{-1}\ln L+ \eta)}\;,
\end{eqnarray}
where the first identity follows from the fact that the sets ${\mathfrak{T}}^{(N)}_{{\bf C},\ell}$
are disjoint, and where in the last inequality we invoked~(\ref{cardT_method1}). 
Furthermore we observe 
\begin{eqnarray} \nonumber 
\sum_{ \Lambda^{(N)}_{\vec{k},\ell} \in {\mathfrak{T}}^{(N)}_{\bf C}}  \Lambda^{(N)}_{\vec{k},\ell}&=&\sum_{\ell=1}^{L} \sum_{\Lambda^{(N)}_{\vec{k},\ell} \in {\mathfrak{T}}^{(N)}_{{\bf C},\ell}}  \Lambda^{(N)}_{\vec{k},\ell} \nonumber \\  \nonumber 
&=& \sum_{\ell=1}^{L} \sum_{{\vec{k}} \in {\mathfrak{X}}^{(N)}_{\ell}}  \Lambda^{(N)}_{\vec{k},\ell} \\ \nonumber 
&=& \sum_{\ell=1}^{L}  \sum_{{\vec{k}} \in {\mathfrak{X}}^{(N)}_{\ell}} \mathscr{S}_\ell^{1-N} \lambda_{\vec{k},\ell}^{(N)}\\
&\geq&
  1 - e^{-\frac{2N\eta^2}{\alpha^2(\rho)} } \;,
  \label{popsT_method101}
\end{eqnarray}
where the last passage follows directly from~(\ref{popsT_method1}). 
We can therefore apply {\bf Proposition~\ref{prop:typicality}} with $s_0$
as in Eq.~(\ref{defs0B}) obtaining  that 
 there exist  $\eta_{\star}\in \; ]0,1[$ and a function
$N_{\star}(\eta)$ mapping $\eta \in \; ]0,\eta_{\star}[$ into 
$\mathbb{R}^+$, 
such that given $\eta \in\;  ]0,\eta_{\star}[$ and $N\geq N_{\star}(\eta)$, we have 
\begin{eqnarray} 
&& \frac{{\cal E}(\rho^{(N)}_{{\bf C}}; H^{(N)})}{N} \label{ergotropia_correlatanew} \\
&& \quad \geq E(\rho; H)-\mathfrak{E}_{s_{\bf C}} 
 - (2\sqrt{2.01\frak{C}_{\max}  \eta}
+ \epsilon_d 
  e^{-\frac{2N\eta^2}{\alpha^2(\rho)}}) \;,\nonumber
\end{eqnarray}
where we recalled (\ref{propertynew}) to set $m=L$ and transform 
$s_0$ into $s_{\bf C}$ of Eq.~(\ref{DEFSC}). 
{\bf Proposition~\ref{prop:method2}} now finally follows by 
identifying again the constants  $C$ and $\alpha$ 
with $\frak{C}_{\max}$ and 
$\alpha(\rho)$ respectively, 
and by
observing that for each $m$ and $\eta$, 
by choosing $N$ sufficiently large Eq.~(\ref{ergotropia_correlatanew}) is a trivial lower bound for the $\mathcal{E}^{(N,m)}_{\max}(\rho; H)$.
$\square$ 
\\

{\bf Remark 3:} We now present the heuristic argument in support of the fact that,
as mentioned in Sec.~\ref{sec:sum}, it is reasonable to think that the functional dependency of $s_{\bf C}$ of Eq.~(\ref{DEFSC}) of~{\bf Proposition~\ref{prop:method2}} can be improved for small levels of correlation.
To see this observe that the following identity holds true
\begin{eqnarray}
\sum_{\ell=1}^{L} \mathscr{S}_\ell \; s[\mathfrak{K}_\ell]  &=& 
- \sum_{\ell=1}^{L} \sum_{k = 1}^{\# \mathfrak{K}_\ell}{\lambda_{k,\ell}}\ln{\frac{\lambda_{k,\ell}}{\mathscr{S}_\ell}} \nonumber \\
&=&- \sum_{\ell=1}^{L} \sum_{k = 1}^{\# \mathfrak{K}_\ell}{\lambda_{k,\ell}}\ln{{\lambda_{k,\ell}}} + \sum_{\ell=1}^{L} \sum_{k=1}^{\# \mathfrak{K}_\ell} \lambda_{k,\ell} \ln{\mathscr{S}_\ell} \nonumber \\
&=&
S(\rho)  + \sum_{\ell=1}^{L} \mathscr{S}_\ell \ln{\mathscr{S}_\ell} \;.
\label{s_piccolo2_media}
\end{eqnarray}
Because of~(\ref{normaliza}), equation~(\ref{s_piccolo2_media}) can be see as a weighted mean of the entropic quantities $s[\mathfrak{K}_\ell]$ when they are averaged with weights $\mathscr{S}_\ell$. 
Therefore, letting
\begin{equation}
s_{\max} := \max_{\ell} s[\mathfrak{K}_\ell]\;, 
\end{equation}
it is always true that
\begin{equation}
s_{\max}  \geq S(\rho) + \sum_{\ell=1}^{L} \mathscr{S}_\ell \ln{\mathscr{S}_\ell}\geq S(\rho) - \ln L\;, 
\label{bound_smax}
\end{equation}
with the inequalities~(\ref{bound_smax}) becoming equalities in the case in which all the partial sums $\mathscr{S}_\ell$ were equal, i.e. if
\begin{equation}
 \mathscr{S}_\ell = {1}/{L}  \;, \qquad \forall \ell\in \{1,\cdots,L\} \;.
\label{tutteleSuguali}
\end{equation}
The values of $\mathscr{S}_\ell$ are determined by the spectrum $\{ \lambda_i; i =1,\cdots, \mbox{r}[\rho]\}$  of the state $\rho$ and by our choice of the partition  ${\mathfrak{P}}$ of 
Eq.~(\ref{PARTITION}).
When $L \ll \mbox{r}[\rho]$, and if the eigenvalues of $\rho$ are sufficiently evenly distributed, we expect it to be possible  to group them  in $L$ subsets such that the total populations $\mathscr{S}_\ell$ in each subset are approximately equal in order to 
fulfil~(\ref{tutteleSuguali}) with good approximation. Accordingly we expect that in such conditions we can write 
\begin{equation}
s_{\max} \simeq S(\rho) - \ln L\;. 
\label{smax_bestcase}
\end{equation}
Thanks to this we can now replace~(\ref{massimo_smax}), with 
\begin{equation}
s[\mathfrak{K}_\ell] \leq  s_{\max}  \simeq S(\rho) - \ln L\;,
\label{massimo_smax_ineq}
\end{equation} 
and hence (\ref{cardT_method1}) with 
\begin{eqnarray}
\# {\mathfrak{X}}^{(N)} \leq \sum_{\ell=1}^L \# {\mathfrak{X}}^{(N)}_\ell &\leq& \sum_{\ell=1}^L e^{N(s[\mathfrak{K}_\ell] + \eta)} \nonumber \\
&\lesssim& L e^{N (  S(\rho) - \ln L+ \eta)}\;,
\label{cardT_method1new1}
\end{eqnarray}
and (\ref{cardT_method10101}) with 
\begin{eqnarray}\label{cardT_method10101new}
\# {\mathfrak{T}}^{(N)}_{\bf C} \lesssim e^{N (  S(\rho) - \frac{N-1}{N}\ln L+ \eta)}\;,
\end{eqnarray}
Following the final passages of the proof we thus arrive to the conclusion that 
Eq.~(\ref{ergotropia_correlatanew}) holds with $s_{\bf C}$ replaced by 
the term  $s_{\bf C}|_{\rm(heu)}$ of Eq.~(\ref{sCheu}).

\subsection{Proof of Proposition~\ref{prop:grandi_blr}} \label{dfds} 
To prove {\bf Proposition~\ref{prop:grandi_blr}} we focus on the family of states $\rho^{(N)}_{\bf A}$ introduced in Sec.~\ref{quarta}.

From Eq.~(\ref{ineqbondAeq1}) it follows that the quantity $|{\mathfrak{P}}|^{(2)}$ fulfils the inequality 
\begin{eqnarray}
\mbox{r}[\rho] \leq &|{\mathfrak{P}}|^{(2)}&  \leq \mbox{r}^2[\rho]  \;,  \label{ineqbondAeq12} 
\end{eqnarray} 
with the lower and upper value being attained respectively by fixing the number $L$ of elements of the partition choosing the partition  ${\mathfrak{P}}$ equal to $\mbox{r}[\rho]$ and $1$. Therefore for each given 
\begin{eqnarray} \label{intervallom} 
m\in \{  \mbox{r}[\rho],\cdots,\mbox{r}^2[\rho]\}\;,
\end{eqnarray}   we can identify a special partition ${\mathfrak{P}}$~(\ref{PARTITION}), such that
\begin{eqnarray}|{\mathfrak{P}}|^{(2)} \leq m\;.
\end{eqnarray} 
Invoking then Eq.~(\ref{ineqbound12}), we can claim that given $m$ as in Eq.~(\ref{intervallom}) the associated 
the density matrix  $\rho^{(N)}_{\bf A}$ of Eq.~(\ref{DEFMIXGHZ}) is an element of
$\mathfrak{S}_T^{(N,m)}(\rho)$, so that the following lower bound holds 
\begin{eqnarray} \label{upperbmgrandiA} 
\mathcal{E}^{(N,m)}_{\max}(\rho; H) &\geq& 
\frac{ \mathcal{E}(\rho^{(N)}_{\bf A}; H^{(N)}) }{N} \\  \nonumber 
&=& E(\rho; H) - \frac{\min_{U \in {\bf{U}}(d^N)} E(U\rho^{(N)}_{\bf A}U^\dagger; H^{(N)}) }{N}
\end{eqnarray} 
see Eq.~(\ref{problema_rhoN_Mfissato}) and (\ref{correlated_ergotropy}).
We remind that the minimization on the right-hand-side of (\ref{correlated_ergotropy}) can be explicitly performed 
producing a closed expression in terms of the spectra of $H^{(N)}$ and $\rho_{\bf A}^{(N)}$~\cite{Mirsky1975, Allahverdyan2004}. 
This yields to simplified formula 
\begin{equation}
\label{correlated_ergotropy}
\min_{U \in {\bf{U}}(d^N)} E(U\rho^{(N)}_{\bf A}U^\dagger; H^{(N)}) =\sum_{j=1}^{d^N} \epsilon_j^{(N)}
 \lambda^{(N,\downarrow)}_{j} \;,
\end{equation}
where $\epsilon_j^{(N)}$ are  the eigenvalues of $H^{(N)}$ that, as in the case of $H$ we organize in increasing order, i.e. 
\begin{eqnarray} \label{ENERGYSPECTRUMN} 
\epsilon_{j+1}^{(N)} \geq \epsilon_j^{(N)}\;, 
\end{eqnarray}  
while $\lambda_j^{(N,\downarrow)}$ are the eigenvalues of $\rho_{\bf A}^{(N)}$, which instead,
as indicated by the arrow,  are 
assumed to arranged in decreasing order, i.e. 
\begin{eqnarray} \label{spettrorhoN} 
\lambda_{j+1}^{(N,\downarrow)} \geq \lambda_j^{(N,\downarrow)}\;.
\end{eqnarray}  
 A closed look at Eq.~(\ref{DEFMIXGHZ}) reveals that these last quantities can be written as 
	\begin{eqnarray}
	\label{spettroghz}
	\lambda_j^{(N,\downarrow)}
	=
	\begin{dcases}
	\mathscr{S}_j^{(\downarrow)} & \mbox{ if } j \leq L \; ;  \\
	0 & \mbox{otherwise} \; ,
	\end{dcases}
	\end{eqnarray}
	where $\mathscr{S}_j^{(\downarrow)}$ are the partial sums~(\ref{sommeparziali}) rearranged in decreasing order, i.e. $\mathscr{S}_1^{(\downarrow)} \geq \mathscr{S}_2^{(\downarrow)} \geq \dots \geq \mathscr{S}_L^{(\downarrow)}$. Therefore we can write 
	\begin{eqnarray}
	\label{upperbmgrandiA3} 
	\min_{U \in {\bf{U}}(d^N)} E(U\rho^{(N)}_{\bf A}U^\dagger; H^{(N)}) &=&	\sum_{j=1}^{L}\epsilon_j^{(N)} \mathscr{S}_j^{(\downarrow)}  \;. 
	\end{eqnarray}
	Now we notice that the structure of the eigenvalues of the Hamiltonian $H^{(N)}$, given by~(\ref{autovalori_HN}), warrants that
	\begin{eqnarray}
	\epsilon_j^{(N)} \leq \epsilon_j \qquad \forall j \in [1,d] \; .
	\label{energiedepresse}
	\end{eqnarray}
	Since by construction $L \leq d$, the inequality~(\ref{energiedepresse}) is also true for $j \in [1,L]$,  and we can use it in~(\ref{upperbmgrandiA3}) obtaining
	\begin{eqnarray}
	\label{upperbmgrandiA4} 
	\min_{U \in {\bf{U}}(d^N)} E(U\rho^{(N)}_{\bf A}U^\dagger; H^{(N)}) \leq
	\sum_{j=1}^{L} \epsilon_j \mathscr{S}_j^{(\downarrow)} \; .
	\end{eqnarray} 
	Since the eigenvalues $\{ \mathscr{S}_j \}_{j = 1\dots L}$ are, by definition~(\ref{sommeparziali}), partial sums of the eigenvalues $\{ \lambda_j \}_{j = 1\dots d}$ of the single-site density matrix $\rho$, we can write (see the appendix B of~\cite{Alimuddin2020}):
	\begin{equation}
	\label{upperbmgrandiA5} 
	\min_{U \in {\bf{U}}(d^N)} E(U\rho^{(N)}_{\bf A}U^\dagger; H^{(N)}) \leq
	\sum_{j=1}^{L} \epsilon_j \mathscr{S}_j^{(\downarrow)} 
	\leq \sum_{j=1}^{d} \epsilon_j \lambda_j^{(\downarrow)} \;. 
	\end{equation} 
	Noticing finally that 
	\begin{eqnarray}
	 \sum_{j=1}^{d} \epsilon_j \lambda_j^{(\downarrow)} &=& \min_{U \in {\bf{U}}(d)} E(U\rho U^\dagger; H)  \nonumber \\
	 &=& {E}(\rho,H) - {\cal E}(\rho,H)\;,
	\end{eqnarray} 
	we finally arrive at~(\ref{upperbmgrandi}) replacing~(\ref{upperbmgrandiA5}) into~(\ref{upperbmgrandiA}).
	$\square$

\subsection{Proof of {\bf Corollary~\ref{prop:productstates}}} \label{sec:claimsnew}

In view of the identity~(\ref{lowerb}), 
 {\bf Corollary~\ref{prop:productstates}} can be seen as a refinement of 
{\bf Proposition \ref{prop:method1}} for $m=1$. Indeed we can derive the statement
following the same passages of Sec.~\ref{sec:method1} and observing that $\rho^{\otimes N}$ is the unique element of  family $\rho_{\bf B}^{(N)}$ we get when setting  $L=1$ (see Eq.~(\ref{purest_separable_state})). 
In this case Eqs.~(\ref{s_piccolo}) and (\ref{massimo_smax}) get replaced by
$s[\mathfrak{K}_1] =S(\rho)$ which in turn allow us to replace  Eqs.~(\ref{cardT_method1})
and (\ref{cardT_method101})  with 
\begin{equation}
\# {\mathfrak{T}}^{(N)}_{\bf B}=
\# {\mathfrak{X}}^{(N)} \leq  e^{N (  S(\rho)+ \eta)}\;.
\label{cardT_method1L1}
\end{equation}
Invoking hence (\ref{popsT_method1}) that still remains valid,
we can  conclude that now 
the set ${\mathfrak{T}}_{\bf B}^{(N)}$ fulfils the hypotheses~(\ref{ipotesi_cardT}) and~(\ref{ipotesi_popsT}) of {\bf Proposition~\ref{prop:typicality}}, with 
$s_0= S(\rho)$ hence leading to
\begin{eqnarray} 
&& \frac{{\cal E}(\rho^{\otimes N}; H^{(N)})}{N} \label{ergotropia_correlataL1} \\
&& \quad \geq E(\rho; H)-\mathfrak{E}_{S(\rho)} 
 - (2\sqrt{2.01\frak{C}_{{S(\rho)}}  \eta}
+ \epsilon_d 
  e^{-\frac{2N\eta^2}{\alpha^2(\rho)}}) \;,\nonumber
\end{eqnarray}
that corresponds to (\ref{CORO1}) by identifying $C$ and $\alpha$ with 
$\frak{C}_{{S(\rho)}}$ and $\alpha(\rho)$ respectively. 
$\square$

\section{Conclusions}\label{CONCLUSIONS}

We derived some analytic lower bound for the work that, in the best case, can be extracted with unitary transformations from a translationally invariant correlated many-body system, using as a measure of correlation the minimum bond link rank (BLR) necessary to represent the state as a matrix product operator.
When the number $N$ of copies of the system is finite, non-classical correlations are required to extract as work the full energy of the system. However, in the macroscopic limit $N \to \infty$, we found that this quantum feature disappears, and that it is possible to create many-body states with classically correlated state which have a relative low BRL (equal at most to to the rank $r$ of the local state, out of a maximum of $r^2$).
Our bounds do not depend on the entropy $S(\rho)$ on the state, so they are worse for states of low entropy. However, heuristic consideration suggest that, at least for small correlations strengths, the bounds can be improved with an explicit dependence on $S(\rho)$.

We conjecture that, for $N \geq 3$, the BLR of the family of states employed in our analysis is equal to the upper bounds that we found by explicit construction. If true, this could allow to derive also upper bound for the ergotropy of a translationally invariant state with a given correlation strength.
Another possible improvement of our work could be repeating the analysis with a measure of correlation more sophisticated the TI-MPO bond link rank, like some form of correlation entropy~\cite{Schindler2020, Rolandi2020}.

\acknowledgments 
We would like to thank  Giacomo De Palma for comments and discussions.
This work is supported by MIUR (Ministero dell’Istruzione, dell’Universit\`a e della Ricerca) via project PRIN 2017 {\it Taming complexity via Quantum Strategies a Hybrid Integrated Photonic approach} (QUSHIP) Id. 2017SRNBRK.

\appendix

\section{Chernoff inequality}
\label{sec:Chernoff}
If we extract $N$ times a random variable $X \in [a, b]$, the Chernoff bound~\cite{Chernoff1952} (or equivalently Hoeffding's inequality~\cite{Hoeffding1963}) tell us that
\begin{equation}
\mathbb{P} \left( \sum_{i=1}^N X_i \leq N( \mathbb{E}[\mathcal{V}] + \varepsilon ) \right) \geq 1- e^{-\frac{2N \varepsilon^2}{(b-a)^2}}\;.
\label{Chernoff_uptail}
\end{equation}
In this paper we will apply the following useful consequence of the Chernoff inequality:

\begin{lemma}
\label{lemma:chernoff_popolazioni}
Let $\mathfrak{X} = \{ x_i \}_i$ be a finite collection of positive real numbers $x_i \in ]0,1]$, such that $\sum_i x_i = 1$, and $-\sum_i x_i \ln x_i = s_0$. 
Let $\mathfrak{X}^{\otimes N}$ denote the set of $N$-ples $\vec{x}:=(x_{i_1}, \dots x_{i_N})$. Then, for any $\eta > 0$, the subset ${\mathfrak{X}}^{(N)} \subseteq \mathfrak{X}^{\otimes N}$ defined by
\begin{equation}
{\mathfrak{X}}^{(N)} := \left\{ \vec{x} = (x_{i_1}, \dots, x_{i_N}) \middle\vert \prod_{k=1}^Nx_{i_k} \geq e^{-N(s_0 + \eta) } \right \}\;,
\label{def_insiemetipico}
\end{equation}
has cardinality
\begin{eqnarray}
\# {\mathfrak{X}}^{(N)}\leq e^{N(s_0 + \eta) }\;,
\label{min_cardinalita_tipica}
\end{eqnarray}
and satisfies the property
\begin{eqnarray}
\sum_{ \vec{x} \in {\mathfrak{X}^{(N)}} } \prod_{k=1}^N x_{i_k} \geq 1 - e^{-2 N\eta^2/ \alpha^2} \; ,
\label{pops_tipica}
\end{eqnarray}
where
\begin{equation}
\alpha: = \max_{x\in\mathfrak{X}}\ln x - \min_{x\in\mathfrak{X}}\ln x = \ln \left( x_{max} / x_{min} \right)\;.
\label{def_alphax}
\end{equation}
\end{lemma}
{\em Proof:} 
The bound~(\ref{min_cardinalita_tipica}) on the cardinality follows trivially from the definition~(\ref{def_insiemetipico}) and the fact that 
\begin{eqnarray}
\sum_{\vec{x} \in {\mathfrak{X}^{(N)}} } \prod_{k=1}^N x_{i_k} \leq 
\sum_{ \vec{x} \in \mathfrak{X}^{\otimes N}} \prod_{k=1}^N x_{i_k}  = 
1 \; .
\end{eqnarray}
To prove~(\ref{pops_tipica})
 it is sufficient to notice that the quantity
$-\ln \prod_{k=1}^N x_{i_k} = -\sum_{k=1}^N \ln x_{i_k}$
can be regarded as the sum of $N$ extractions of the random variable $X :=- \ln x$, that is, of the variable that with probability $x_i$ takes the value $X_i = -\ln x_i$. 
Then the thesis~(\ref{pops_tipica}) follows straightforwardly from~(\ref{Chernoff_uptail}). 
$\square$

\section{Gibbs states} \label{sec:Gibbs}

The Gibbs states of a single-site of our model are the density matrices
\begin{equation}
\omega_\beta := \frac{e^{-\beta H}}{Z_\beta} \,\;, \quad Z_{\beta} := \Tr[e^{-\beta H} ] \; ,
\end{equation}
where the parameter $\beta\geq 0$ can be called the \emph{inverse temperature} of the system, in analogy with the classical case.
They are diagonal in the eigenbasis $\{ \ket{\epsilon_i} ; i\in [1,d]\}$
of $H$, 
\begin{equation}
\label{Gibbs}
\omega_\beta = \sum_{i=1}^d \hat{\lambda}_i(\beta)  \ket{\epsilon_i}\bra{\epsilon_i}\;,
\end{equation}
with  population given by 
\begin{eqnarray} \hat{\lambda}_i(\beta): = Z^{-1}_\beta e^{-\beta\epsilon_i}\;. \end{eqnarray} 
The quantity $Z_{\beta}$ is usually called the \emph{partition function} of the system and  
 allow us to establish a natural correspondences between the mean energy of 
 $\omega_\beta$ (a quantity that we shall refer to as the equilibrium energy of the model),  its entropy, and the inverse temperature $\beta$.
 Specifically we have 
 \begin{eqnarray}
\label{def_Eeq}
E_\beta &:=& E(\omega_\beta; H) = \Tr[\omega_\beta H]  = - \frac{\partial}{\partial \beta} \ln Z_\beta\;, \\ 
S_\beta&:=& S(\omega_\beta) = - \Tr[ \omega_\beta \ln \omega_\beta] = 
\beta E_\beta + \ln Z_\beta \nonumber \\
&=&  -\beta  \frac{\partial}{\partial \beta} \ln Z_\beta +  \ln Z_\beta\;, \label{def_Seq} 
\end{eqnarray}
which lead to the identity 
\begin{eqnarray}\label{IDE} 
\frac{\partial E_\beta}{\partial S_\beta}= 1/\beta\;. 
\end{eqnarray} 
The first derivative with respect to $\beta$ of the equilibrium energy~(\ref{def_Eeq}) define the heat capacity functional of the model, specifically 
\begin{eqnarray}
C_\beta :=-  \frac{\partial E_\beta}{\partial \beta} = \frac{\partial^2}{\partial^2 \beta} \ln Z_\beta \; ,
\label{def_captermica}
\end{eqnarray} 
which enters in  the following Taylor expansions formulas
\begin{eqnarray}
\label{E_primoordine}
E_{\beta'} &=& E_{\beta} - (\beta' - \beta) C_\beta+  \mathcal{O}\left(  (\beta' - \beta)^2 \right)\;,
\\ 
\ln Z_{\beta'} &=& 
\ln Z_{\beta} - (\beta' - \beta) E_\beta + \frac{1}{2}   (\beta' - \beta)^2 C_\beta
\nonumber \\
&& \qquad + \mathcal{O}\left((\beta' - \beta)^3\right) \;, \label{lnZsecondordine}
\end{eqnarray}
(the minus sign in Eq.~(\ref{def_captermica}) accounts for the fact that  that $\beta$ is an inverse temperature).
The functional $\ln Z_{\beta}$ can be shown to be decreasing and convex implying the  inequality 
\begin{eqnarray}
\ln Z_{\beta'} &>& \ln Z_{\beta} + (\beta' - \beta) \frac{\partial \ln Z_{\beta}}{\partial \beta} \nonumber \\
&=& \ln Z_{\beta} - (\beta' - \beta) E_\beta\;,  \label{lnZprimoordine} 
\end{eqnarray}
valid for $\beta' < \beta$. 
This property  also ensures also
 the positivity of $C_\beta$ which in turns implies 
 that both  $E_\beta$ and $S_\beta$ are monotonically decreasing functions of $\beta$
 in agreement with Eq.~(\ref{IDE}). 
Exploiting these one-to-one correspondences with $\beta$, we can naturally associate to each entropy value $s$ a  thermal energy value  $\mathfrak{E}_s$, 
a heat capacity $\frak{C}_s$, and an inverse
temperature $\beta_s$  via the identities 
\begin{eqnarray}
\label{def_frakE}
\left\{ \begin{array}{l} 
\frak{E}_s := E_\beta   \;, \\ \\
\frak{C}_s := C_\beta   \;, \\ \\
\beta_s : = \beta\;,  
\end{array} \right. 
 \iff s = S(\omega_\beta)\;.
\end{eqnarray}
More generally, given $\rho\in \mathfrak{S}$ a generic single-site 
state, we define its associated thermal energy $\frak{E}(\rho)$, an effective heat capacity $\frak{C}(\rho)$, and inverse effective temperature $\beta(\rho)$ via the identities
\begin{eqnarray}
\label{def_thermal}
\left\{ \begin{array}{l} 
\frak{E}(\rho) := E_\beta   \;, \\ \\
\frak{C}(\rho) := C_{\beta} \;, \\  \\ 
\beta(\rho) : = \beta\;,  
\end{array} \right. 
 \iff S(\rho) = S(\omega_\beta)\;.
\end{eqnarray}
Thanks to these definitions 
we can rewrite the corresponding total ergotropy~(\ref{asymptotic_ergotropyy}) as
 \begin{eqnarray}
\mathcal{E}_{\text{tot}}(\rho; H)
&=& E(\rho; H) - \frak{E}(\rho)\;. 
\label{asymptotic_ergotropyyr}
\end{eqnarray}

We conclude  by noticing that the above construction can be trivially generalized to the non-interacting 
$N$  sites model.
Specifically from~(\ref{HMDEF}) it follows that 
the associated Gibbs configurations are tensor product of single sites Gibbs states, i.e. 
 $\omega^{\otimes N}_{\beta}$. 
 Introducing $\ket{\epsilon_{\vec i}} := \ket{\epsilon_{i_1}} \otimes \dots \otimes \ket{\epsilon_{i_N}}$ the eigenvectors of $H^{(N)}$ we then observe that the following identity hold
\begin{eqnarray}
\label{autovalori_HN}
H^{(N)} \ket{\epsilon_{\vec i}} = \epsilon_{\vec{i}} \ket{\epsilon_{\vec i}} \; , \qquad 
\omega^{\otimes N}_{\beta} \ket{\epsilon_{\vec i}} = \hat\lambda_{\vec{i}} \ket{\epsilon_{\vec i}} \; , 
\end{eqnarray}
where
\begin{eqnarray}
\epsilon_{\vec{i}} :=  \sum_{k=1}^{N} \epsilon_{i_k} \;, \qquad 
\hat{\lambda}_{\vec{i}}(\beta) :=  \prod_{k=1}^{N} \hat\lambda_{i_k}(\beta)  \;,
\label{epsilonelambda_Gibbs}
\end{eqnarray}
which from Eq.~(\ref{Gibbs})  imply \begin{equation}
\label{rettaGibbs}
\ln\hat\lambda_{\vec{i}}(\beta) = N\ln Z_{\beta} + \beta \epsilon_{\vec{i}} \; .
\end{equation}

\subsection{A useful Lemma} \label{applemma}

In this section we  provide an estimation of the number of 
eiegenvalues of  $H^{(N)}$ with energy just above a given threshold linked via Eq.~(\ref{def_frakE}) 
 to single-site entropy values. Specifically, given 
 $s_0>0$, $N$ integer, and $\xi>0$
	define 
	\begin{eqnarray}
	\label{cond_energiainsiemeA}
		{\mathfrak{A}_{\xi}^{(N)}} :=\left\{ \epsilon_{\vec{i}}   \middle\vert N\mathfrak{E}_{s_0} \leq  \epsilon_{\vec{i}} \leq N(\mathfrak{E}_{s_0} + 2\xi)\right\} \;, 
	\end{eqnarray}
	the subset of the eigenvalues of the Hamiltonian $H^{(N)}$   whose energy share per site is  $2\xi$-close to 
	$\mathfrak{E}_{s_0}$. Then the following property holds:
\begin{lemma}
	\label{prop:insiemeA}	Given $\zeta <1$ a constant 
	strictly smaller than 1, for all $s_0>0$   	
	there exists   $\xi_*\in \;  ]0,1[$ such that for all $\xi\in\;  ]0,\xi_*[$, we can identify $N_*(\xi)$ integer such that 
for all $N\geq N_*(\xi)$ 
	the  cardinality of ${\mathfrak{A}_{\xi}^{(N)}}$ is bounded by the inequality 
	\begin{eqnarray}
\label{cond_numelementiA}
\# {\mathfrak{A}_{\xi}^{(N)}} \geq e^{N \left( s_0 + \frac{\zeta \xi^2}{{2}\frak{C}_{s_0}} \right)}  \; .
\end{eqnarray}	
\end{lemma}
{\it Proof:--}  We remind that according to the notation introduced in Eq.~(\ref{def_frakE}) 
$\mathfrak{E}_{s_0}$ is the mean energy of $E_{\beta_{s_0}}$ of a single-site Gibbs state $\omega_{\beta_{s_0}}$
with entropy $s_0$, ${\beta_{s_0}}$ being the corresponding inverse temperature defined as in 
 Eq.~(\ref{def_thermal}).
For $s_0$ assigned and $\xi>0$ sufficiently small consider  the inverse temperature $\beta'$ defined by the identity 
\begin{eqnarray} \label{primaprma} 
E_{\beta'} = E_{\beta_{s_0}} + \xi = \mathfrak{E}_{s_0} + \xi\;, 
\end{eqnarray}
which, by construction is slightly smaller than  $\beta_{s_0}$.
By virtue of Eq.~(\ref{cond_energiainsiemeA}) and 
~(\ref{rettaGibbs}) we have
\begin{eqnarray}
\label{definsiemeA_2}
 \epsilon_{\vec{i}} \in {\mathfrak{A}_{\xi}^{(N)}} &\iff& \left\lvert \epsilon_{\vec{i}} - NE_{\beta'} \right\rvert 
 \leq  N\xi 
\nonumber \\
&\iff& \left\lvert -\ln\hat\lambda_{\vec{i}}(\beta') - NS_{\beta'} \right\rvert \leq N\beta' \xi\;,
\end{eqnarray}
which in particular implies 
\begin{eqnarray}
\label{definsiemeA_213} 
\epsilon_{\vec{i}} \in {\mathfrak{A}_{\xi}^{(N)}}  \Longrightarrow \hat\lambda_{\vec{i}}(\beta') \geq e^{ - N(S_{\beta'} + \beta' \xi)} \;.
\end{eqnarray}
Remember then that the expected values of $-\ln\hat\lambda_{\vec{i}}(\beta')$ for the state $\omega^{\otimes N}_{\beta'}$ are:
\begin{eqnarray}
\mathbb{E}[-\ln\hat{\lambda}_{\vec{i}}(\beta')] &=&- \Tr[\omega^{\otimes N}_{\beta'} \ln \omega^{\otimes N}_{\beta'}] = NS_{\beta'} \;.\label{ExpSGibbs} 
\end{eqnarray}
Identifying hence the variables~$x_i$ of {\bf Lemma~\ref{lemma:chernoff_popolazioni}} of 
Appendix~\ref{sec:Chernoff} with the population~$\hat{\lambda}_i(\beta')$ of the Gibbs state 
$\omega_{\beta'}$, and $s_0$ with the associated entropy $S_{\beta'}$, from 
Eq.~(\ref{pops_tipica}) we can 
claim that 
in the state $\omega_{\beta'}^{\otimes N}$, the set ${\mathfrak{A}_{\xi}^{(N)}}$ hosts a total population of at least
\begin{equation}
\sum_{\vec{i}| \epsilon_{\vec{i}} \in {\mathfrak{A}_{\xi}^{(N)}}} \hat\lambda_{\vec{i}}(\beta') \geq 1 - e^{-\frac{2N\xi^2}{\epsilon^2_d} }\;,  \label{PoptotaleinA}
\end{equation}
where according to Eq.~(\ref{ENERGYSPECTRUM})  $\epsilon_d$ is the maximum eigenvalue of $H$.
Equation~(\ref{definsiemeA_2}) also implies 
that
\begin{equation}
\epsilon_{\vec{i}} \in {\mathfrak{A}_{\xi}^{(N)}}  \Longrightarrow 
\hat\lambda_{\vec{i}}(\beta') \leq e^{-N(S_{\beta'} - \beta' \xi)}  \; . \label{maxPopinA}
\end{equation}
From~(\ref{PoptotaleinA}) and~(\ref{maxPopinA}), it then follows that the cardinality of the set of eigenvalues ${\mathfrak{A}_{\xi}^{(N)}}$ is at least
\begin{equation}
\# {\mathfrak{A}_{\xi}^{(N)}} \geq e^{N(S_{\beta'} - \beta' \xi)} \left( 1 - e^{-\frac{2N\xi^2}{\epsilon^2_d} } \right)\;. 
\label{cardA}
\end{equation}
 From the second order expansion~(\ref{lnZsecondordine}) we get 
\begin{eqnarray} \nonumber 
e^{N(S_{\beta'} - \beta' \xi)} &=& e^{N\left(s_0 + \frac{1}{2} \frak{C}_{s_0} (\beta_0 - \beta')^2 + \mathcal{O}\left((\beta_0 - \beta')^3\right)\right)} \\ 
&=& e^{N \left( s_0 + \frac{\xi^2}{2\frak{C}_{s_0}} +  \mathcal{O}\left(\xi^3\right) \right)}\;, 
\label{slack_autovalori_xi}
\end{eqnarray}
where in the second identity we invoked~(\ref{E_primoordine}).
The term in $\mathcal{O}\left(\xi^3\right)$ could be positive or negative, depending on the sign of $\frac{d
{C}(\beta)}{d \beta}$ for $\beta=\beta_{s_0}$. Notice however that 
the inequality~(\ref{lnZprimoordine}) assures that
\begin{eqnarray}
e^{N(S_{\beta'} - \beta' \xi)} &=& e^{N \left( \ln Z_{\beta'} + \beta'  \mathfrak{E}_{s_0} \right) } 
\label{nuovaeq1} \\\nonumber 
&>& e^{N \left[ \ln Z_{\beta_0}+ (\beta_0 - \beta') \mathfrak{E}_{s_0}+ \beta' \mathfrak{E}_{s_0}\right] }
=
e^{Ns_0}, 
\end{eqnarray}
hence implying that 
 $\frac{\xi^2}{2\frak{C}_{s_0}} +  \mathcal{O}\left(\xi^3\right)$ in~(\ref{slack_autovalori_xi}) 
  is globally positive. 
Replacing~(\ref{slack_autovalori_xi}) into~(\ref{cardA}), we now get
\begin{eqnarray}
\# {\mathfrak{A}_{\xi}^{(N)}} &\geq& \exp \Big[ N \left( s_0 +  \tfrac{\xi^2}{2\frak{C}_{s_0}} +  \mathcal{O}\left(\xi^3\right) \right) \nonumber \\
&&\qquad + \ln\left(  1 -  \exp[{-\tfrac{2N\xi^2}{\epsilon^2_d}}] \right) \Big]\;. \label{questa} 
\end{eqnarray}
The thesis then follows by noticing that for fixed $\xi$ in the limit of large $N$  the 
exponent on the right-hand-side of (\ref{questa}) approaches  
$N \left( s_0 + \frac{\xi^2}{2\frak{C}_{s_0}} + \mathcal{O}\left(\xi^3\right) \right)$ which for enough $\xi$ 
small can be forced to be larger than $N \left( s_0 + \zeta \frac{\xi^2}{ \frak{C}_{s_0}}  \right)$.
To see this explicitly observe for instance that
 for each fixed $\xi>0$ there exists $N_*(\xi)$  integer such that
for all $N\geq N_*(\xi)$ we can ensure 
\begin{eqnarray} \label{hold} 
 \frac{\ln(  1 - \exp[{-\tfrac{2N\xi^2}{\epsilon^2_d}}])}{N} \geq -  \left(\frac{1-\zeta}{2}\right) \frac{ \xi^2}{2\frak{C}_{s_0}}\;,
\end{eqnarray}  
a condition that allows  us to replace  (\ref{questa}) with 
\begin{equation}
\# {\mathfrak{A}_{\xi}^{(N)}} \geq e^{  N \left( s_0 + \frac{1+ \zeta}{2}  \tfrac{\xi^2}{2\frak{C}_{s_0}}  
+  \mathcal{O}\left(\xi^3\right) \right)} \;.  
  \label{quella} 
\end{equation}
Now observe that taking  $\xi>0$ smaller than some critical value $\xi^{(1)}_*$ which depends upon $s_0$ and $\zeta$, we can 
impose 
\begin{eqnarray}
 \frac{1+\zeta}{2}  \frac{\xi^2}{2\frak{C}_{s_0}}  +  \mathcal{O}\left(\xi^3\right)  \geq   \zeta \frac{\xi^2}{2\frak{C}_{s_0}}  \;, 
\end{eqnarray} 
 hence transforming~(\ref{quella}) into (\ref{cond_numelementiA}). 
We conclude noticing that the value $\xi_*$  is  obtained by taking the smallest among 
$\xi^{(1)}_*$ and the threshold $\epsilon_d-\mathfrak{E}_{s_0}$ needed to ensure that 
 the effective inverse temperature $\beta'$ introduced in Eq.~(\ref{primaprma}) is properly
defined, i.e. $\xi_*:= \min\{\xi^{(1)}_*, \epsilon_d-\mathfrak{E}_{s_0},1 \}$.
$\square$
\\

{\bf Remark 4:} It is worth stressing that  the parameter $\xi_*$ introduced in  {\bf {Lemma}\ref{prop:insiemeA}} is a function of $s_0$, $\zeta$, and, due to 
 the presence of  $\frak{C}_{s_0}$ and $\frak{E}_{s_0}$ in Eqs.~(\ref{questa}) and (\ref{primaprma}), of the Hamiltonian $H$, i.e. $\xi_* = \xi_*(s_0,\zeta, H)$; similar considerations hold  also for $N_*(\xi)$ which, 
besides depending upon $\xi \leq \xi_*$, it is also a function of 
$s_0$, $\zeta$, and $H$, i.e. 
 $N_*(\xi) = N_*(\xi, s_0,\zeta, H)$.
In particular, as discussed in {\bf Remark 5} below, a not necessarily optimal choice of $N_*(\xi,  s_0,\zeta, H)$ which is however 
sufficient to ensure~(\ref{hold}),~is   \begin{eqnarray} \label{defNast} 
N_*(\xi) = {K_* }/{\xi^2} \;,
\end{eqnarray} 
with $K_* = K_*(s_0,\zeta, H)$ a factor that depends upon $s_0$, $\zeta$, and $H$ -- see Eq.~(\ref{defK}) for details.
Notice also that  in our analysis, the choice of the parameter $\zeta$ in  $]0,1[$  is free, however the higher we take it the larger becomes 
$N_*(\xi)$ reducing the range of $N$ for which
Eq.~(\ref{cond_numelementiA}) applies. 
In an effort to reduce the number of parameters, in the remaining of the paper we shall fix such constant equal to
 \begin{eqnarray} 
 \zeta = 2/2.01 = 0.995025\;, \label{fissozeta} 
 \end{eqnarray} 
we stress however that none of the results that follow depend crucially on such a choice. 
\\

{\bf Remark 5:} In order to find an estimation of  $N_*(\xi, s_0,H)$ so that Eq.~(\ref{hold}) holds for all $N\geq N_*(\xi, s_0,H)$,
let us rewrite such inequality as
\begin{eqnarray} \label{hold1} 
 \frac{\ln(  1 - e^{-x}) }{x} \geq -  
  \left(\frac{1-\zeta}{4}\right) \frac{ \epsilon^2_d}{\frak{C}_{s_0}}\;, 
\end{eqnarray}  
with $x:= \tfrac{2N\xi^2}{\epsilon^2_d}$.
Notice then that for 
\begin{eqnarray} x\geq \ln(\frac{e}{e-1})\;, \label{cond111} \end{eqnarray} we can write 
\begin{eqnarray} \label{hold2} 
 \frac{\ln(  1 - e^{-x}) }{x} \geq -\frac{1}{x}  \;,
\end{eqnarray}  
whose right-hand-side is larger than the right-hand-side of~(\ref{hold1}) 
for 
\begin{eqnarray}
x\geq \left(\frac{4}{1-\zeta}\right)
\frac{\frak{C}_{s_0}}{ \epsilon^2_d}  \label{cond222} \;. 
\end{eqnarray} 
Therefore enforcing $x$ to fulfil both (\ref{cond111}) and (\ref{cond222}), i.e. imposing
\begin{eqnarray}
N\geq \frac{K_*(s_0,\zeta,H) }{\xi^2} \;, 
\end{eqnarray} 
with 
\begin{equation}\label{defK}  K_*(s_0,\zeta,H):=\frac{1}{2}  \max\left\{ {\epsilon_d^2} \ln(\frac{e}{e-1}), \frac{4{\frak{C}_{s_0}}}{1-\zeta}  \right\}\;, 
\end{equation} 
we can ensure that (\ref{hold1}) (i.e. (\ref{hold})) applies, hence proving Eq.~(\ref{defNast}). 
As mentioned in the main text this choice for $N_*(\xi, s_0,\zeta,H)$ is arguably not optimal as it
relay on the correct but drastic simplification~(\ref{hold2}).

\section{Generalization of  $\rho^{(N)}_{{\bf B}}$  to non-uniform partitions} \label{APPGENA} 

Here we generalize the construction of Sec.~\ref{sec:method1_stato} to the case in which the partition
${\mathfrak{P}}$ is not necessarily uniform. 
Also in this case we introduce $\rho^{(N)}_{{\bf B}}$ via the expression~(\ref{targetB}) with the matrices
$L^2 \times L^2$ matrices $A^{ij}_ {\bf{B}}$ expressed as in Eq.~(\ref{DEFAB}). In this case however we observe that for $i\neq j$, 
depending on the cardinalities of 
the  associated subsets ${\mathfrak{K}}_{\ell^{[i]}}$  and ${\mathfrak{K}}_{\ell^{[j]}}$, the indexes $k^{[i]}$ and 
$k^{[j]}$ can run over sets of different sizes:  the rectangular matrix
$\delta_{k^{[i]},k^{[j]}}$ appearing in~(\ref{DEFAB}) should hence be interpret as the  natural generalization of 
of
 Kronecker delta which take same values of the latter on the common subsets of the indices and which is zero everywhere else; this prescription ensures that all the identities from Eqs.~(\ref{rhoN_method1}) to 
 (\ref{identita0}) still hold. 
 In deriving the equivalent formula of (\ref{rhoN_method11}) we need however some extra precaution.
 Again the problem is related to the fact that if 
 the partition~${\mathfrak{P}}$ is not uniform then the subsets 
  ${\mathfrak{K}}_{\ell}$ have different cardinalities. 
 For compensate for this fact,  we extend these sets adding extra zero elements to push their effective cardinality to the maximum value $\#_{\max}$; formally this is obtained by replacing ${{\mathfrak{K}}}_{\ell}$ with the new
  set  
  \begin{eqnarray}\label{notazionenew} 
 \tilde{{\mathfrak{K}}}_{\ell}:= \{ \tilde{\lambda}_{k,\ell}; k=1,\cdots, \#_{\max} \}\;,\end{eqnarray}
 with 
 \begin{eqnarray} 
 \tilde{\lambda}_{k,\ell} :=\left\{ 
\begin{array}{ll} 
{\lambda}_{k,\ell} \;, &  \quad \forall k=1,\cdots, \# {\mathfrak{K}}_{\ell}\;, \\ \\
0 \;, & \quad  \forall k= \#{{\mathfrak{K}}}_{\ell}+1, \cdots, \#_{\max}\;. 
\end{array} 
\right. 
 \end{eqnarray} 
 Similarly we define a new set of orthonormal  single-site vectors $\{  |\tilde{\lambda}_{k,\ell} \rangle ; 
 k=1,\cdots, \#_{\max}\}$ with the prescription that the first  $\# {\mathfrak{K}}_{\ell}$ elements 
 fulfil the condition 
  \begin{eqnarray} 
 |\tilde{\lambda}_{k,\ell} \rangle = 
|{\lambda}_{k,\ell}\rangle \;,\qquad \forall k=1,\cdots, \# {\mathfrak{K}}_{\ell}\;, \end{eqnarray} 
while the remaining $\#_{\max}-{\mathfrak{K}}_{\ell}$ can be chosen freely. 
 Given then an $N$-uple $\vec{k}= \left( k_{1} , k_{2} , \dots k_{N} \right)$ where each component
 can now assume up to $\#_{\max}$ distinct values, we 
introduce the quantities 
 \begin{eqnarray}
\tilde{\lambda}^{(N)}_{\vec{k}, \ell} &:=& \prod_{a=1}^{N} \tilde{\lambda}_{k_{a
},\ell} \;,
\end{eqnarray} 
 and
 the $N$ site states  
\begin{eqnarray}
\ket{\tilde{\lambda}^{(N)}_{\vec{k}, \ell}}  &:=& \ket{ \tilde{\lambda}_{k_{1
},\ell} \; \tilde{\lambda}_{k_{2
},\ell}\; \dots \; \tilde{\lambda}_{k_{N
},\ell} }    \;.
\end{eqnarray} 
Notice that the $\tilde{\lambda}^{(N)}_{\vec{k}, \ell}$ 
 correspond to the positive terms of Eq.~(\ref{autovalori_method1}) if 
$k_{a}\leq \#{\mathfrak{K}}_{\ell}$ for all $a=1,\cdots, N$, and are instead equal to zero otherwise;
notice also that for each chosen $\vec{k}= \left( k_{1} , k_{2} , \dots k_{N} \right)$ we can ensure that
there exists at least one value of $\ell$ such that $\tilde{\lambda}^{(N)}_{\vec{k}, \ell}>0$
(this follows from the fact that 
  since $\#_{\max}$ is the greatest of all $\# {\mathfrak{K}}_{\ell}$, 
there is at least one value of $\ell$ for which $\tilde{{\mathfrak{K}}}_{\ell}={\mathfrak{K}}_{\ell}$): 
therefore we can conclude that the terms
\begin{equation}
\tilde{\Lambda}^{(N)}_{\vec{k}} := \sum_{\ell=1}^{L} 
\frac{\tilde{\lambda}^{(N)}_{\vec{k}, \ell} }{\mathscr{S}_\ell^{N-1} } 
 \;,
\label{autovalori_method1asdfasdf}
\end{equation}
are not null for all $\vec{k}$.

With the help of these definitions  can now replace~(\ref{rhoN_method11}) and (\ref{NORM}) with 
 \begin{eqnarray}
\rho^{(N)}_{{\bf B}} \label{rhoN_method119}&=&   \sum_{\vec{k}}
\sum_{\ell,\ell'=1}^L   \sqrt{
	\tfrac{\tilde{\lambda}^{(N)}_{\vec{k}, \ell}}{\mathscr{S}_{\ell}^{N-1}} }\sqrt{\tfrac{\tilde{\lambda}^{(N)}_{\vec{k}, \ell'}}{\mathscr{S}_{\ell'}^{N-1}}
} \ket{ \tilde{\lambda}^{(N)}_{\vec{k}, \ell}}\bra{\tilde{\lambda}^{(N)}_{\vec{k}, \ell'}} \nonumber \\
&=& \sum_{\vec{k}} \tilde{\Lambda}_{\vec{k}}^{(N)} \;  \label{NORM22} 
|\tilde{\Psi}^{(N)}_{\vec{k}}\rangle \langle \tilde{\Psi}^{(N)}_{\vec{k}}| \;, 
\end{eqnarray}
with 
\begin{equation} 
|\tilde{\Psi}^{(N)}_{\vec{k}}\rangle  :=\tfrac{1}{\sqrt{\tilde{\Lambda}_{\vec{k}}^{(N)}}}  \sum_{\ell=1}^{L} 
\sqrt{
	\tfrac{\tilde{\lambda}^{(N)}_{\vec{k}, \ell}}{\mathscr{S}_{\ell}^{N-1}} }\; 
 \ket{ \tilde{\lambda}^{(N)}_{\vec{k}, \ell}  } \;, 
\end{equation} 
being orthonormal elements of $\mathfrak{S}^{(N)}$. 
It is worth stressing that also in this case the state
 (\ref{NORM22}) is 
properly normalized thanks to the fact that 
\begin{eqnarray} \nonumber 
\sum_{\vec{k}} \tilde{\Lambda}_{\vec{k}}^{(N)}  &=&  \sum_{\vec{k}}
\sum_{\ell=1}^{L} \mathscr{S}_\ell^{1-N} \left(\prod_{a=1}^{N} \tilde{\lambda}_{k_{a
},\ell}\right) 
\\ \nonumber 
&=&\sum_{\ell=1}^{L} \mathscr{S}_\ell^{1-N} \left(\sum_{k=1}^{\#_{\max}} \tilde{\lambda}_{k,\ell}\right)^N \nonumber \\ 
&=&\sum_{\ell=1}^{L} \mathscr{S}_\ell^{1-N} \left(\sum_{k=1}^{\#{\mathfrak{K}}_{\ell}} \lambda_{k,\ell}\right)^N = \sum_{\ell=1}^{L} \mathscr{S}_\ell = 1\;.  \nonumber 
\end{eqnarray} 
As anticipated in the main text,
while the BLR $\mathfrak{br} [\rho^{(N)}_{{\bf B}}]$  is upper bounded by $L^2$ 
 the rank of the state $\rho^{(N)}_{{\bf B}}$ is given by the total number of 
 $N$-uple $\vec{k}$, leading to Eq.~(\ref{rankrhob1}).

%


\begin{thebibliography}{29}%
	\makeatletter
	\providecommand \@ifxundefined [1]{%
		\@ifx{#1\undefined}
	}%
	\providecommand \@ifnum [1]{%
		\ifnum #1\expandafter \@firstoftwo
		\else \expandafter \@secondoftwo
		\fi
	}%
	\providecommand \@ifx [1]{%
		\ifx #1\expandafter \@firstoftwo
		\else \expandafter \@secondoftwo
		\fi
	}%
	\providecommand \natexlab [1]{#1}%
	\providecommand \enquote  [1]{``#1''}%
	\providecommand \bibnamefont  [1]{#1}%
	\providecommand \bibfnamefont [1]{#1}%
	\providecommand \citenamefont [1]{#1}%
	\providecommand \href@noop [0]{\@secondoftwo}%
	\providecommand \href [0]{\begingroup \@sanitize@url \@href}%
	\providecommand \@href[1]{\@@startlink{#1}\@@href}%
	\providecommand \@@href[1]{\endgroup#1\@@endlink}%
	\providecommand \@sanitize@url [0]{\catcode `\\12\catcode `\$12\catcode
		`\&12\catcode `\#12\catcode `\^12\catcode `\_12\catcode `\%12\relax}%
	\providecommand \@@startlink[1]{}%
	\providecommand \@@endlink[0]{}%
	\providecommand \url  [0]{\begingroup\@sanitize@url \@url }%
	\providecommand \@url [1]{\endgroup\@href {#1}{\urlprefix }}%
	\providecommand \urlprefix  [0]{URL }%
	\providecommand \Eprint [0]{\href }%
	\providecommand \doibase [0]{http://dx.doi.org/}%
	\providecommand \selectlanguage [0]{\@gobble}%
	\providecommand \bibinfo  [0]{\@secondoftwo}%
	\providecommand \bibfield  [0]{\@secondoftwo}%
	\providecommand \translation [1]{[#1]}%
	\providecommand \BibitemOpen [0]{}%
	\providecommand \bibitemStop [0]{}%
	\providecommand \bibitemNoStop [0]{.\EOS\space}%
	\providecommand \EOS [0]{\spacefactor3000\relax}%
	\providecommand \BibitemShut  [1]{\csname bibitem#1\endcsname}%
	\let\auto@bib@innerbib\@empty
	\bibitem [{\citenamefont {Thomson}(1852)}]{Thomson1852}%
	\BibitemOpen
	\bibfield  {author} {\bibinfo {author} {\bibfnamefont {W.}~\bibnamefont
			{Thomson}},\ }\href {\doibase 10.1080/14786445208647064} {\bibfield
		{journal} {\bibinfo  {journal} {The London, Edinburgh, and Dublin
				Philosophical Magazine and Journal of Science}\ }\textbf {\bibinfo {volume}
			{4}},\ \bibinfo {pages} {8} (\bibinfo {year} {1852})}\BibitemShut {NoStop}%
	\bibitem [{\citenamefont {Fermi}\ \emph {et~al.}(1955)\citenamefont {Fermi},
		\citenamefont {Pasta}, \citenamefont {Ulam},\ and\ \citenamefont
		{Tsingou}}]{Fermi1955}%
	\BibitemOpen
	\bibfield  {author} {\bibinfo {author} {\bibfnamefont {E.}~\bibnamefont
			{Fermi}}, \bibinfo {author} {\bibfnamefont {P.}~\bibnamefont {Pasta}},
		\bibinfo {author} {\bibfnamefont {S.}~\bibnamefont {Ulam}}, \ and\ \bibinfo
		{author} {\bibfnamefont {M.}~\bibnamefont {Tsingou}},\ }\href {\doibase
		10.2172/4376203} {\emph {\bibinfo {title} {Studies of the nonlinear
				problems}}},\ \bibinfo {type} {Tech. Rep.}\ (\bibinfo {year}
	{1955})\BibitemShut {NoStop}%
	\bibitem [{\citenamefont {Allahverdyan}\ \emph {et~al.}(2004)\citenamefont
		{Allahverdyan}, \citenamefont {Balian},\ and\ \citenamefont
		{Nieuwenhuizen}}]{Allahverdyan2004}%
	\BibitemOpen
	\bibfield  {author} {\bibinfo {author} {\bibfnamefont {A.~E.}\ \bibnamefont
			{Allahverdyan}}, \bibinfo {author} {\bibfnamefont {R.}~\bibnamefont
			{Balian}}, \ and\ \bibinfo {author} {\bibfnamefont {T.~M.}\ \bibnamefont
			{Nieuwenhuizen}},\ }\href {\doibase 10.1209/epl/i2004-10101-2} {\bibfield
		{journal} {\bibinfo  {journal} {Europhysics Letters ({EPL})}\ }\textbf
		{\bibinfo {volume} {67}},\ \bibinfo {pages} {565} (\bibinfo {year}
		{2004})}\BibitemShut {NoStop}%
	\bibitem [{\citenamefont {Niedenzu}\ \emph {et~al.}(2019)\citenamefont
		{Niedenzu}, \citenamefont {Huber},\ and\ \citenamefont
		{Boukobza}}]{Niedenzu2019}%
	\BibitemOpen
	\bibfield  {author} {\bibinfo {author} {\bibfnamefont {W.}~\bibnamefont
			{Niedenzu}}, \bibinfo {author} {\bibfnamefont {M.}~\bibnamefont {Huber}}, \
		and\ \bibinfo {author} {\bibfnamefont {E.}~\bibnamefont {Boukobza}},\ }\href
	{\doibase 10.22331/q-2019-10-14-195} {\bibfield  {journal} {\bibinfo
			{journal} {{Quantum}}\ }\textbf {\bibinfo {volume} {3}},\ \bibinfo {pages}
		{195} (\bibinfo {year} {2019})}\BibitemShut {NoStop}%
	\bibitem [{\citenamefont {Alicki}\ and\ \citenamefont
		{Fannes}(2013)}]{AlickiFannes2013}%
	\BibitemOpen
	\bibfield  {author} {\bibinfo {author} {\bibfnamefont {R.}~\bibnamefont
			{Alicki}}\ and\ \bibinfo {author} {\bibfnamefont {M.}~\bibnamefont
			{Fannes}},\ }\href@noop {} {\bibfield  {journal} {\bibinfo  {journal} {Phys.
				Rev. E}\ }\textbf {\bibinfo {volume} {87}},\ \bibinfo {pages} {042123}
		(\bibinfo {year} {2013})}\BibitemShut {NoStop}%
	\bibitem [{\citenamefont {Huber}\ \emph {et~al.}(2015)\citenamefont {Huber},
		\citenamefont {Perarnau-Llobet}, \citenamefont {Hovhannisyan}, \citenamefont
		{Skrzypczyk}, \citenamefont {Kl\"{o}ckl}, \citenamefont {Brunner},\ and\
		\citenamefont {Ac{\'{\i}}n}}]{Huber2015}%
	\BibitemOpen
	\bibfield  {author} {\bibinfo {author} {\bibfnamefont {M.}~\bibnamefont
			{Huber}}, \bibinfo {author} {\bibfnamefont {M.}~\bibnamefont
			{Perarnau-Llobet}}, \bibinfo {author} {\bibfnamefont {K.~V.}\ \bibnamefont
			{Hovhannisyan}}, \bibinfo {author} {\bibfnamefont {P.}~\bibnamefont
			{Skrzypczyk}}, \bibinfo {author} {\bibfnamefont {C.}~\bibnamefont
			{Kl\"{o}ckl}}, \bibinfo {author} {\bibfnamefont {N.}~\bibnamefont {Brunner}},
		\ and\ \bibinfo {author} {\bibfnamefont {A.}~\bibnamefont {Ac{\'{\i}}n}},\
	}\href {\doibase 10.1088/1367-2630/17/6/065008} {\bibfield  {journal}
		{\bibinfo  {journal} {New J. Phys.}\ }\textbf {\bibinfo {volume}
			{17}},\ \bibinfo {pages} {065008} (\bibinfo {year} {2015})}\BibitemShut
	{NoStop}%
	\bibitem [{\citenamefont {Francica}\ \emph {et~al.}(2017)\citenamefont
		{Francica}, \citenamefont {Goold}, \citenamefont {Plastina},\ and\
		\citenamefont {Paternostro}}]{Francica2017}%
	\BibitemOpen
	\bibfield  {author} {\bibinfo {author} {\bibfnamefont {G.}~\bibnamefont
			{Francica}}, \bibinfo {author} {\bibfnamefont {J.}~\bibnamefont {Goold}},
		\bibinfo {author} {\bibfnamefont {F.}~\bibnamefont {Plastina}}, \ and\
		\bibinfo {author} {\bibfnamefont {M.}~\bibnamefont {Paternostro}},\ }\href
	{\doibase 10.1038/s41534-017-0012-8} {\bibfield  {journal} {\bibinfo
			{journal} {npj Quantum Information}\ }\textbf {\bibinfo {volume} {3}}
		(\bibinfo {year} {2017}),\ 10.1038/s41534-017-0012-8}\BibitemShut {NoStop}%
	\bibitem [{\citenamefont {Oppenheim}\ \emph {et~al.}(2002)\citenamefont
		{Oppenheim}, \citenamefont {Horodecki}, \citenamefont {Horodecki},\ and\
		\citenamefont {Horodecki}}]{OPPE2002}%
	\BibitemOpen
	\bibfield  {author} {\bibinfo {author} {\bibfnamefont {J.}~\bibnamefont
			{Oppenheim}}, \bibinfo {author} {\bibfnamefont {M.}~\bibnamefont
			{Horodecki}}, \bibinfo {author} {\bibfnamefont {P.}~\bibnamefont
			{Horodecki}}, \ and\ \bibinfo {author} {\bibfnamefont {R.}~\bibnamefont
			{Horodecki}},\ }\href {\doibase 10.1103/physrevlett.89.180402} {\bibfield
		{journal} {\bibinfo  {journal} {Phys. Rev. Lett.}\ }\textbf {\bibinfo
			{volume} {89}} (\bibinfo {year} {2002}),\
		10.1103/physrevlett.89.180402}\BibitemShut {NoStop}%
	\bibitem [{\citenamefont {Vitagliano}\ \emph {et~al.}(2018)\citenamefont
		{Vitagliano}, \citenamefont {Kl\"{o}ckl}, \citenamefont {Huber},\ and\
		\citenamefont {Friis}}]{VITA2019}%
	\BibitemOpen
	\bibfield  {author} {\bibinfo {author} {\bibfnamefont {G.}~\bibnamefont
			{Vitagliano}}, \bibinfo {author} {\bibfnamefont {C.}~\bibnamefont
			{Kl\"{o}ckl}}, \bibinfo {author} {\bibfnamefont {M.}~\bibnamefont {Huber}}, \
		and\ \bibinfo {author} {\bibfnamefont {N.}~\bibnamefont {Friis}},\ }in\ \href
	{\doibase 10.1007/978-3-319-99046-0_30} {\emph {\bibinfo {booktitle}
			{Fundamental Theories of Physics}}}\ (\bibinfo  {publisher} {Springer
		International Publishing},\ \bibinfo {year} {2018})\ pp.\ \bibinfo {pages}
	{731--750}\BibitemShut {NoStop}%
	\bibitem [{\citenamefont {Goold}\ \emph {et~al.}(2016)\citenamefont {Goold},
		\citenamefont {Huber}, \citenamefont {Riera}, \citenamefont {del Rio},\ and\
		\citenamefont {Skrzypczyk}}]{GOOLD2017}%
	\BibitemOpen
	\bibfield  {author} {\bibinfo {author} {\bibfnamefont {J.}~\bibnamefont
			{Goold}}, \bibinfo {author} {\bibfnamefont {M.}~\bibnamefont {Huber}},
		\bibinfo {author} {\bibfnamefont {A.}~\bibnamefont {Riera}}, \bibinfo
		{author} {\bibfnamefont {L.}~\bibnamefont {del Rio}}, \ and\ \bibinfo
		{author} {\bibfnamefont {P.}~\bibnamefont {Skrzypczyk}},\ }\href {\doibase
		10.1088/1751-8113/49/14/143001} {\bibfield  {journal} {\bibinfo  {journal}
			{J. Phys. A: Math. Th.l}\ }\textbf {\bibinfo
			{volume} {49}},\ \bibinfo {pages} {143001} (\bibinfo {year}
		{2016})}\BibitemShut {NoStop}%
	\bibitem [{\citenamefont {Bera}\ \emph {et~al.}(2017)\citenamefont {Bera},
		\citenamefont {Riera}, \citenamefont {Lewenstein},\ and\ \citenamefont
		{Winter}}]{BERA2017}%
	\BibitemOpen
	\bibfield  {author} {\bibinfo {author} {\bibfnamefont {M.~N.}\ \bibnamefont
			{Bera}}, \bibinfo {author} {\bibfnamefont {A.}~\bibnamefont {Riera}},
		\bibinfo {author} {\bibfnamefont {M.}~\bibnamefont {Lewenstein}}, \ and\
		\bibinfo {author} {\bibfnamefont {A.}~\bibnamefont {Winter}},\ }\href
	{\doibase 10.1038/s41467-017-02370-x} {\bibfield  {journal} {\bibinfo
			{journal} {Nat. Comm.}\ }\textbf {\bibinfo {volume} {8}} (\bibinfo
		{year} {2017}),\ 10.1038/s41467-017-02370-x}\BibitemShut {NoStop}%
	\bibitem [{\citenamefont {Manzano}\ \emph {et~al.}(2018)\citenamefont
		{Manzano}, \citenamefont {Plastina},\ and\ \citenamefont
		{Zambrini}}]{MANZ2019}%
	\BibitemOpen
	\bibfield  {author} {\bibinfo {author} {\bibfnamefont {G.}~\bibnamefont
			{Manzano}}, \bibinfo {author} {\bibfnamefont {F.}~\bibnamefont {Plastina}}, \
		and\ \bibinfo {author} {\bibfnamefont {R.}~\bibnamefont {Zambrini}},\ }\href
	{\doibase 10.1103/physrevlett.121.120602} {\bibfield  {journal} {\bibinfo
			{journal} {Phys. Rev. Lett.}\ }\textbf {\bibinfo {volume} {121}}
		(\bibinfo {year} {2018}),\ 10.1103/physrevlett.121.120602}\BibitemShut
	{NoStop}%
	\bibitem [{\citenamefont {Andolina}\ \emph {et~al.}(2019)\citenamefont
		{Andolina}, \citenamefont {Keck}, \citenamefont {Mari}, \citenamefont
		{Campisi}, \citenamefont {Giovannetti},\ and\ \citenamefont
		{Polini}}]{ANDO2019}%
	\BibitemOpen
	\bibfield  {author} {\bibinfo {author} {\bibfnamefont {G.~M.}\ \bibnamefont
			{Andolina}}, \bibinfo {author} {\bibfnamefont {M.}~\bibnamefont {Keck}},
		\bibinfo {author} {\bibfnamefont {A.}~\bibnamefont {Mari}}, \bibinfo {author}
		{\bibfnamefont {M.}~\bibnamefont {Campisi}}, \bibinfo {author} {\bibfnamefont
			{V.}~\bibnamefont {Giovannetti}}, \ and\ \bibinfo {author} {\bibfnamefont
			{M.}~\bibnamefont {Polini}},\ }\href {\doibase
		10.1103/physrevlett.122.047702} {\bibfield  {journal} {\bibinfo  {journal}
			{Phys. Rev. Lett.}\ }\textbf {\bibinfo {volume} {122}} (\bibinfo
		{year} {2019}),\ 10.1103/physrevlett.122.047702}\BibitemShut {NoStop}%
	\bibitem [{\citenamefont {Verstraete}\ \emph {et~al.}(2004)\citenamefont
		{Verstraete}, \citenamefont {Garc\'ia-Ripoll},\ and\ \citenamefont
		{Cirac}}]{Cirac2004MPO}%
	\BibitemOpen
	\bibfield  {author} {\bibinfo {author} {\bibfnamefont {F.}~\bibnamefont
			{Verstraete}}, \bibinfo {author} {\bibfnamefont {J.~J.}\ \bibnamefont
			{Garc\'ia-Ripoll}}, \ and\ \bibinfo {author} {\bibfnamefont {J.~I.}\
			\bibnamefont {Cirac}},\ }\href {\doibase 10.1103/PhysRevLett.93.207204}
	{\bibfield  {journal} {\bibinfo  {journal} {Phys. Rev. Lett.}\
		}\textbf {\bibinfo {volume} {93}},\ \bibinfo {pages} {207204} (\bibinfo
		{year} {2004})}\BibitemShut {NoStop}%
	\bibitem [{\citenamefont {Baumgratz}\ \emph {et~al.}(2014)\citenamefont
		{Baumgratz}, \citenamefont {Cramer},\ and\ \citenamefont
		{Plenio}}]{Baumgratz2014}%
	\BibitemOpen
	\bibfield  {author} {\bibinfo {author} {\bibfnamefont {T.}~\bibnamefont
			{Baumgratz}}, \bibinfo {author} {\bibfnamefont {M.}~\bibnamefont {Cramer}}, \
		and\ \bibinfo {author} {\bibfnamefont {M.~B.}\ \bibnamefont {Plenio}},\
	}\href {\doibase 10.1103/PhysRevLett.113.140401} {\bibfield  {journal}
		{\bibinfo  {journal} {Phys. Rev. Lett.}\ }\textbf {\bibinfo {volume} {113}},\
		\bibinfo {pages} {140401} (\bibinfo {year} {2014})}\BibitemShut {NoStop}%
	\bibitem [{\citenamefont {Francica}\ \emph {et~al.}(2020)\citenamefont
		{Francica}, \citenamefont {Binder}, \citenamefont {Guarnieri}, \citenamefont
		{Mitchison}, \citenamefont {Goold},\ and\ \citenamefont
		{Plastina}}]{Francica2020}%
	\BibitemOpen
	\bibfield  {author} {\bibinfo {author} {\bibfnamefont {G.}~\bibnamefont
			{Francica}}, \bibinfo {author} {\bibfnamefont {F.~C.}\ \bibnamefont
			{Binder}}, \bibinfo {author} {\bibfnamefont {G.}~\bibnamefont {Guarnieri}},
		\bibinfo {author} {\bibfnamefont {M.~T.}\ \bibnamefont {Mitchison}}, \bibinfo
		{author} {\bibfnamefont {J.}~\bibnamefont {Goold}}, \ and\ \bibinfo {author}
		{\bibfnamefont {F.}~\bibnamefont {Plastina}},\ }\href {\doibase
		10.1103/PhysRevLett.125.180603} {\bibfield  {journal} {\bibinfo  {journal}
			{Phys. Rev. Lett.}\ }\textbf {\bibinfo {volume} {125}},\ \bibinfo {pages}
		{180603} (\bibinfo {year} {2020})}\BibitemShut {NoStop}%
	\bibitem [{\citenamefont {\c{C}akmak}(2020)}]{Cakmak2020}%
	\BibitemOpen
	\bibfield  {author} {\bibinfo {author} {\bibfnamefont {B.}~\bibnamefont
			{\c{C}akmak}},\ }\href {\doibase 10.1103/PhysRevE.102.042111} {\bibfield
		{journal} {\bibinfo  {journal} {Phys. Rev. E}\ }\textbf {\bibinfo {volume}
			{102}},\ \bibinfo {pages} {042111} (\bibinfo {year} {2020})}\BibitemShut
	{NoStop}%
	\bibitem [{\citenamefont {Touil}\ \emph {et~al.}(2021)\citenamefont {Touil},
		\citenamefont {\c{C}akmak},\ and\ \citenamefont {Deffner}}]{Touil2021}%
	\BibitemOpen
	\bibfield  {author} {\bibinfo {author} {\bibfnamefont {A.}~\bibnamefont
			{Touil}}, \bibinfo {author} {\bibfnamefont {B.}~\bibnamefont {\c{C}akmak}}, \
		and\ \bibinfo {author} {\bibfnamefont {S.}~\bibnamefont {Deffner}},\
	}\href@noop {} {\enquote {\bibinfo {title} {Second law of thermodynamics for
				quantum correlations},}\ } (\bibinfo {year} {2021}),\ \Eprint
	{http://arxiv.org/abs/arXiv:2102.13606} {arXiv:2102.13606} \BibitemShut
	{NoStop}%
	\bibitem [{\citenamefont {Kliesch}\ \emph {et~al.}(2014)\citenamefont
		{Kliesch}, \citenamefont {Gross},\ and\ \citenamefont
		{Eisert}}]{Kliesch2014}%
	\BibitemOpen
	\bibfield  {author} {\bibinfo {author} {\bibfnamefont {M.}~\bibnamefont
			{Kliesch}}, \bibinfo {author} {\bibfnamefont {D.}~\bibnamefont {Gross}}, \
		and\ \bibinfo {author} {\bibfnamefont {J.}~\bibnamefont {Eisert}},\ }\href
	{\doibase 10.1103/physrevlett.113.160503} {\bibfield  {journal} {\bibinfo
			{journal} {Phys. Rev. Lett.}\ }\textbf {\bibinfo {volume} {113}}
		(\bibinfo {year} {2014}),\ 10.1103/physrevlett.113.160503}\BibitemShut
	{NoStop}%
	\bibitem [{\citenamefont {Navascues}\ and\ \citenamefont
		{Vertesi}(2018)}]{Navascues2018}%
	\BibitemOpen
	\bibfield  {author} {\bibinfo {author} {\bibfnamefont {M.}~\bibnamefont
			{Navascues}}\ and\ \bibinfo {author} {\bibfnamefont {T.}~\bibnamefont
			{Vertesi}},\ }\href {\doibase 10.22331/q-2018-01-31-50} {\bibfield  {journal}
		{\bibinfo  {journal} {Quantum}\ }\textbf {\bibinfo {volume} {2}},\ \bibinfo
		{pages} {50} (\bibinfo {year} {2018})}\BibitemShut {NoStop}%
	\bibitem [{\citenamefont {Parker}\ \emph {et~al.}(2020)\citenamefont {Parker},
		\citenamefont {Cao},\ and\ \citenamefont {Zaletel}}]{Parker2020}%
	\BibitemOpen
	\bibfield  {author} {\bibinfo {author} {\bibfnamefont {D.~E.}\ \bibnamefont
			{Parker}}, \bibinfo {author} {\bibfnamefont {X.}~\bibnamefont {Cao}}, \ and\
		\bibinfo {author} {\bibfnamefont {M.~P.}\ \bibnamefont {Zaletel}},\ }\href
	{\doibase 10.1103/physrevb.102.035147} {\bibfield  {journal} {\bibinfo
			{journal} {Phys. Rev. B}\ }\textbf {\bibinfo {volume} {102}} (\bibinfo
		{year} {2020}),\ 10.1103/physrevb.102.035147}\BibitemShut {NoStop}%
	\bibitem [{\citenamefont {Perarnau-Llobet}\ and\ \citenamefont
		{Uzdin}(2019)}]{PerarnauLlobet2019}%
	\BibitemOpen
	\bibfield  {author} {\bibinfo {author} {\bibfnamefont {M.}~\bibnamefont
			{Perarnau-Llobet}}\ and\ \bibinfo {author} {\bibfnamefont {R.}~\bibnamefont
			{Uzdin}},\ }\href {\doibase 10.1088/1367-2630/ab36a9} {\bibfield  {journal}
		{\bibinfo  {journal} {New J. Phys.}\ }\textbf {\bibinfo {volume}
			{21}},\ \bibinfo {pages} {083023} (\bibinfo {year} {2019})}\BibitemShut
	{NoStop}%
	\bibitem [{\citenamefont {Correa}\ \emph {et~al.}(2015)\citenamefont {Correa},
		\citenamefont {Mehboudi}, \citenamefont {Adesso},\ and\ \citenamefont
		{Sanpera}}]{Correa2015}%
	\BibitemOpen
	\bibfield  {author} {\bibinfo {author} {\bibfnamefont {L.~A.}\ \bibnamefont
			{Correa}}, \bibinfo {author} {\bibfnamefont {M.}~\bibnamefont {Mehboudi}},
		\bibinfo {author} {\bibfnamefont {G.}~\bibnamefont {Adesso}}, \ and\ \bibinfo
		{author} {\bibfnamefont {A.}~\bibnamefont {Sanpera}},\ }\href {\doibase
		10.1103/physrevlett.114.220405} {\bibfield  {journal} {\bibinfo  {journal}
			{Phys. Rev. Lett.}\ }\textbf {\bibinfo {volume} {114}} (\bibinfo
		{year} {2015}),\ 10.1103/physrevlett.114.220405}\BibitemShut {NoStop}%
	\bibitem [{\citenamefont {Mirsky}(1975)}]{Mirsky1975}%
	\BibitemOpen
	\bibfield  {author} {\bibinfo {author} {\bibfnamefont {L.}~\bibnamefont
			{Mirsky}},\ }\href {\doibase 10.1007/bf01647331} {\bibfield  {journal}
		{\bibinfo  {journal} {Monatshefte f\"{u}r Mathematik}\ }\textbf {\bibinfo
			{volume} {79}},\ \bibinfo {pages} {303} (\bibinfo {year} {1975})}\BibitemShut
	{NoStop}%
	\bibitem [{\citenamefont {Alimuddin}\ \emph {et~al.}(2020)\citenamefont
		{Alimuddin}, \citenamefont {Guha},\ and\ \citenamefont
		{Parashar}}]{Alimuddin2020}%
	\BibitemOpen
	\bibfield  {author} {\bibinfo {author} {\bibfnamefont {M.}~\bibnamefont
			{Alimuddin}}, \bibinfo {author} {\bibfnamefont {T.}~\bibnamefont {Guha}}, \
		and\ \bibinfo {author} {\bibfnamefont {P.}~\bibnamefont {Parashar}},\ }\href
	{\doibase 10.1103/physreve.102.012145} {\bibfield  {journal} {\bibinfo
			{journal} {Phys. Rev. E}\ }\textbf {\bibinfo {volume} {102}} (\bibinfo
		{year} {2020}),\ 10.1103/physreve.102.012145}\BibitemShut {NoStop}%
	\bibitem [{\citenamefont {Schindler}\ \emph {et~al.}(2020)\citenamefont
		{Schindler}, \citenamefont {{\v{S}}afr{\'{a}}nek},\ and\ \citenamefont
		{Aguirre}}]{Schindler2020}%
	\BibitemOpen
	\bibfield  {author} {\bibinfo {author} {\bibfnamefont {J.}~\bibnamefont
			{Schindler}}, \bibinfo {author} {\bibfnamefont {D.}~\bibnamefont
			{{\v{S}}afr{\'{a}}nek}}, \ and\ \bibinfo {author} {\bibfnamefont
			{A.}~\bibnamefont {Aguirre}},\ }\href {\doibase 10.1103/physreva.102.052407}
	{\bibfield  {journal} {\bibinfo  {journal} {Phys. Rev. A}\ }\textbf
		{\bibinfo {volume} {102}} (\bibinfo {year} {2020}),\
		10.1103/physreva.102.052407}\BibitemShut {NoStop}%
	\bibitem [{\citenamefont {Rolandi}\ and\ \citenamefont
		{Wilming}(2020)}]{Rolandi2020}%
	\BibitemOpen
	\bibfield  {author} {\bibinfo {author} {\bibfnamefont {A.}~\bibnamefont
			{Rolandi}}\ and\ \bibinfo {author} {\bibfnamefont {H.}~\bibnamefont
			{Wilming}},\ }\href@noop {} {\enquote {\bibinfo {title} {Extensive R\'enyi
				entropies in matrix product states},}\ } (\bibinfo {year} {2020}),\ \Eprint
	{http://arxiv.org/abs/arXiv:2008.11764} {arXiv:2008.11764} \BibitemShut
	{NoStop}%
	\bibitem [{\citenamefont {Chernoff}(1952)}]{Chernoff1952}%
	\BibitemOpen
	\bibfield  {author} {\bibinfo {author} {\bibfnamefont {H.}~\bibnamefont
			{Chernoff}},\ }\href {\doibase 10.1214/aoms/1177729330} {\bibfield  {journal}
		{\bibinfo  {journal} {The Annals of Mathematical Statistics}\ }\textbf
		{\bibinfo {volume} {23}},\ \bibinfo {pages} {493} (\bibinfo {year}
		{1952})}\BibitemShut {NoStop}%
	\bibitem [{\citenamefont {Hoeffding}(1963)}]{Hoeffding1963}%
	\BibitemOpen
	\bibfield  {author} {\bibinfo {author} {\bibfnamefont {W.}~\bibnamefont
			{Hoeffding}},\ }\href {\doibase 10.1080/01621459.1963.10500830} {\bibfield
		{journal} {\bibinfo  {journal} {J. Am. Stat. Assoc.}\ }\textbf {\bibinfo {volume} {58}},\ \bibinfo {pages} {13}
		(\bibinfo {year} {1963})}\BibitemShut {NoStop}%
\end{thebibliography}

\end{document}